\RequirePackage{fix-cm}
\documentclass[twocolumn,epjc3]{svjour3}  
\smartqed  
\RequirePackage{graphicx}
\journalname{Eur. Phys. J. C}
\usepackage{amssymb}
\newcommand{\be}{\begin{eqnarray}}
\newcommand{\ee}{\end{eqnarray}}
\newcommand{\bfq}{{\bf q}_{\perp}}

\newcommand{\bfk}{{\bf k}_{\perp}}
\newcommand{\bfkpr}{{\bf k}_{\perp}^\prime}

\newcommand{\bfb}{{\bf b}_{\perp}}

\usepackage{color}

\begin{document}
\title{Generalized parton distributions and transverse  densities  in a light-front quark-diquark model for the nucleons}
\author{ Chandan Mondal \and Dipankar Chakrabarti 
}                     
%
%
\institute{Department of Physics, Indian Institute of Technology Kanpur, Kanpur-208016, India.}
\date{Received: date / Revised version: date}
%
\maketitle



\begin{abstract}
We present a study of the generalized parton distributions (GPDs) for the  quarks in a proton in both momentum and position spaces using the light-front wave functions (LFWFs) of a quark-diquark model for the nucleon predicted by the soft-wall model of AdS/ QCD. The results are compared with  the soft-wall AdS/ QCD model of proton GPDs for zero skewness. We also calculate the GPDs for nonzero skewness. 
We observe that the GPDs have a diffraction pattern in longitudinal position space, as seen before in  other models.  Then we  present a comparative study of the nucleon charge and anomalous magnetization densities in the transverse plane. Flavor decompositions of the form factors and transverse densities are also discussed.

\end{abstract}


\vskip0.2in
\noindent
\section{Introduction}\label{intro}

Hadronic structure and their properties being nonperturbative in nature are always very difficult to  evaluate from QCD first principle and there have been numerous  attempts to gain insight into hadrons by studying QCD inspired models.
The quark-diquark  model,  where a nucleon is considered to be a bound state of a single quark and a scalar or vector diquark state,  is proven to reproduce many interesting properties of nucleons and has been extensively used to investigate the proton structure.  Recently, a light front quark-diquark model for the nucleons has been proposed in Ref. \cite{Gut},
 where the light front wave functions are modeled by the wave functions obtained  from a soft-wall model in light front AdS/QCD. The light front wave functions (LFWF) are derived by matching the electromagnetic form factors of hadrons in the light front QCD and soft-wall model of AdS/QCD. The model is consistent with Drell-Yan-West relation relating the high $Q^2$ behavior of the nucleon form factors and  the large $x$ behavior of the structure functions.
 Recently, LFWF of baryon and the light baryon spectrum have been described by extending the superconformal quantum mechanics to the light front and embedding it in AdS space\cite{SQM}. The LFWF  for  the rho meson in AdS/QCD has been successfully applied to predict the diffractive rho meson electroproduction\cite{rho}.
  
 In this paper, we study the proton structure and  evaluate the  Generalized Parton Distributions(GPDs),  transverse charge and magnetization densities in the light front quark-diquark model.   
Contrary to ordinary parton distribution function, GPDs are functions of three variables, namely, longitudinal momentum faction $x$ of the quark or gluon,   square of the total momentum transferred ($t$)  and the skewness $\zeta$, which represents the longitudinal momentum transferred in the process and provide interesting information about the spin and orbital angular momentum of the constituents,  as well as the spatial structure, of the nucleons(see \cite{rev} for reviews on GPDs). 
The GPDs appear in the exclusive processes like deeply virtual Compton scattering (DVCS) or vector meson productions and 
they reduce to the ordinary parton distributions in the forward limit.  Their first moments are related to the form factors and 
 the second moment of the of the 
sum of the GPDs are related to the angular momentum by a sum rule proposed by Ji\cite{ji97}. Being off-forward matrix elements, the GPDs have no probabilistic interpretation.
But for zero skewness, the  Fourier transforms of the GPDs  with respect to the transverse momentum transfer ($\Delta_\perp$) give  the impact parameter dependent GPDs which satisfy the positivity condition and can be interpreted as distribution functions \cite{burk}. The transverse impact parameter dependent GPDs  provide us with  the information about partonic distributions in the impact parameter or the transverse position space for a given longitudinal momentum ($x$).  The impact parameter $b_\perp$ gives the separation of the struck quark from the center of momentum.  In parallel to the efforts to understand the GPDs by theoretical modeling, different experiments are also measuring deeply virtual Compton scattering and deeply virtual meson production to gain insight  and experimentally constrain the GPDs\cite{expts}.

 We   evaluate the proton GPDs for both zero and nonzero skewness and compare with the results in a soft-wall AdS/QCD model \cite{CM1}(for hard-wall and soft-wall AdS/QCD models of hadrons, see \cite{BT2,AC}) . For zero skewness, the GPDs are investigated in the impact parameter or transverse position space. The LF diquark results for GPD $H(x,b_\perp)$ for u-quark is almost the same as AdS/QCD results whereas there is a little difference for d-quark. But the LF diquark model results for $E(x,b_\perp)$ for both $u$ and $d$ quarks are different from AdS/QCD results.  For nonzero skewness, the GPDs in longitudinal impact parameter space show a diffraction pattern. 
It is interesting to note  that similar diffraction patterns were observed in simple QED model for DVCS amplitude\cite{BDHAV} and  GPDs\cite{CMM1} and in a phenomenological model of proton GPDs\cite{CMM2}.

 Electric charge and magnetization densities in the transverse plane also provide  insights into the structure of nucleons.  The charge and magnetization densities in the transverse plane are defined as the Fourier transform of the electromagnetic form factors.  The form factors involve initial and final states with different momenta and the three dimensional Fourier transforms cannot be interpreted as densities whereas the transverse densities(i.e., Fourier transformed only for transverse momenta) defined at fixed light front time are free from this difficulty and have proper density interpretation\cite{miller,venkat}. We calculate the transverse charge and anomalous magnetization densities for both proton and neutron  in the light-front diquark model and compare with the two different global parameterizations proposed by Kelly \cite{kelly04} and Bradford $et$ $al$ \cite{brad}. We present results for both unpolarized and transversely polarized nucleons. We also present a comparison with the AdS/QCD results\
cite{CM3} for the transverse charge and 
magnetization densities.

The paper is organized as follows. In Section \ref{model}, we give a brief introductions about the nucleon LFWFs of quark-diquark model as well as the electromagnetic flavor form factors. We show the results for proton GPDs of $u$ and $d$ quarks in momentum space in Section \ref{gpd}. Then we discuss the GPDs in the transverse as well as the longitudinal impact parameter space in Sections \ref{impact} and \ref{long_impact}. We present the results of the charge and anomalous magnetization densities in the transverse plane in section \ref{density}. Finally we provide a brief summary and conclusions in Section \ref{con}.   For GPDs with nonzero skewness, we present a comparison of the quark-diquark model results with a Double Distribution(DD) model in the appendix.

\vskip0.2in
\noindent
\section {Light-front quark-diquark model for the nucleon}\label{model}
In quark-scalar diquark model, the three valence quarks of nucleon are considered as an effectively composite system composed of a fermion and a neutral scalar bound state of diquark based on one loop quantum fluctuations. In the light-cone formalism for a spin $\frac{1}{2}$ composite system the Dirac and Pauli form factors $F_1(q^2)$ and $F_2(q^2)$ are identified to the helicity-conserving and helicity-flip matrix elements of the $J^+$ current \cite{BD}
\be
\langle P+q, \uparrow|\frac{J^+(0)}{2P^+}|P, \uparrow\rangle &=&F_1(q^2),\\
\langle P+q, \uparrow|\frac{J^+(0)}{2P^+}|P, \downarrow\rangle &=&-(q^1-iq^2)\frac{F_2(q^2)}{2M_n},
\ee
here $M_n$ is the nucleon mass. Writing proton as a two particle bound state of a quark and a scalar diquark in the light front quark-diquark model, the Dirac and Pauli form factors for the quarks  can be written in the light-front representation \cite{BD,BHMI} as
\be
&&F_1^q(Q^2) = \int_0^1dx \int \frac{d^2\bfk}{16\pi^3}~\bigg[\psi_{+q}^{+*}(x,\bfk')\psi_{+q}^+(x,\bfk) \nonumber\\
&&+\psi_{-q}^{+*}(x,\bfk')\psi_{-q}^+(x,\bfk)\bigg],\\
\nonumber\\
&&F_2^q(Q^2) = -\frac{2M_n}{q^1-iq^2}\int_0^1dx \int \frac{d^2\bfk}{16\pi^3}~\nonumber\\
&&\bigg[\psi_{+q}^{+*}(x,\bfk')\psi_{+q}^-(x,\bfk)+\psi_{-q}^{+*}(x,\bfk')\psi_{-q}^-(x,\bfk)\bigg],
\ee
where $\bfk'=\bfk+(1-x)\bfq$. $\psi_{\lambda_q q}^{\lambda_N}(x,\bfk)$ are the LFWFs with specific nucleon helicities $\lambda_N=\pm$ and for the struck quark $\lambda_q=\pm$, where plus and minus correspond to $+\frac{1}{2}$ and $-\frac{1}{2}$ respectively. We consider the frame where $q=(0,0,\bfq)$,  thus $Q^2=-q^2=\bfq^2$. 

We adopt the generic ansatz for the quark-diquark model of the valence Fock state of the nucleon LFWFs at an initial scale $\mu_0=313$~MeV as proposed in \cite{Gut} :  
\be 
\psi_{+q}^+(x,\bfk) &=&  \varphi_q^{(1)}(x,\bfk) 
\,, \nonumber\\
\psi_{-q}^+(x,\bfk) &=& -\frac{k^1 + ik^2}{xM_n}   \, \varphi_q^{(2)}(x,\bfk) \,, \nonumber\\
\psi_{+q}^-(x,\bfk) &=& \frac{k^1 - ik^2}{xM_n}  \, \varphi_q^{(2)}(x,\bfk)
\,, \nonumber\\
\psi_{-q}^-(x,\bfk) &=& \varphi_q^{(1)}(x,\bfk) 
\,, \label{WF} 
\ee 
where $\varphi_q^{(1)}(x,\bfk) $ and $\varphi_q^{(2)}(x,\bfk) $ are   the wave functions predicted by soft-wall AdS/QCD\cite{BT}, modified by introducing the tunable parameters $a_q^{(i)}$ and $b_q^{(i)}$ for quark $q$ \cite{Gut}:
\be
\varphi_q^{(i)}(x,\bfk)&=&N_q^{(i)}\frac{4\pi}{\kappa}\sqrt{\frac{\log(1/x)}{1-x}}x^{a_q^{(i)}}(1-x)^{b_q^{(i)}}\nonumber\\
&&\exp\bigg[-\frac{\bfk^2}{2\kappa^2}\frac{\log(1/x)}{(1-x)^2}\bigg].
\ee
The parameters are tuned to fit the electromagnetic properties of the nucleons.
Following the convention of \cite{Gut}, we fix the normalizations  of  the Dirac and Pauli form factors as
\be
F_1^q(Q^2)=n_q\frac{I_1^q(Q^2)}{I_1^q(0)},~~~~~~~~~F_2^q(Q^2)=\kappa_q\frac{I_2^q(Q^2)}{I_2^q(0)},\label{D_P_FF}
\ee
so that $F_1^q(0)=n_q$  and $F_2^q(0)=\kappa_q$ where $n_u=2,~n_d=1$ and the anomalous magnetic moments for the $u$ and $d$ quarks are $\kappa_u=1.673$ and $\kappa_d=-2.033$.  The advantage of the modified formulae in Eq.(\ref{D_P_FF}) is that, irrespective of the values of the parameters, the normalization conditions for the form factors are automatically satisfied. The structure integrals, $I_i^q(Q^2)$ have the form as
\be
I_1^q(Q^2)&=& \int_0^1 dx x^{2a^{(1)}_q} (1-x)^{1+2b^{(1)}_q}R_q(x,Q^2)\nonumber\\
&&\exp \bigg[-\frac{Q^2}{4\kappa^2}\log(1/x)\bigg],\label{i1}\\
I_2^q(Q^2)&=& 2\int_0^1 dx x^{2a^{(1)}_q-1} (1-x)^{2+2b^{(1)}_q}\sigma_q(x)\nonumber\\
&&\exp \bigg[-\frac{Q^2}{4\kappa^2}\log(1/x)\bigg],\label{i2}
\ee
with
\be
R_q(x,Q^2)&=&1+\sigma_q^2(x)\frac{(1-x)^2}{x^2}\frac{\kappa^2}{M_n^2\log(1/x)}\nonumber\\
&&\bigg[1-\frac{Q^2}{4\kappa^2}\log(1/x)\bigg],\nonumber\\
\sigma_q(x)&=&\frac{N_q^{(2)}}{N_q^{(1)}}x^{a^{(2)}_q - a^{(1)}_q}(1-x)^{b^{(2)}_q-b^{(1)}_q}.
\ee
It is straightforward to write down the flavor decompositions of the Dirac and Pauli form factors of nucleon as
\be
F_i^{p(n)}=e_u F_i^{u(d)}+e_d F_i^{d(u)},~~~~~~~~(i=1,2)
\ee
where $e_u$ and $e_d$ are the charges of $u$ and $d$ quarks in units of positron charge($e$). 
\begin{table}[ht]
\centering 
\begin{tabular}{ c c c } 
\hline 
Parameters&~~~~~~~~~~~~$u$~~~~~~~~~~~~~& ~~~~~~~~~~~$d$~~~~~~~~~~~~  \\ [0.5ex] 
\hline 
$a^{(1)}$ & 0.035 & 0.20\\ 
$b^{(1)}$ & 0.080 & 1.00 \\
\hline
$a^{(2)}$ & 0.75 & 1.25 \\ 
$b^{(2)}$ & -0.60 & -0.20 \\
\hline
$N^{(1)}$ & 29.180 & 33.918 \\
$N^{(2)}$ & 1.459 & 1.413 \\
\hline
\end{tabular} 
\caption{The parameters in the model for $\kappa=0.4066$ GeV} 
\label{table} 
\end{table} 
\begin{table}[ht]
\centering 
\begin{tabular}{ c c c } 
\hline
Quantity&~~~Our results~~~& ~~~Measured data\cite{pdg}~~~  \\ [0.5ex] 
\hline 
$r_E^p$ & 0.7861 fm  & $0.877 \pm 0.005$  fm\\ 
$r_M^p$ & 0.7719 fm & $0.777\pm0.016$ fm \\ 
\hline
$\langle r_E^2\rangle^n$ & -0.085 fm$^2$ & $-0.1161 \pm 0.0022$ fm$^2$ \\

$r_M^n$ & 0.7596 fm  &  $0.862^{+0.009}_{-0.008} $ fm\\
\hline 
\end{tabular} 
\caption{electromagnetic radii of the nucleons}
\label{table2} 
\end{table} 
  On top of the AdS/QCD scale  parameter $\kappa$, the wave functions involve four more parameters $a_q^{(i)}$ and $b_q^{(i)}$ (with $i=1,2$) for each quark.
 In Ref.\cite{Gut},  $\kappa$ is taken to be $0.35$ GeV and the parameters  are evaluated to fit the electromagnetic properties of the nucleon. But the results for the form factors presented in that paper are not converged with respect to the lower limit of $x$ integrations in Eqs.(\ref{i1} and \ref{i2}). The comparisons with experimental data   presented in several plots in Ref.\cite{Gut} are true only for an unrealistically large value of lower limit for the $x$ integrations which drastically change when  integrated from   $x\to 0$.   So, when proper limits in the $x$-integrations are taken, the parameters presented  in \cite{Gut} cannot reproduce the data. In this work, we use a different scale parameter  $\kappa=0.4066$ GeV  which was obtained by fitting the nucleon form factors in AdS/QCD soft-wall model\cite{CM1,CM2}.
 Here, we show that we can reproduce the nucleon form factors with  the  new parameters $a_q^{(i)}$,  $b_q^{(i)}$ and $N^{(i)}$ listed in Table \ref{table}. The results are stable and converged under the integration over $x$.
 The new parameters  reproduce the experimental data quite accurately for a wide range of $Q^2$ values. 
 
\begin{figure*}[htbp]
\begin{minipage}[c]{0.98\textwidth}
\small{(a)}
\includegraphics[width=7.5cm,height=5.5cm,clip]{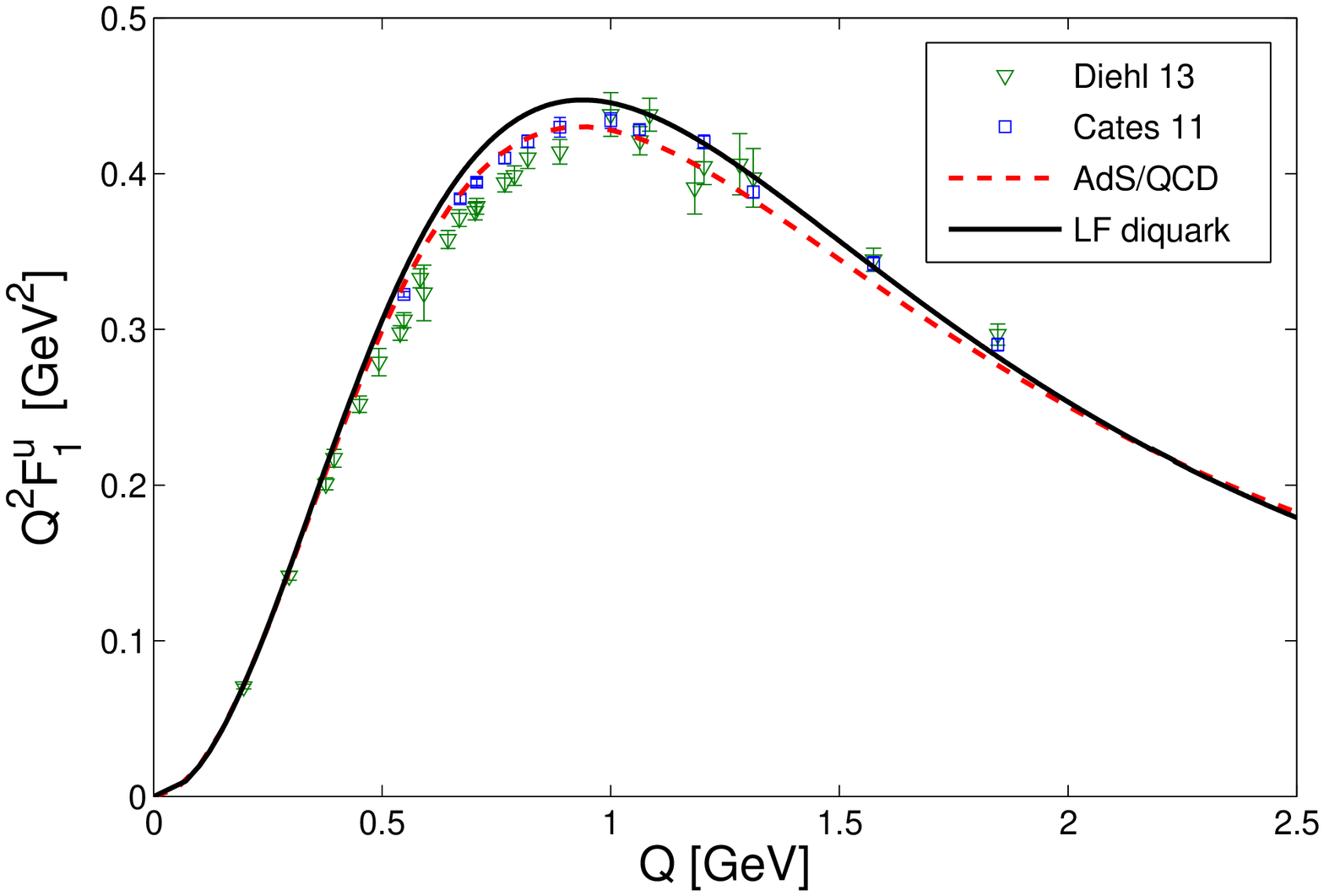}
\hspace{0.1cm}%
\small{(b)}\includegraphics[width=7.5cm,height=5.5cm,clip]{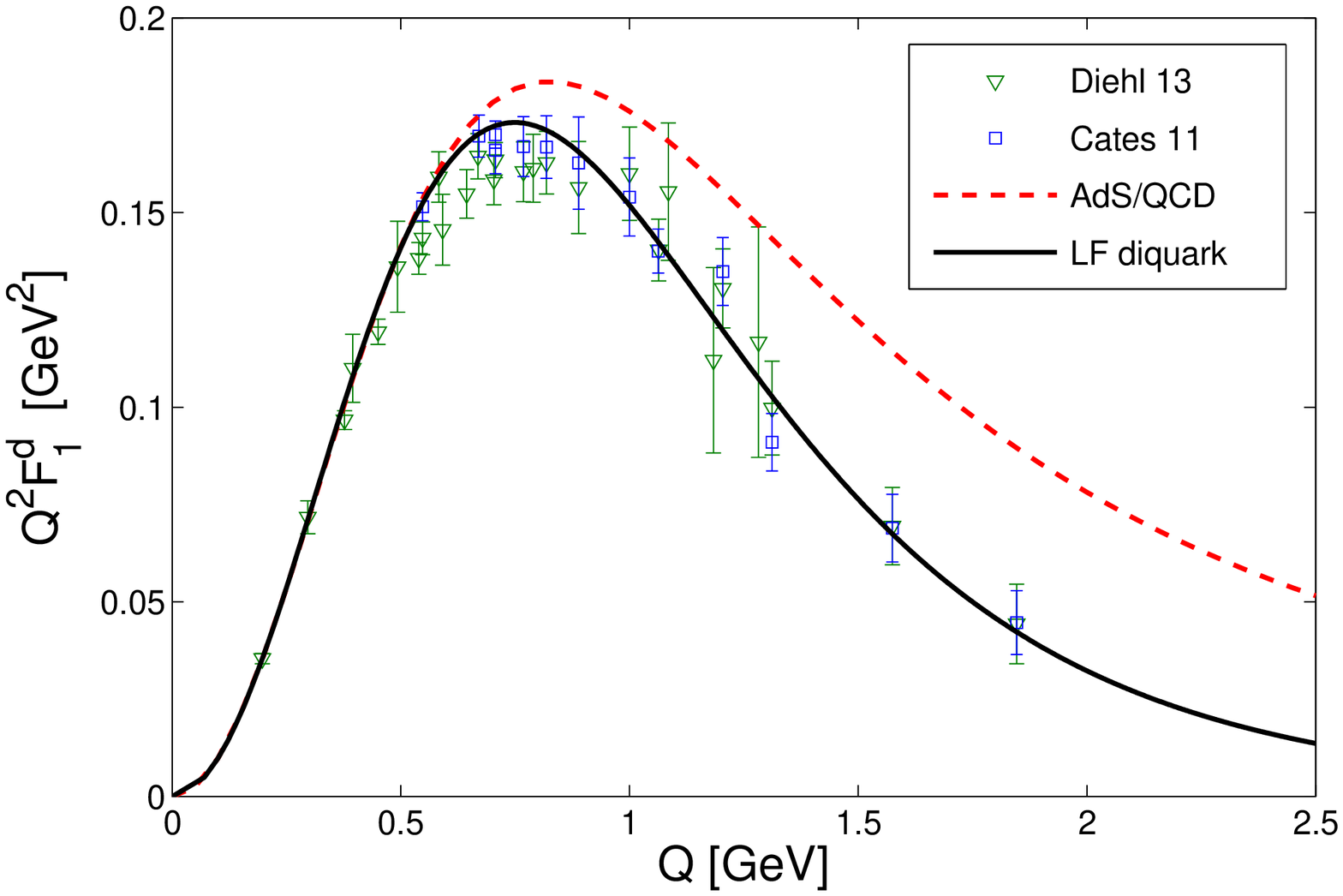}
\end{minipage}
\begin{minipage}[c]{0.98\textwidth}
\small{(c)}
\includegraphics[width=7.5cm,height=5.5cm,clip]{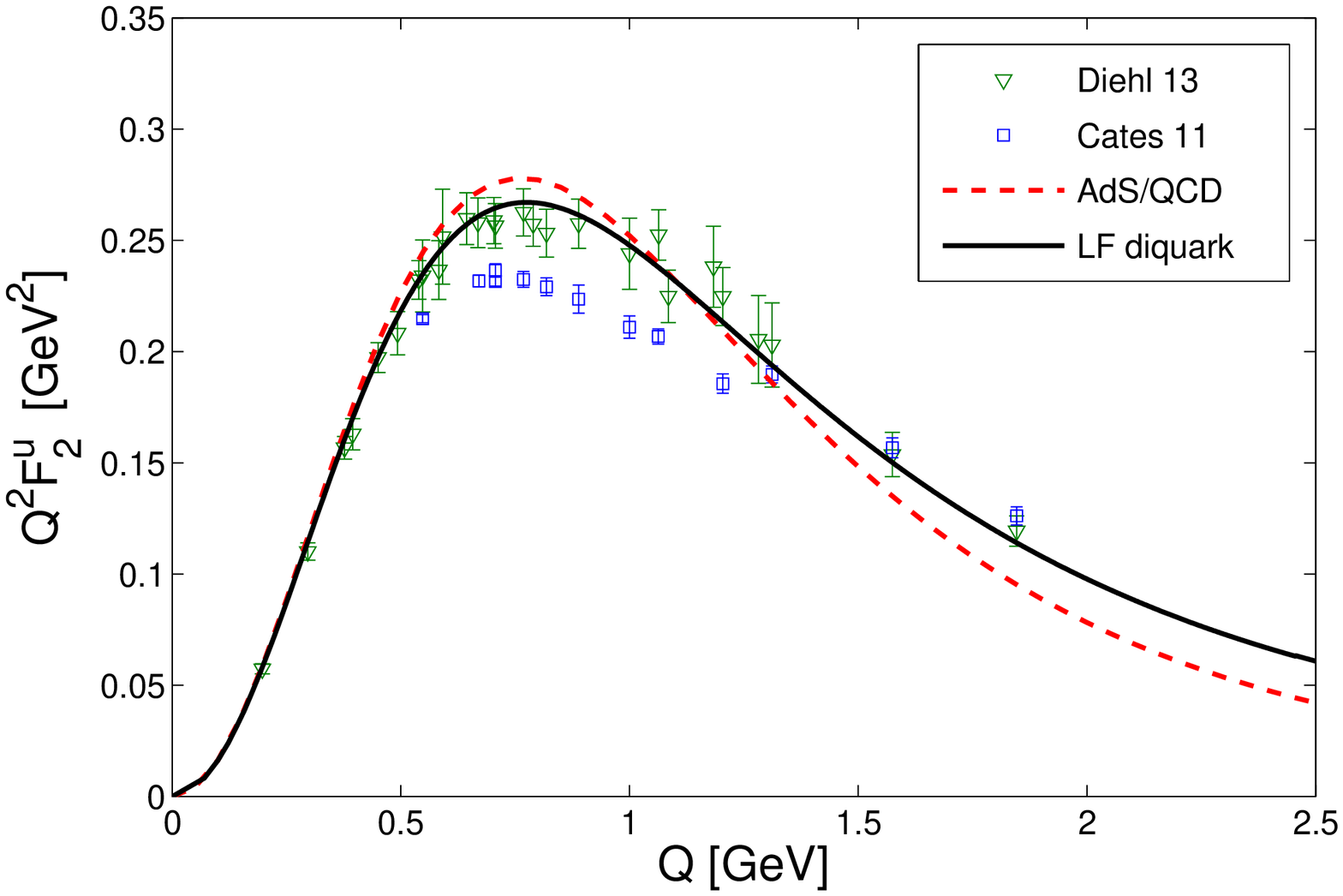}
\hspace{0.1cm}%
\small{(d)}\includegraphics[width=7.5cm,height=5.5cm,clip]{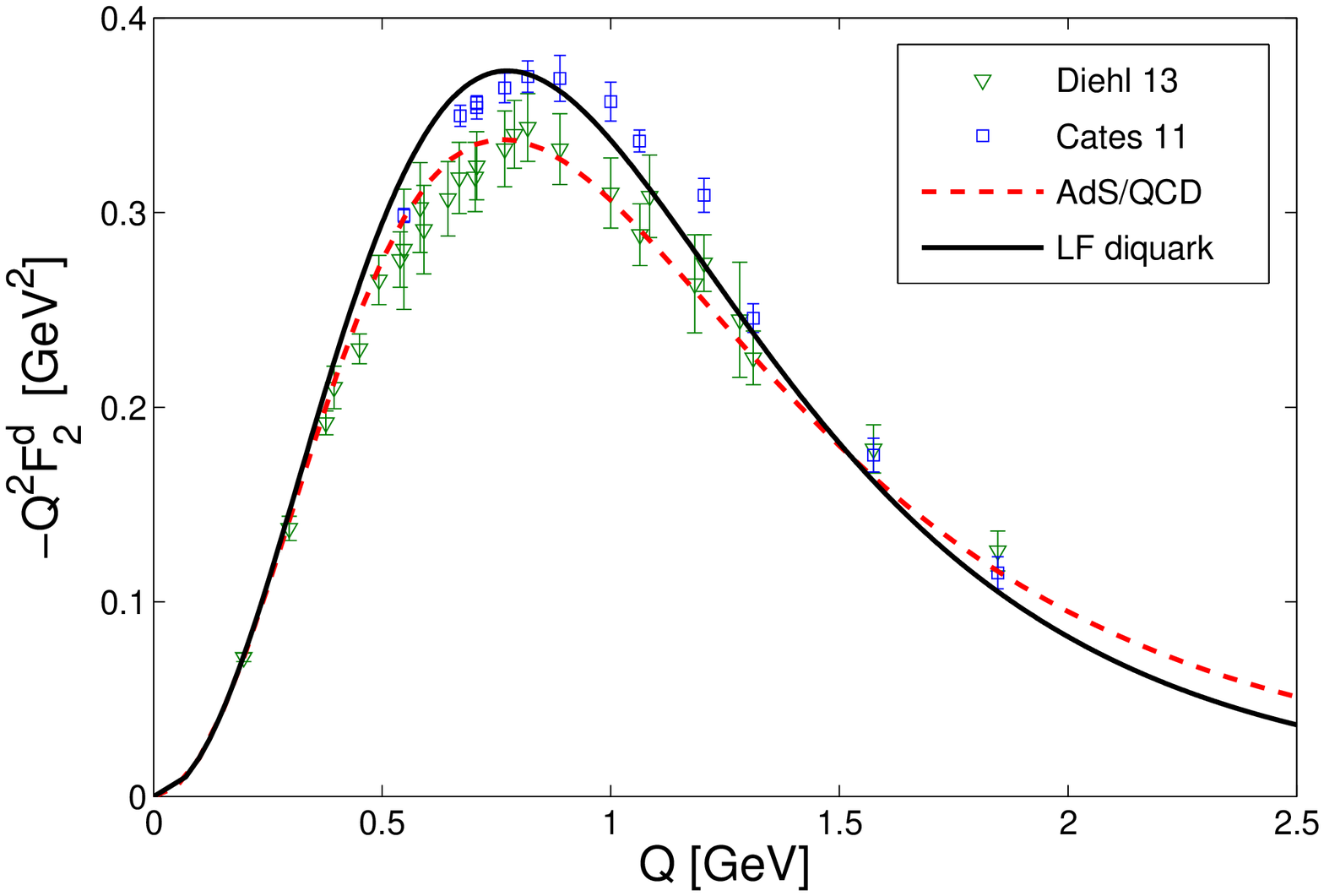}
\end{minipage}
\caption{\label{FF_flavors}(Color online) Plots of  flavor dependent form factors for $u$ and $d$ quarks. The experimental data are taken from \cite{Cates,diehl13}. The red dashed lines represent the soft-wall AdS/QCD model \cite{CM2}.}
\end{figure*} 
\begin{figure*}[htbp]
\begin{minipage}[c]{0.98\textwidth}
\small{(a)}
\includegraphics[width=7.5cm,height=5.5cm,clip]{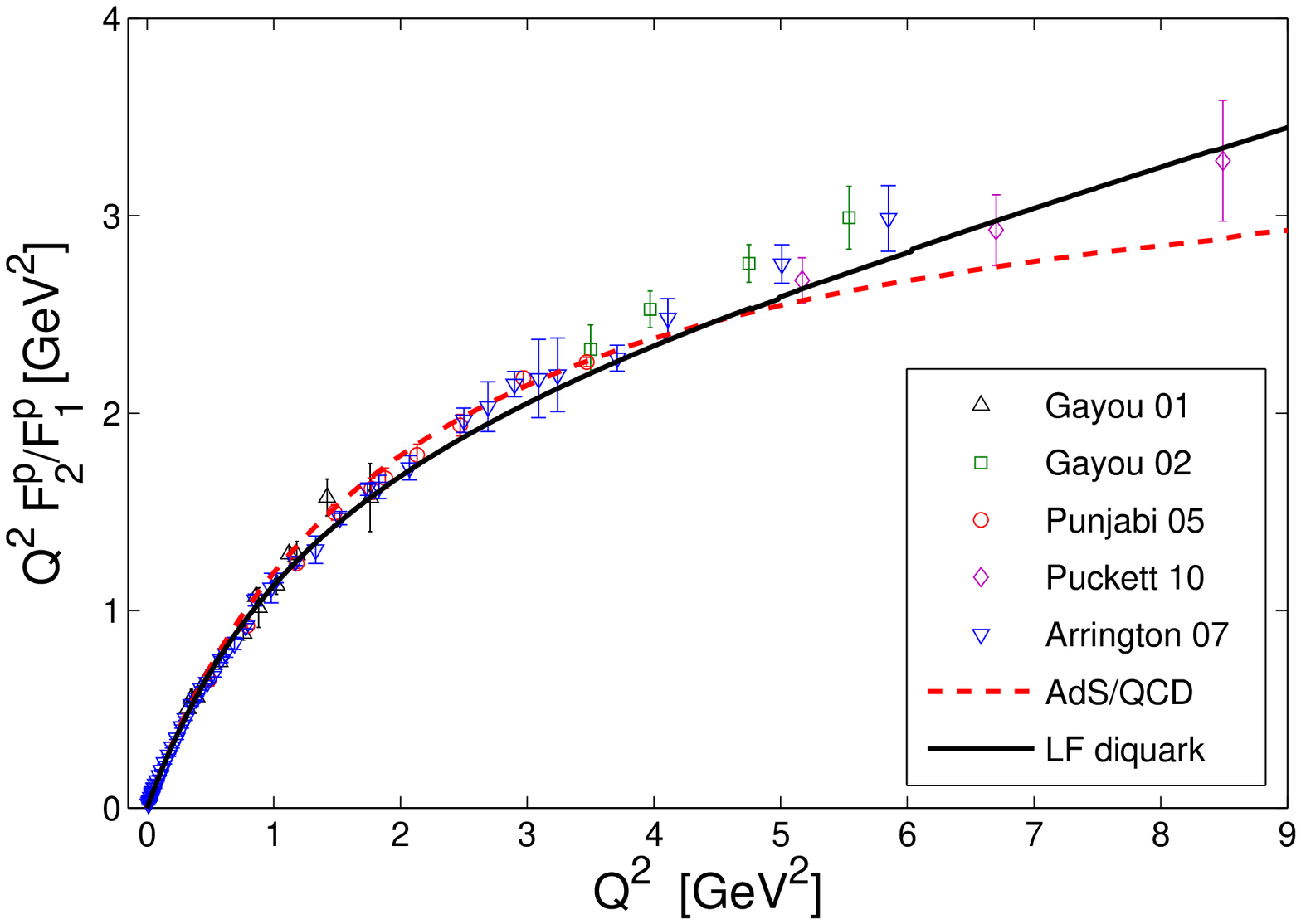}
\hspace{0.1cm}%
\small{(b)}\includegraphics[width=7.5cm,height=5.5cm,clip]{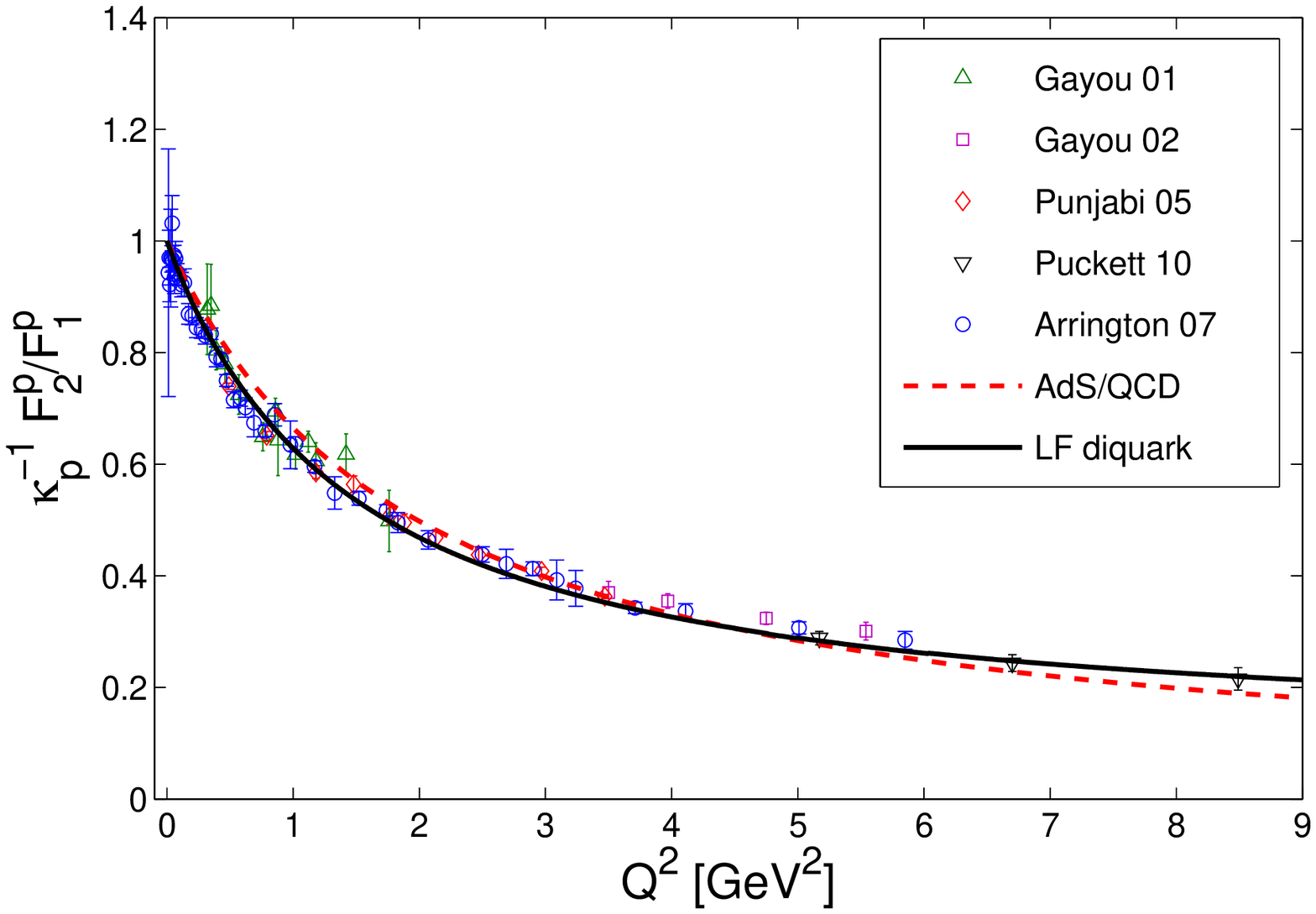}
\end{minipage}
\caption{\label{proton_fit}(Color online)  Light-front quark-diquark model results are fitted with the experimental data.
The plots show the ratio of Pauli and Dirac form factors for the proton, (a) the ratio is multiplied by $Q^2=-q^2=-t$, (b) the ratio is divided by $\kappa_p$. The experimental  data are taken from Refs. \cite{Gay1,Gay2,Arr,Pun,Puck}. The red dashed lines represent the soft-wall AdS/QCD model\cite{CM2}. }
\end{figure*}
\begin{figure*}[htbp]
\centering
\includegraphics[width=10cm,clip]{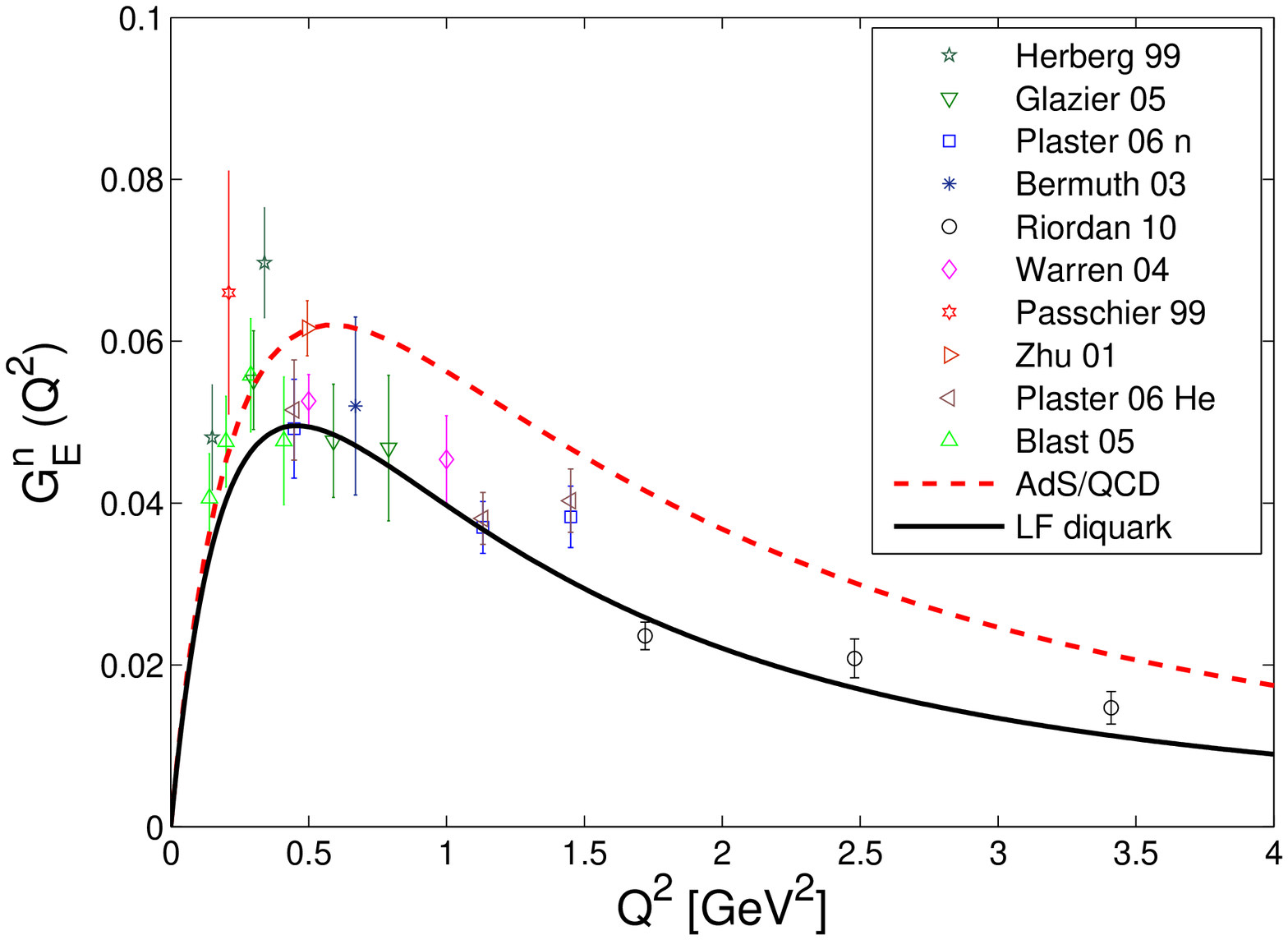}
\caption{\label{G_E_n} (Color Online) The Sach form factro $G_E^n(Q^2)$ for the neutron. The experimental data are taken from Refs. \cite{Herb,Glaz,Plast,Berm,Rior,Warr,Pass,Zhu,Blast}. The red dashed line represents the soft-wall AdS/QCD model prediction.}
\end{figure*}
 The Dirac and Pauli form factors of both $u$ and $d$ quarks  are shown in Fig.\ref{FF_flavors}. The form factors  $F_1^q$ and $F_2^q$ for both $u$ and $d$ quarks in light-front quark-diquark model for the scale parameter $\kappa=0. 4066$ GeV  and   the parameters defined in Table \ref{table} are in excellent agreement with the data.
For $F_1^d$, we can see a clear improvement in the quark-diquark model over the AdS/QCD model.
 It is important  to note  that  other models  fail to reproduce the form factors data for $d$ quark \cite{Qattan}. 
In Fig.\ref{proton_fit}, we have shown the fit of light-front quark-diquark model results with  experimental data of proton form factors. We get excellent agreement with the data. In the same plots, we also show  comparisons of  the light-front quark-diquark model  and the soft-wall AdS/QCD model with the same value of $\kappa$ \cite{CM2}. The results of the light-front quark-diquark model agree with the data better than AdS/QCD, specially at large $Q^2$ values we achieve substantial  improvement.  The Sach form factor $G_E(Q^2)$ for the neutron is  shown in Fig.\ref{G_E_n}. Again, our results agree with the experimental data much better than the AdS/QCD results.
The fitted results for the electromagnetic radii of the nucleons are listed in Table \ref{table2}. The standard formulae for the electromagnetic radii  of nucleon used here are given below:
 \be
 \langle r^2_E\rangle^N&=&-6 \frac{d G_E^N(Q^2)}{dQ^2}{\Big\vert}_{Q^2=0},\\
  \langle r^2_M\rangle^N&=& -\frac{6}{G_M^N(0)} \frac{d G_M^N(Q^2)}{dQ^2}{\Big\vert}_{Q^2=0}
  \ee
  where $N$ stands for nucleon($N=p/n$) and the Sachs form factors are defined as
  \be
  G_E^N(Q^2) &=& F_1^N(Q^2)-\frac{Q^2}{4 M_N^2}F_2^N(Q^2),\\
  G_M^N(Q^2) &=& F_1^N(Q^2)+F_2^N(Q^2).
  \ee

\section{Generalized parton distributions}\label{gpd}

Using the overlap formalism of light front wave functions, we evaluate the GPDs in light front quark-diquark model.
We consider the DGLAP domain , i.e., $\zeta<x<1$ where $\zeta$ is the skewness and $x$ is the light front longitudinal momentum fraction carried by the struck quark. This domain corresponds to the situation where one removes a quark from the initial proton  with light-front momentum fraction $x=\frac{k^+}{P^+}$ and the transverse momentum $\bfk$ and re-insert it into the final state  of the  proton with longitudinal momentum fraction $x-\zeta$ and transverse momentum $\bfk-\bfq$. The contributions to the GPDs for $0<x<\zeta$ come from the particle number changing interactions and cannot be studied in this model. The kinematical domain for GPDs studied here is thus restricted to $\zeta<x<1$ where only diagonal overlaps(2-particle state $\to$ 2-particle state) contribute.  
The  GPDs  $H$ and $E$  are defined through the matrix element of the bilocal vector current on the light-front:
\be
&&\int \frac{dy^-}{8\pi}e^{ixP^+y^-/2}\langle P',\lambda'|~\bar{\psi(0)}\gamma^+\psi(y)~|P, \lambda\rangle \nonumber\\
&=&\frac{1}{2P^+}\bar{U}(P',\lambda')\bigg[H(x,\zeta,t)\gamma^+\nonumber\\
&+&E(x,\zeta,t)\frac{i}{2M_n}\sigma^{+\alpha}q_{\alpha}\bigg]U(P,\lambda).\label{gpds}
\ee
The proton state $\mid P,\lambda\rangle $ is written in
 two particle Fock states with one fermion and a scalar boson in the light front quark-diquark model.  Using the  relations
\be
&&\frac{1}{2\bar{P}^+}\bar{U}(P',\lambda')\gamma^+U(P,\lambda)=\frac{\sqrt{1-\zeta}}{1-\frac{\zeta}{2}}~\delta_{\lambda,\lambda'},\nonumber\\
&&\frac{1}{2\bar{P}^+}\bar{U}(P',\lambda')\frac{i}{2M_n}\sigma^{+\alpha}q_{\alpha}U(P,\lambda)\nonumber\\
&=&-\frac{\zeta^2}{4(1-\frac{\zeta}{2})\sqrt{1-\zeta}}~\delta_{\lambda,\lambda'}+\frac{1}{\sqrt{1-\zeta}}~\frac{\lambda q^1+iq^2}{2M_n}~\delta_{\lambda,-\lambda'}.\nonumber
\\\label{spinor}
\ee
where $\bar{P}=(P+P')/2$ and $\lambda(\lambda^\prime)=\pm\frac{1}{2}$ is the initial(final) proton spin,
 we have the following expressions for  the GPDs in terms  of the LFWFs in the quark-diquark model 
\be
&&\frac{\sqrt{1-\zeta}}{1-\frac{\zeta}{2}}~H^q_v(x,\zeta,t)-\frac{\zeta^2}{4(1-\frac{\zeta}{2})\sqrt{1-\zeta}}~E^q_v(x,\zeta,t)\nonumber\\
&=&\int \frac{d^2\bfk}{16\pi^3}~\bigg[\psi_{+q}^{+*}(x',\bfk')\psi_{+q}^+(x,\bfk)\nonumber\\
&+&\psi_{-q}^{+*}(x',\bfk')\psi_{-q}^+(x,\bfk)\bigg],\label{H} \\
\nonumber\\
{\rm and} &&\frac{1}{\sqrt{1-\zeta}}~\frac{-( q^1-iq^2)}{2M_n}~E^q_v(x,\zeta,t)\nonumber\\
&=&\int \frac{d^2\bfk}{16\pi^3}~\bigg[\psi_{+q}^{+*}(x',\bfk')\psi_{+q}^-(x,\bfk)\nonumber\\
&+&\psi_{-q}^{+*}(x',\bfk')\psi_{-q}^-(x,\bfk)\bigg],\label{E}
\ee
where
\be
x'=\frac{x-\zeta}{1-\zeta},~~~~~~~~~~~~\bfkpr=\bfk-\frac{1-x}{1-\zeta}~\bfq.
\ee
\begin{figure*}[htbp]
\begin{minipage}[c]{0.98\textwidth}
\small{(a)}
\includegraphics[width=7.5cm,height=5.5cm,clip]{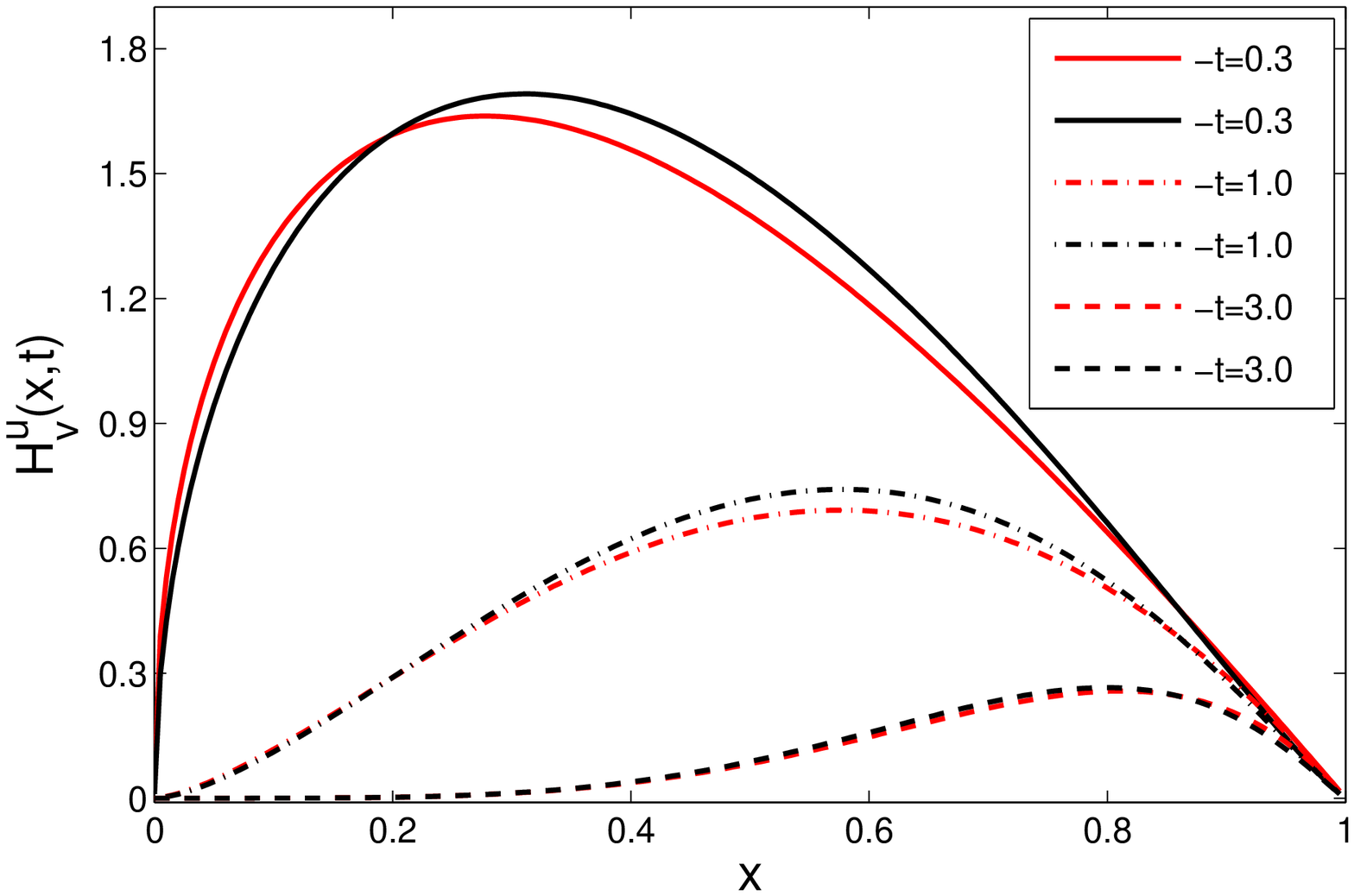}
\hspace{0.1cm}%
\small{(b)}\includegraphics[width=7.5cm,height=5.5cm,clip]{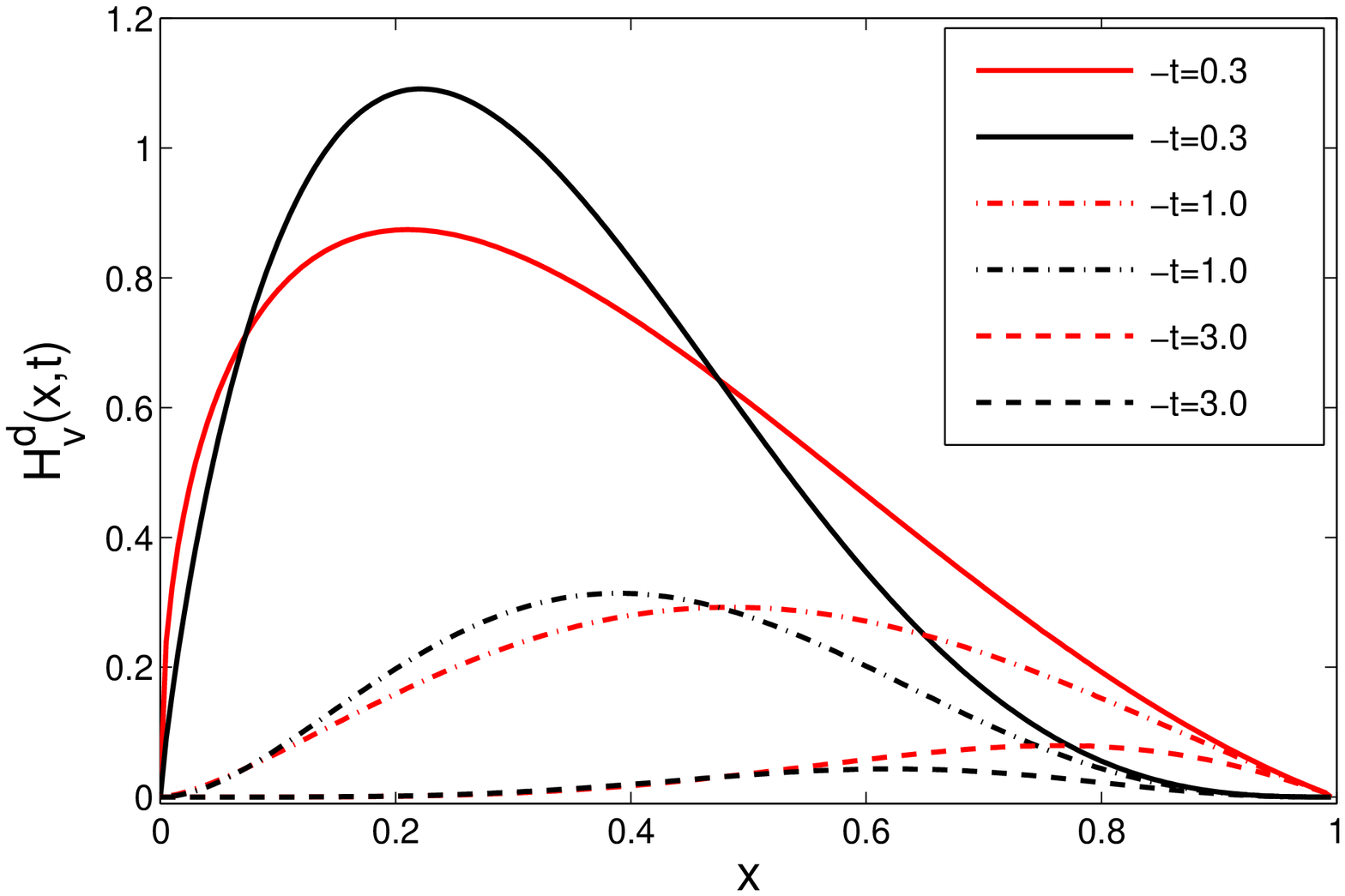}
\end{minipage}
\begin{minipage}[c]{0.98\textwidth}
\small{(c)}
\includegraphics[width=7.5cm,height=5.5cm,clip]{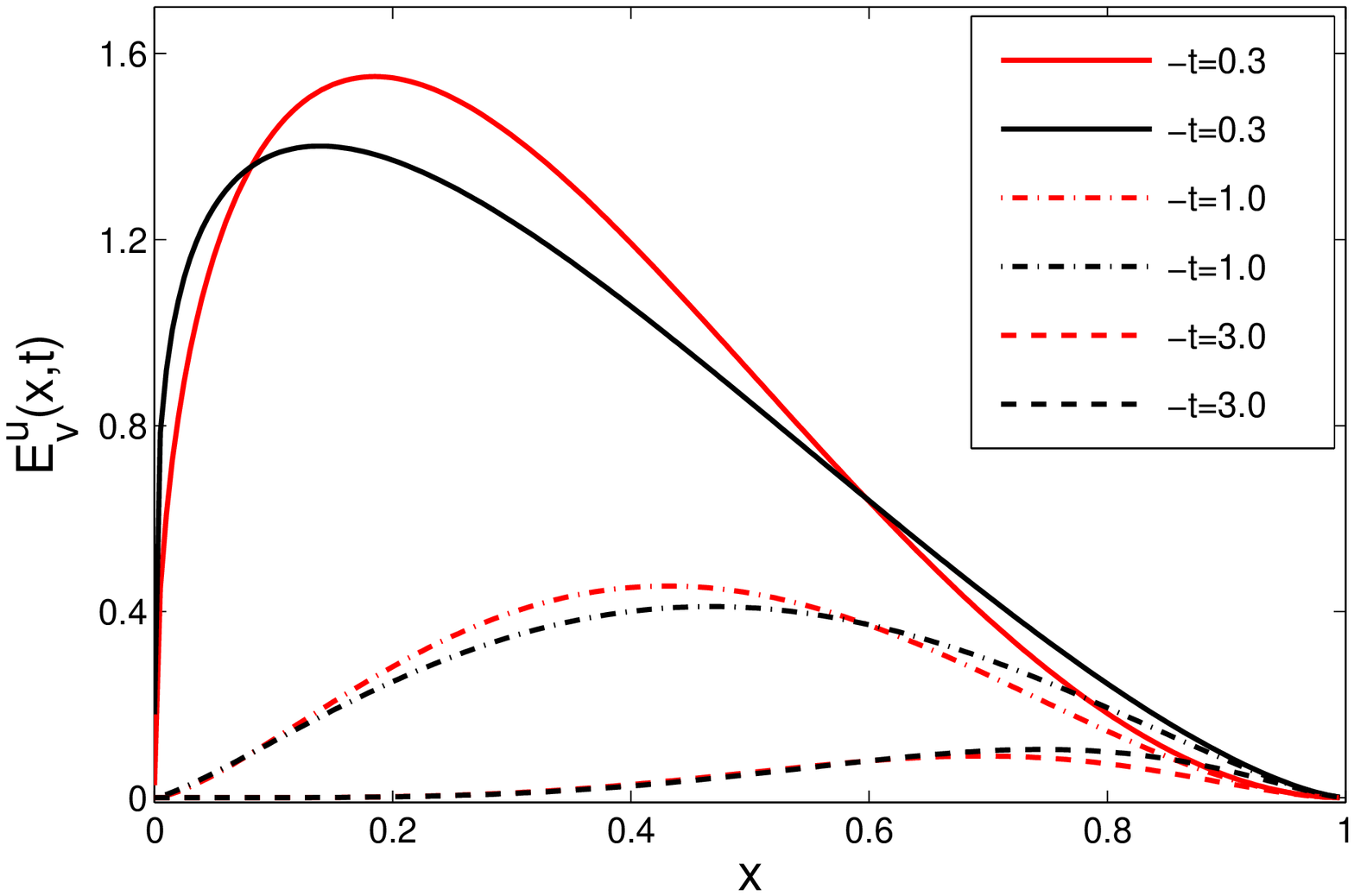}
\hspace{0.1cm}%
\small{(d)}\includegraphics[width=7.5cm,height=5.5cm,clip]{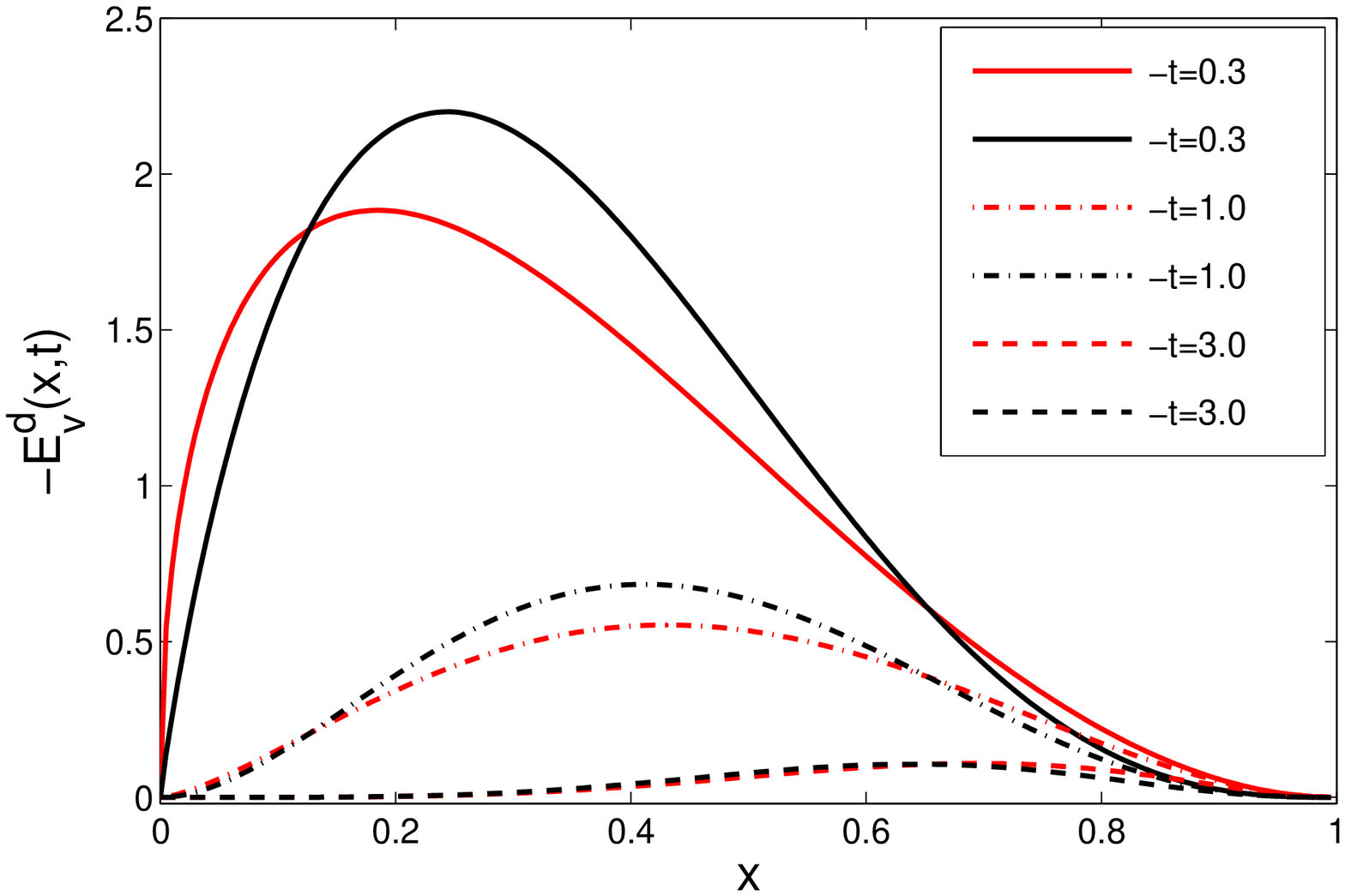}
\end{minipage}
\caption{\label{GPDs_zero}(Color online) Plots of (a) $H^u(x,t)$ vs $x$ (b) the same as in (a) but for $d$ quark (c) $E^u(x,t)$ vs $x$ (d) the same as in (c) but for $d$ quark for fixed values of $-t$. The black lines represent the light-front quark-diquark model and the red lines are for AdS/QCD model \cite{CM1}. }
\end{figure*} 
Substituting the LFWFs (Eq.(\ref{WF})) in Eqs.(\ref{H}) and (\ref{E}) and integrating over $\bfk$, we get the following expressions for GPDs 
\be
E^q_v(x,\zeta,t)&=&\kappa_q~\frac{\mathcal{F}_2^q(x,\zeta,t)}{I^q_2(0)},\label{gpdE}\\
H^q_v(x,\zeta,t)&=& n_q~\frac{1-\frac{\zeta}{2}}{I_1^q(0)\sqrt{1-\zeta}}\bigg[\mathcal{F}_1^q(x,\zeta,t)\nonumber\\
&+&\frac{\zeta^2}{4(1-\frac{\zeta}{2})\sqrt{1-\zeta}}~E^q_v(x,\zeta,t)\bigg].\label{gpdH}
\ee
The   functions $\mathcal{F}_i^q(x,\zeta,t)$ are given by
\be
&&\mathcal{F}_1^q(x,\zeta,t)=\frac{1}{\kappa^2}\sqrt{\frac{\log(1/x)}{(1-x)}}\sqrt{\frac{\log(1/x')}{(1-x')}}\bigg[(xx')^{a_q^{(1)}}
\nonumber\\
&&((1-x)(1-x'))^{b_q^{(1)}}\frac{1}{A}
+\bigg[\frac{N_q^{(2)}}{N_q^{(1)}}\bigg]^2\frac{1}{M_n^2}((xx')^{a_q^{(2)}-1}\nonumber\\
&&((1-x)(1-x'))^{b_q^{(2)}}
\bigg(\frac{1}{A^2}-\frac{(1-x')^2Q^2}{4}\frac{1}{A}\nonumber\\
&+&Q^2\bigg(\frac{\log(1/x')}{2\kappa^2(1-x')A}-\frac{1-x'}{2}\bigg)^2\frac{1}{A}\bigg)\bigg]\nonumber\\
&\times&\exp\bigg[\frac{\log(1/x')}{2\kappa^2}Q^2\bigg(\frac{\log(1/x')}{2\kappa^2(1-x')^2A}-1\bigg)\bigg], \\
&&\nonumber\\
&&\mathcal{F}_2^q(x,\zeta,t)=\frac{2\sqrt{1-\zeta}}{\kappa^2}\frac{N_q^{(2)}}{N_q^{(1)}}\sqrt{\frac{\log(1/x)}{(1-x)}}\sqrt{\frac{\log(1/x')}{(1-x')}}\nonumber\\
&&\bigg[\frac{\log(1/x')}{2\kappa^2(1-x')A}
\bigg(x'^{a_q^{(1)}}x^{a_q^{(2)}-1}(1-x')^{b_q^{(1)}}(1-x)^{b_q^{(2)}}\nonumber\\
&-&x^{a_q^{(1)}}x'^{a_q^{(2)}-1}(1-x)^{b_q^{(1)}}(1-x')^{b_q^{(2)}}\bigg) \nonumber\\
&+& x^{a_q^{(1)}}x'^{a_q^{(2)}-1}(1-x)^{b_q^{(1)}}(1-x')^{b_q^{(2)}+1}\bigg]\frac{1}{A}\nonumber\\
&\times&\exp\bigg[\frac{\log(1/x')}{2\kappa^2}Q^2\bigg(\frac{\log(1/x')}{2\kappa^2(1-x')^2A}-1\bigg)\bigg],
\ee
where $Q^2=-t(1-\zeta)-M_n^2\zeta^2$ and $A$ is a function of $x$ and $x'$:
\be
A=A(x,x')=\frac{\log(1/x)}{2\kappa^2(1-x)^2}+\frac{\log(1/x')}{2\kappa^2(1-x')^2}.
\ee
$I_1^q(0)$ and $I_2^q(0)$ are the integrals  defined in Eqs.(\ref{i1}) and (\ref{i2}) for $Q^2=0$.
The GPDs are normalized as 
\be
\int_0^1 dx H^q(x,0,0)&=&n_q,\nonumber\\
\int_0^1 dx E^q(x,0,0)&=&\kappa_q,
\ee
where $n_q$ denotes the number of $u$ or $d$ valence quarks in the proton and the quark anomalous magnetic moment is denoted by $\kappa_q$.
According to the polynomiality condition, the  $n$-th Mellin moment of a GPD should be a polynomial  with  highest power  $\zeta^n$  at  $t \to 0$\cite{pol1,pol2}. Since the moments require the GPDs for all values of $-1\le x\le 1$, it is not possible to confirm the polynomiality condition  in this model as the GPDs are evaluated only for $0\le\zeta<x$. But we have numerically checked for $n=1,2,3,$ and $4$ that the   moments  $\int_\zeta^1 x^{n-1} H(x,\zeta,t)$  show the behavior consistent with the polynomiality condition in the limit $t\to 0$.

\begin{figure*}[htbp]
\begin{minipage}[c]{0.98\textwidth}
\small{(a)}
\includegraphics[width=7.5cm,height=5.5cm,clip]{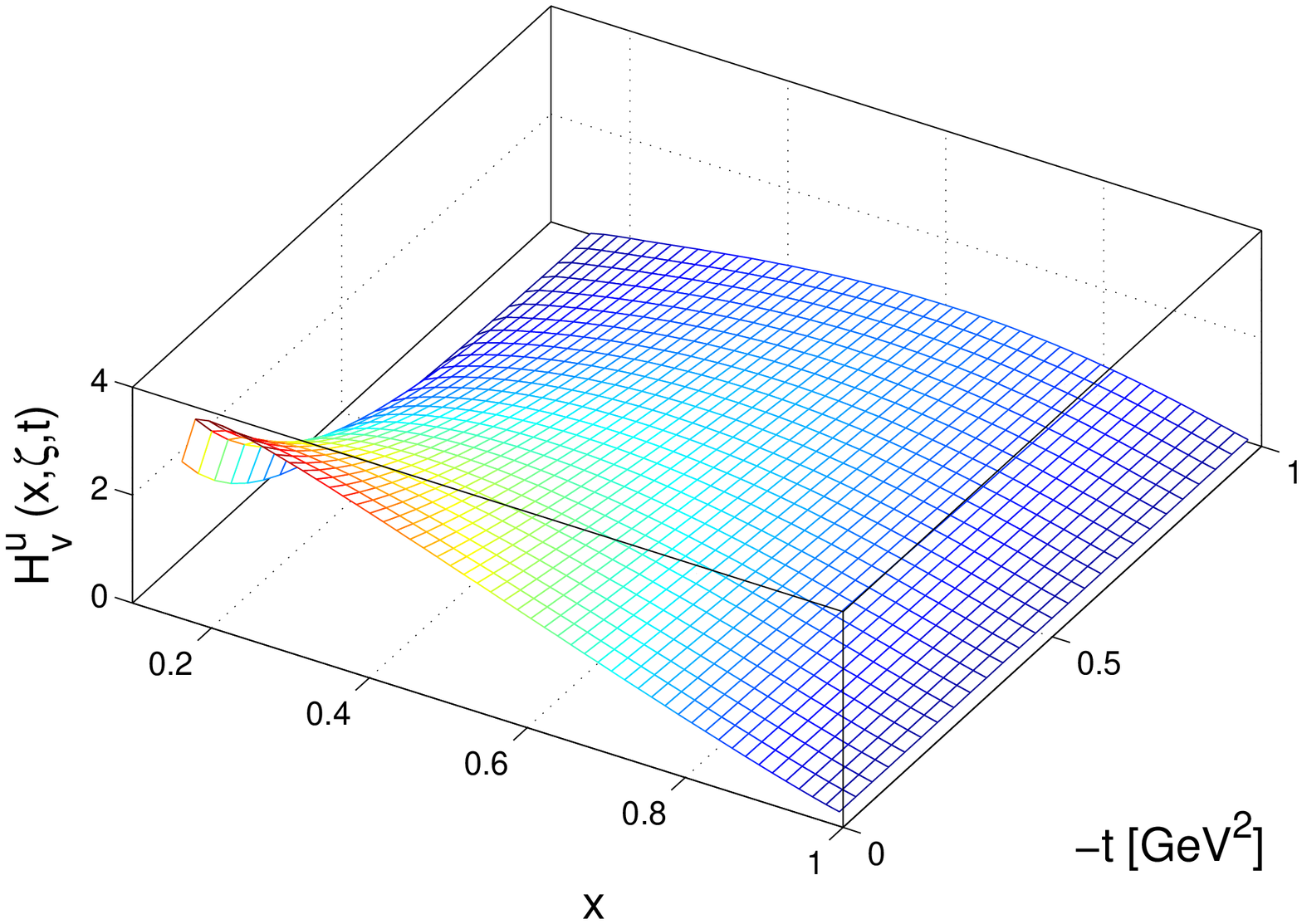}
\hspace{0.1cm}%
\small{(b)}\includegraphics[width=7.5cm,height=5.5cm,clip]{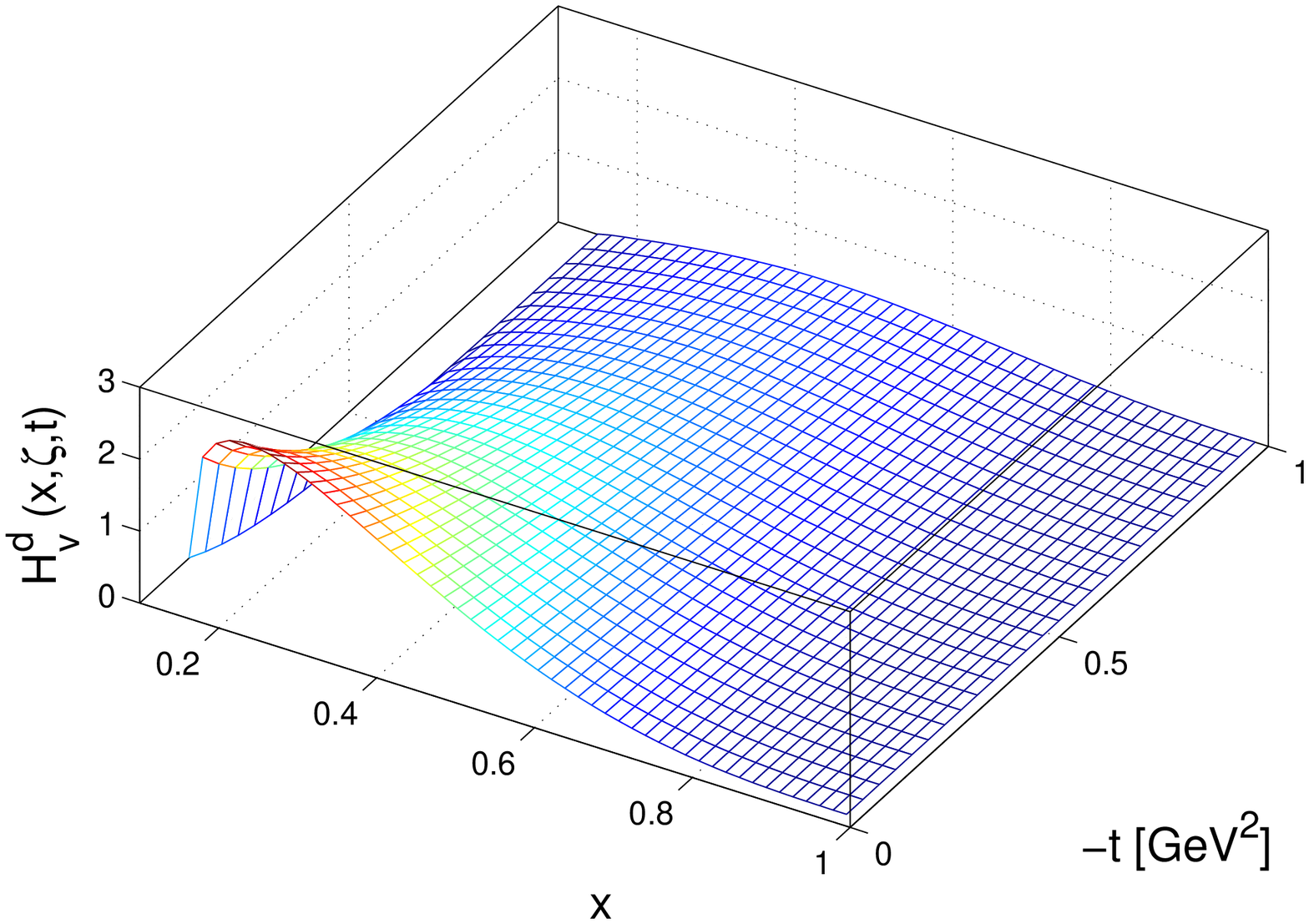}
\end{minipage}
\begin{minipage}[c]{0.98\textwidth}
\small{(c)}
\includegraphics[width=7.5cm,height=5.5cm,clip]{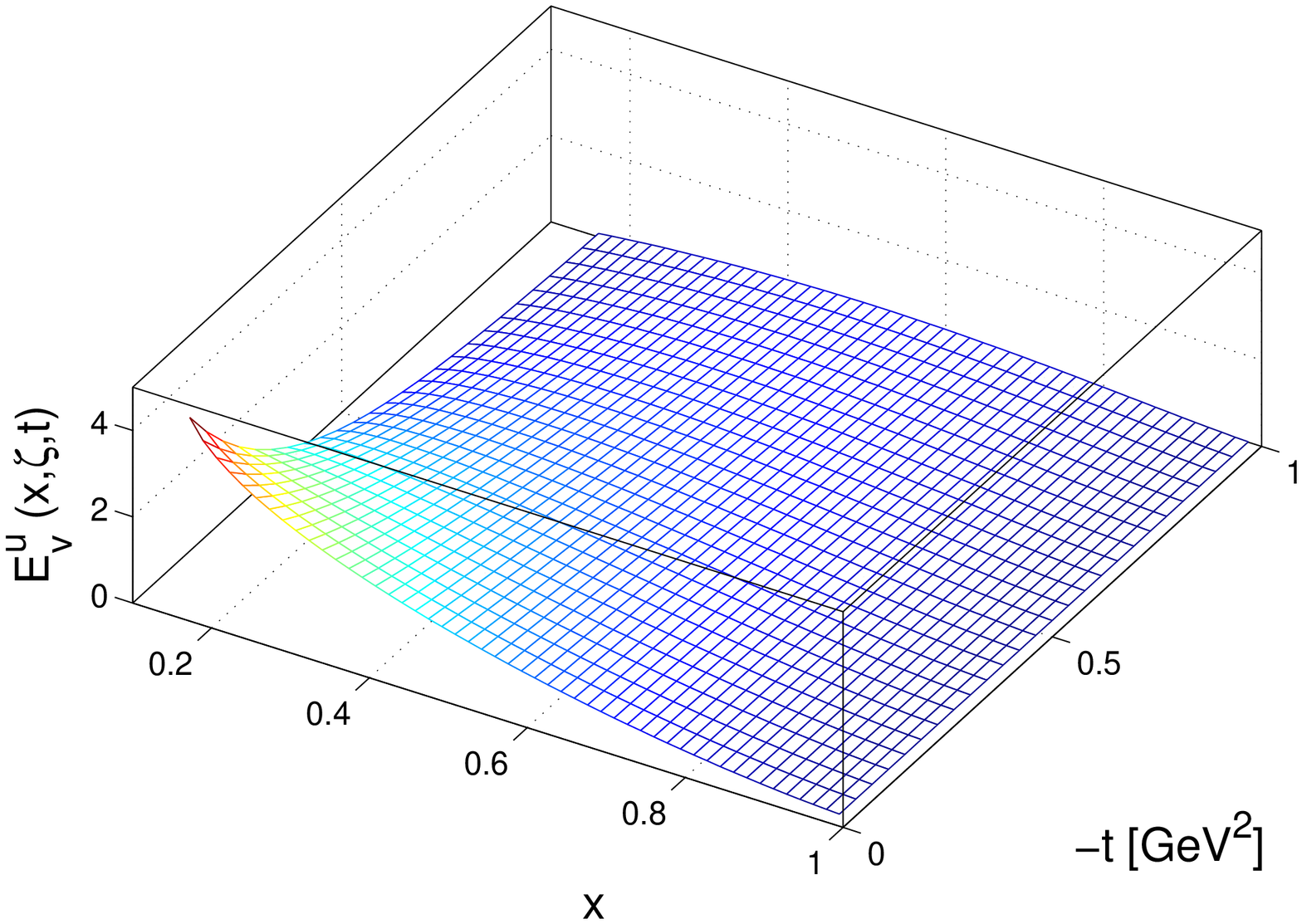}
\hspace{0.1cm}%
\small{(d)}\includegraphics[width=7.5cm,height=5.5cm,clip]{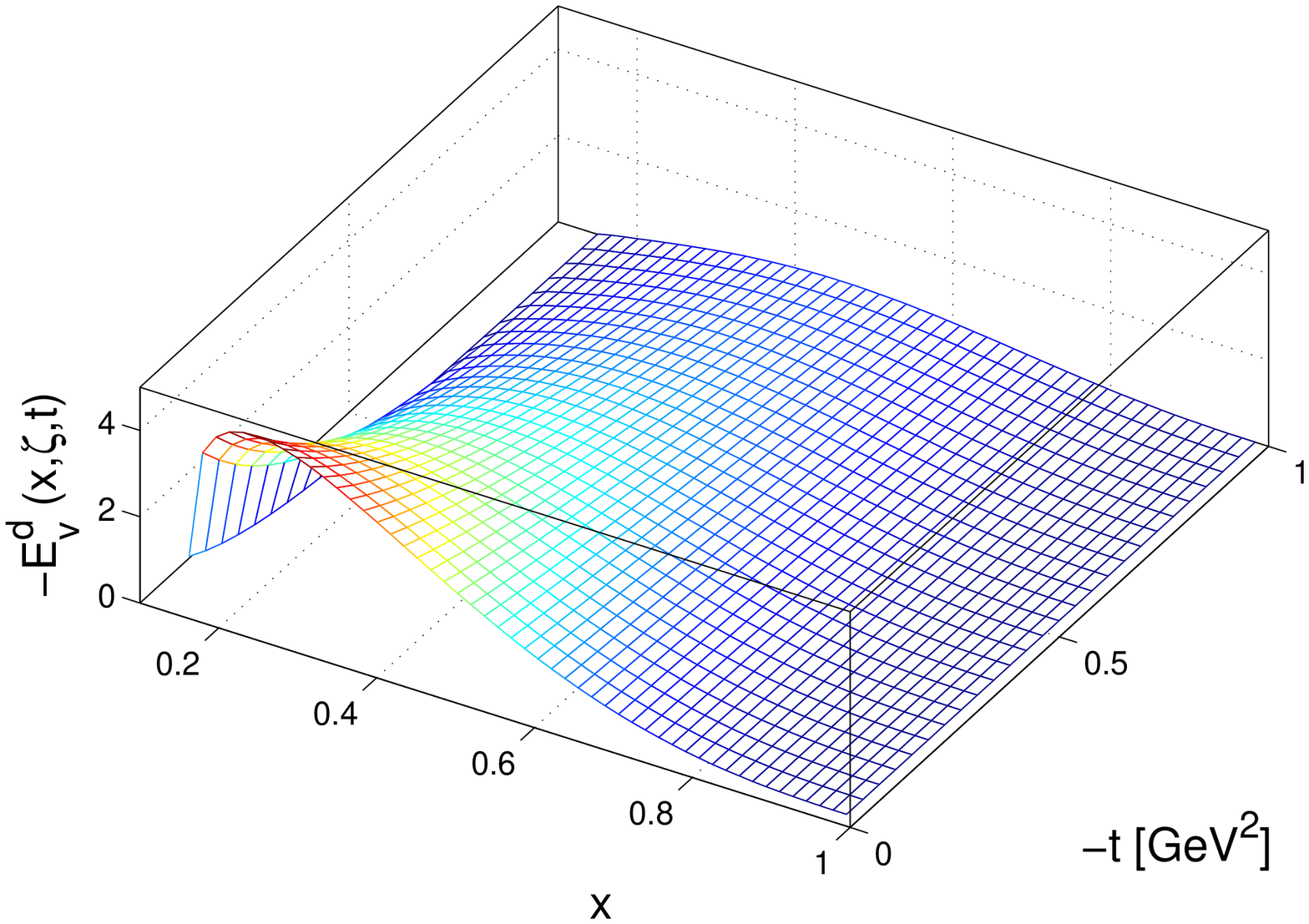}
\end{minipage}
\caption{\label{GPDs_t}(Color online) Plots of (a) $H^u(x,\zeta,t)$ vs $x$ and $-t$; (b) the same as in (a) but for $d$ quark; (c) $E^u(x,\zeta,t)$ vs $x$ and $-t$; (d) the same as in (c) but for $d$ quark for fixed $\zeta=0.15$.   For $\zeta=0.15$,  the minimum value of $-t \approx 0.024$ GeV$^2$. }
\end{figure*} 
The GPDs for zero skewness($\zeta=0$) in light-front quark-diquark model are compared with the AdS/QCD results \cite{CM1} in Fig.\ref{GPDs_zero}. In  Figs. \ref{GPDs_zero}(a) and \ref{GPDs_zero}(b), we show the GPD $H(x,t)$ as a  function of $x$ for different values of $-t$ for u and d quarks. Similar plots of $E(x,t)$ for $u$ and $d$ quarks are shown in Figs.\ref{GPDs_zero}(c) and \ref{GPDs_zero}(d). The overall nature of both the models is same for $u$ quark while there are some  disagreements in the GPD $H(x,t)$ for the $d$ quark.  Since the $d$-quark form factors  are not well described in AdS/QCD, this disagreements are expected.  The GPD $H(x,t)$ falls off faster as $x$ increases for $d$ quark compare to $u$ quark in both the model. Unlike $H(x,t)$, the fall-off of the GPD $E(x,t)$ at large $x$ is similar for both $u$ and $d$ quarks with increasing $x$.

\begin{figure*}[htbp]
\begin{minipage}[c]{0.98\textwidth}
\small{(a)}
\includegraphics[width=7.5cm,height=5.5cm,clip]{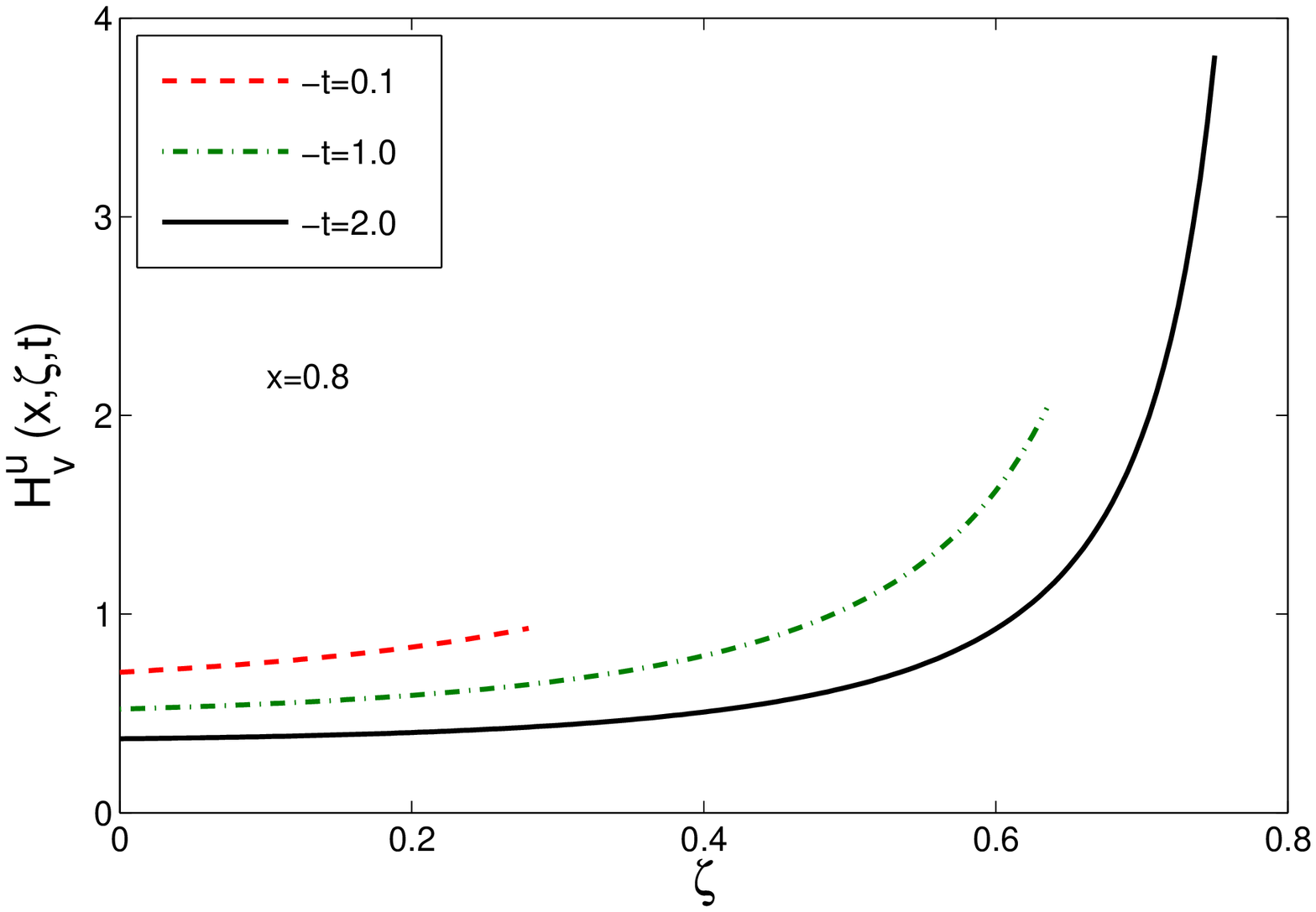}
\hspace{0.1cm}%
\small{(b)}\includegraphics[width=7.5cm,height=5.5cm,clip]{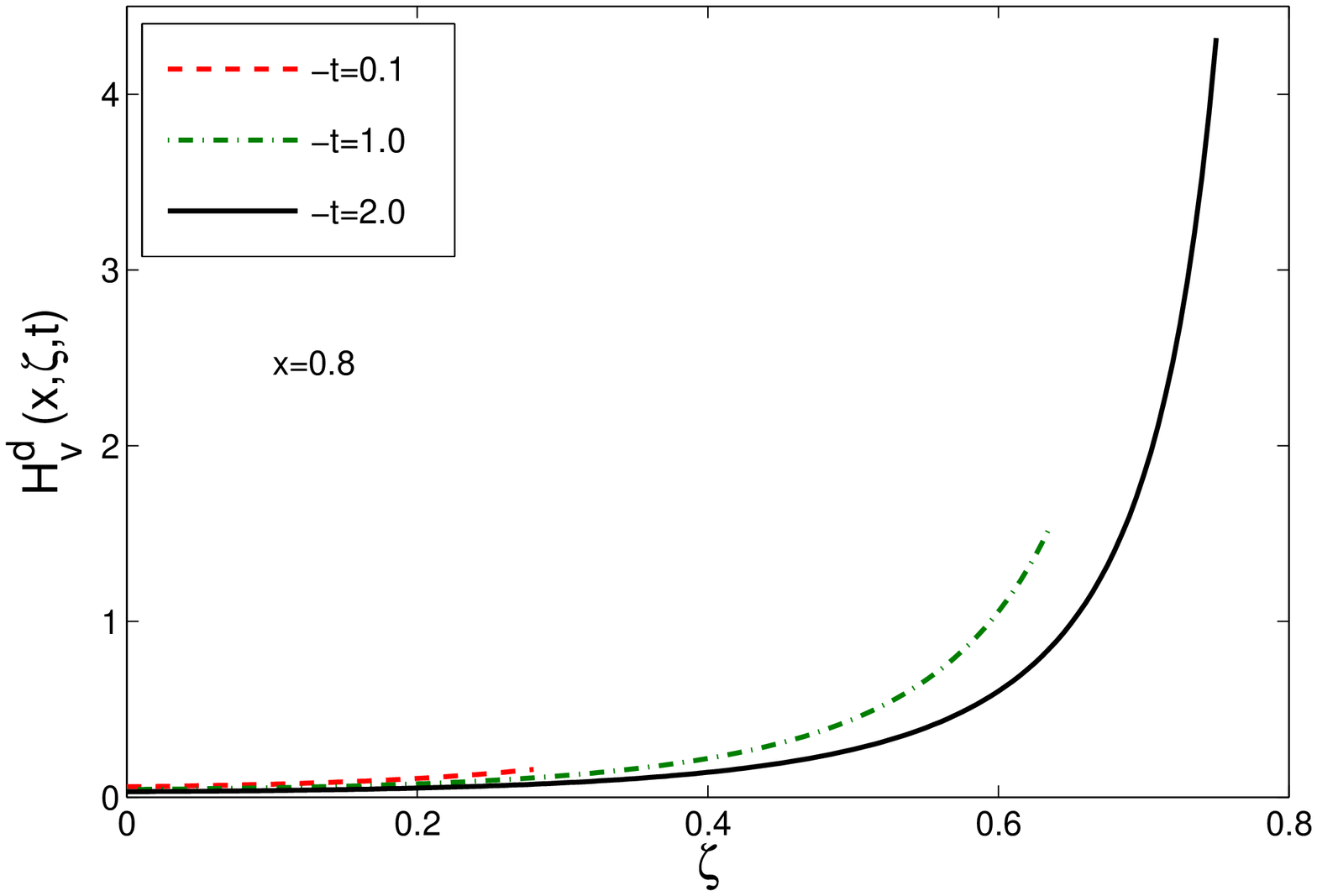}
\end{minipage}
\begin{minipage}[c]{0.98\textwidth}
\small{(c)}
\includegraphics[width=7.5cm,height=5.5cm,clip]{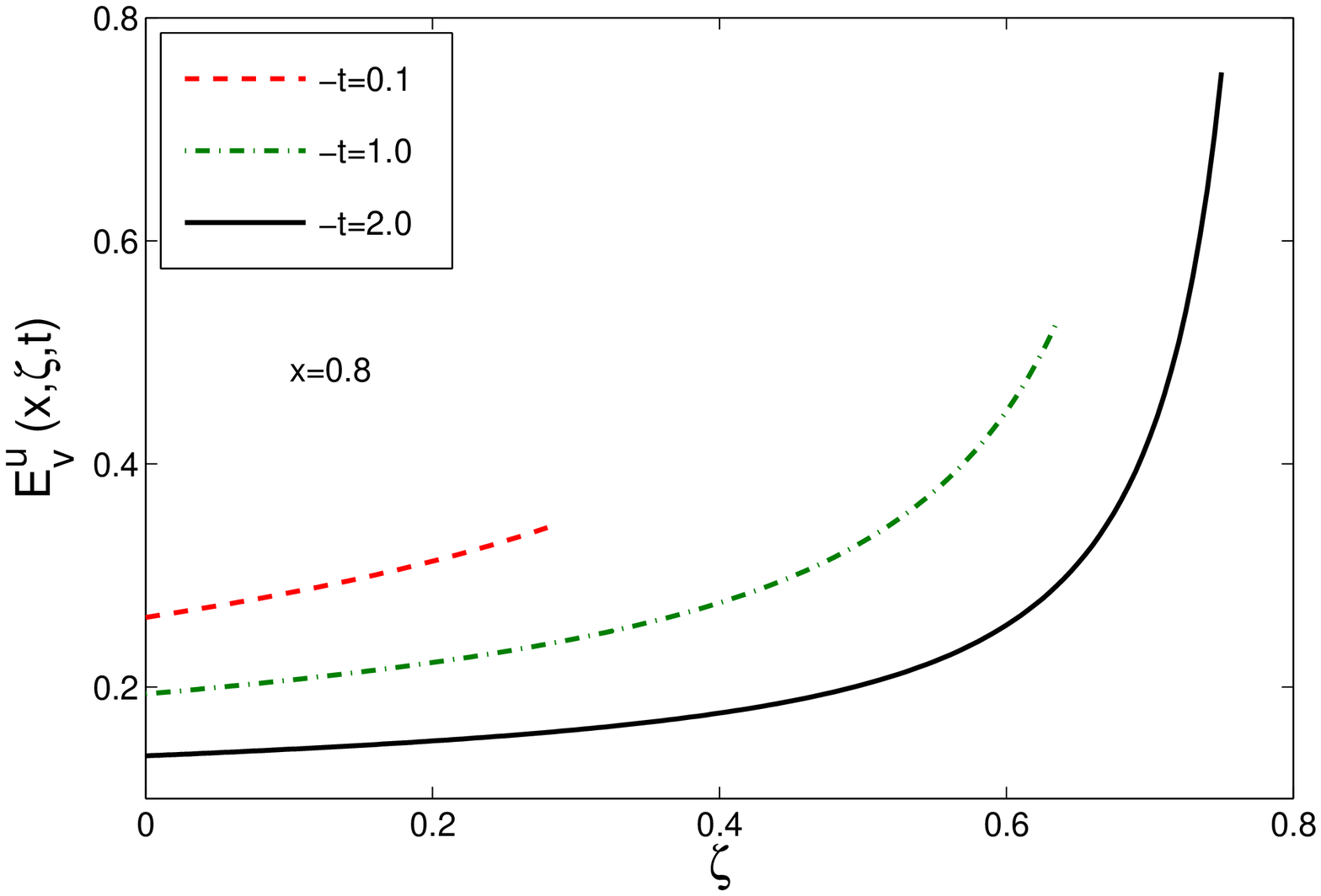}
\hspace{0.1cm}%
\small{(d)}\includegraphics[width=7.5cm,height=5.5cm,clip]{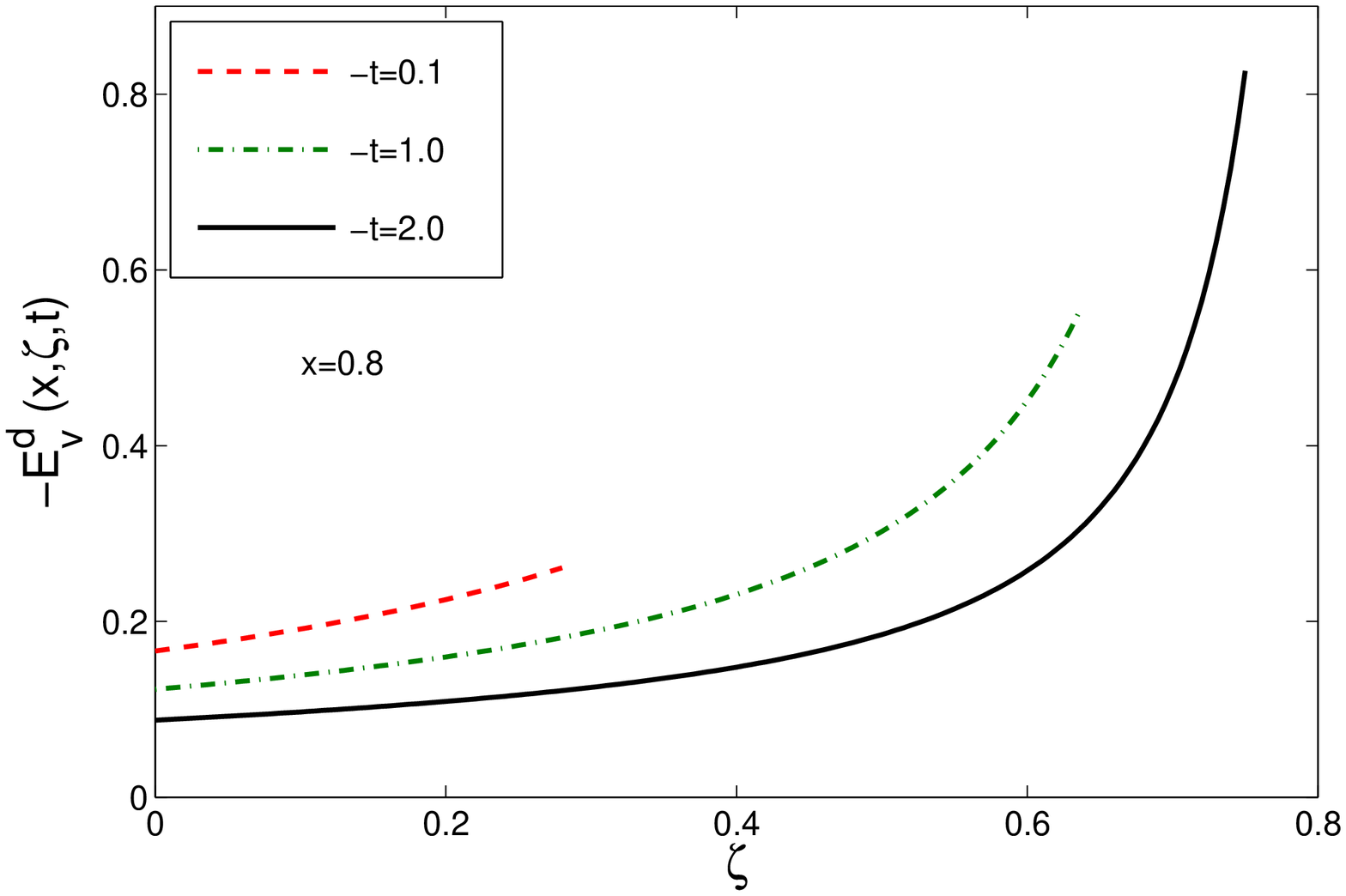}
\end{minipage}
\caption{\label{GPDs_vs_zeta}(Color online) Plots of (a) $H^u(x,\zeta,t)$ vs $\zeta$ ; (b) the same as in (a) but for $d$ quark; (c) $E^u(x,\zeta,t)$ vs $\zeta$; (d) the same as in (c) but for $d$ quark for fixed $x=0.8$. }
\end{figure*} 
\begin{figure*}[htbp]
\begin{minipage}[c]{0.98\textwidth}
\small{(a)}
\includegraphics[width=7.5cm,height=5.5cm,clip]{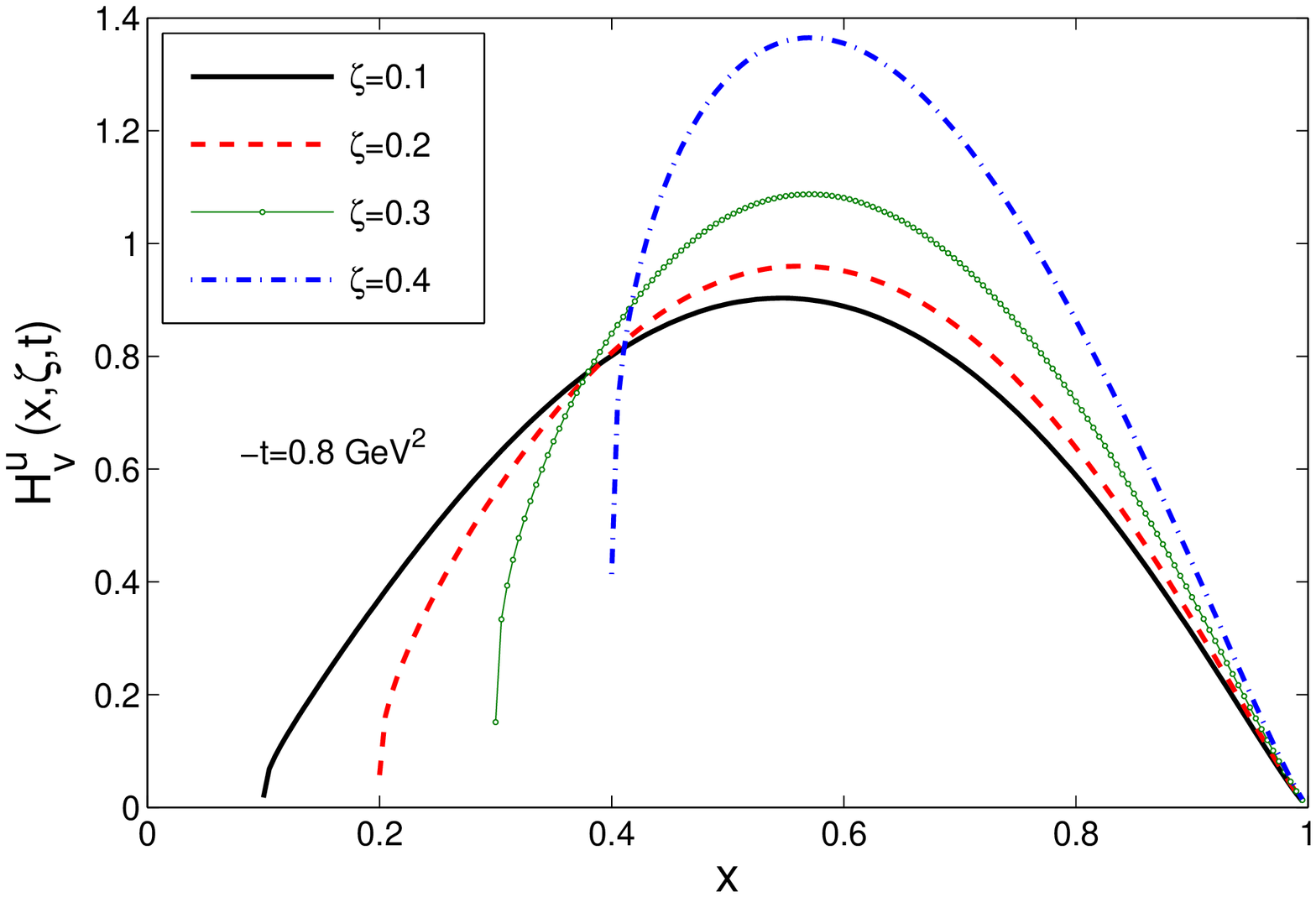}
\hspace{0.1cm}%
\small{(b)}\includegraphics[width=7.5cm,height=5.5cm,clip]{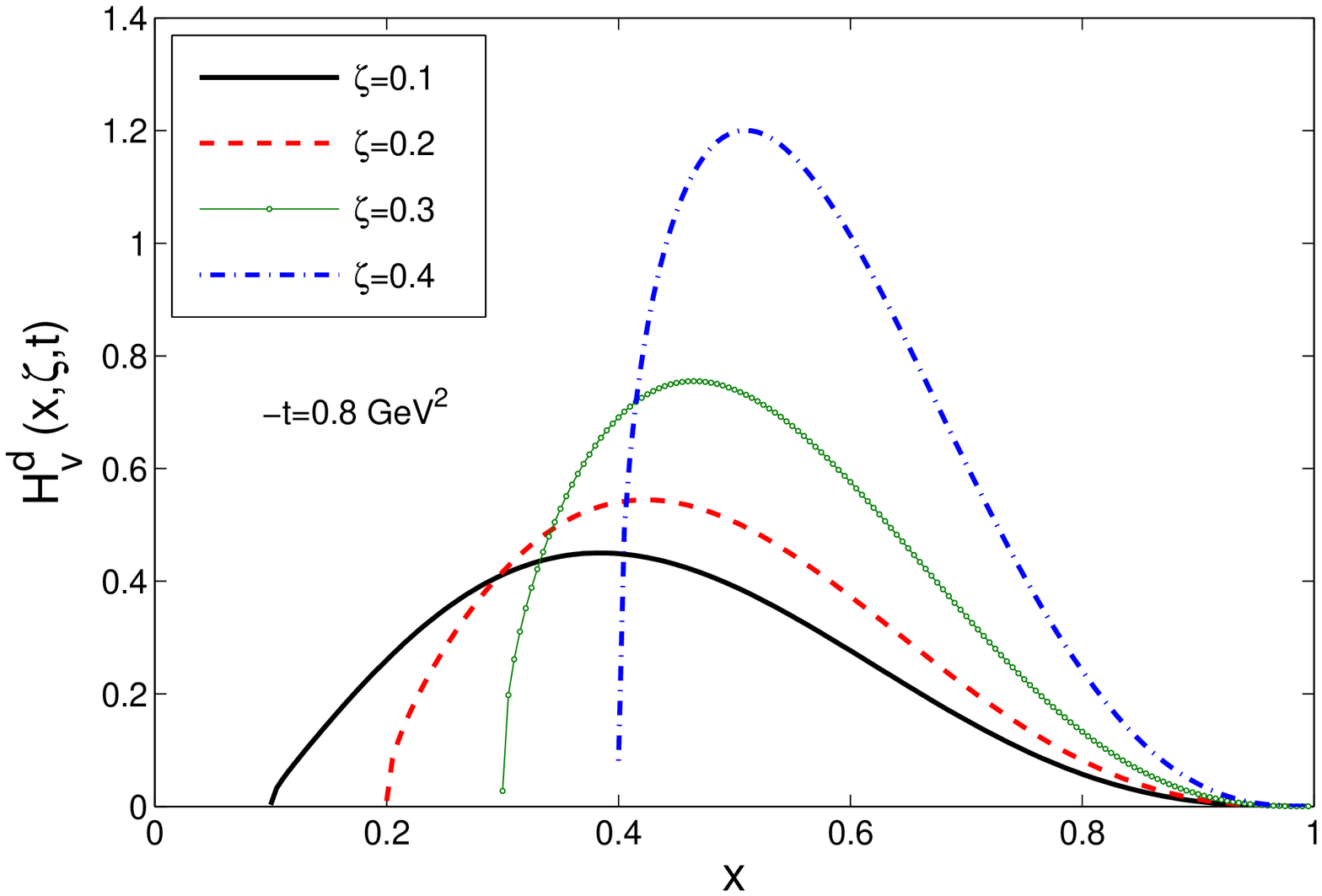}
\end{minipage}
\begin{minipage}[c]{0.98\textwidth}
\small{(c)}
\includegraphics[width=7.5cm,height=5.5cm,clip]{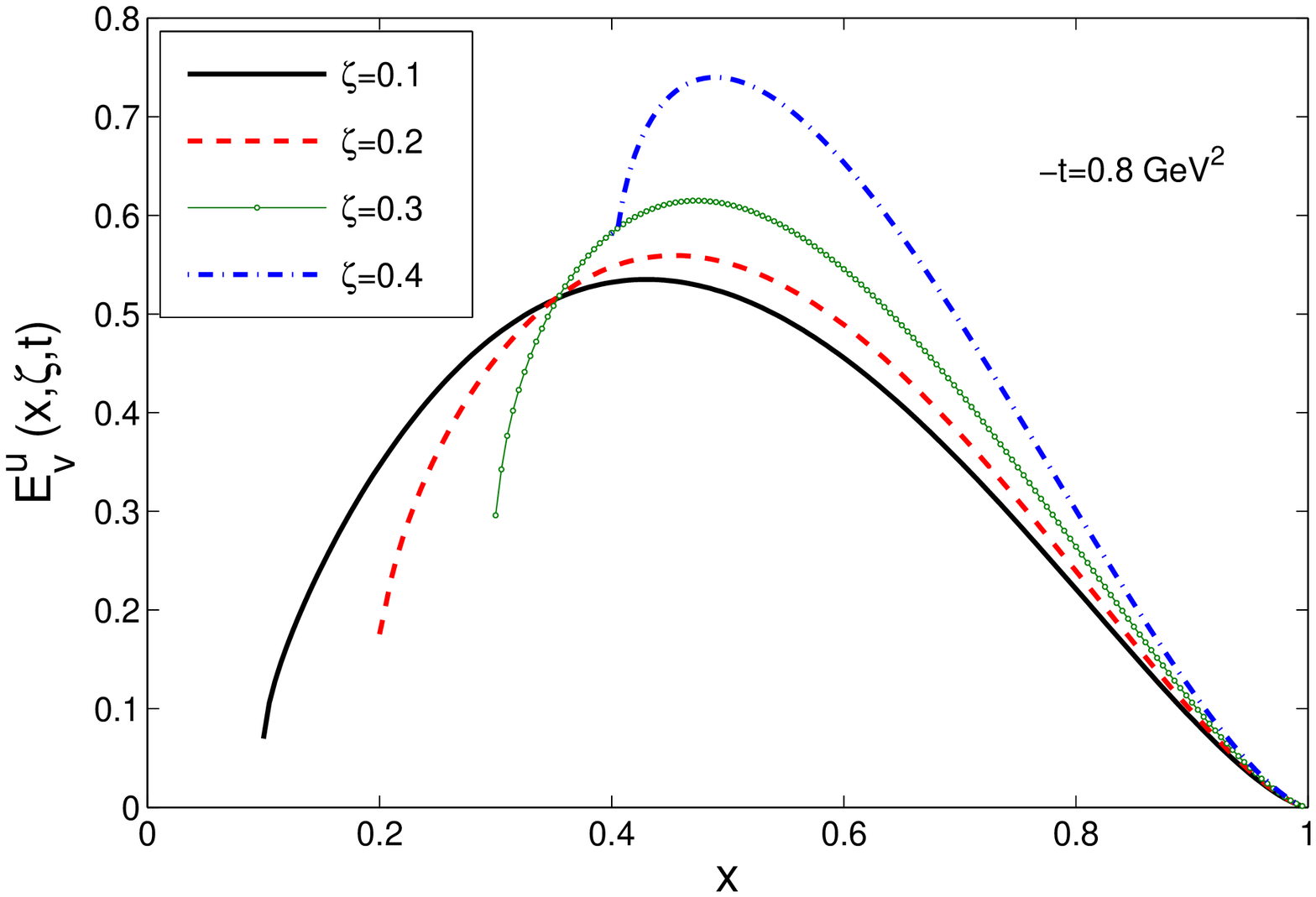}
\hspace{0.1cm}%
\small{(d)}\includegraphics[width=7.5cm,height=5.5cm,clip]{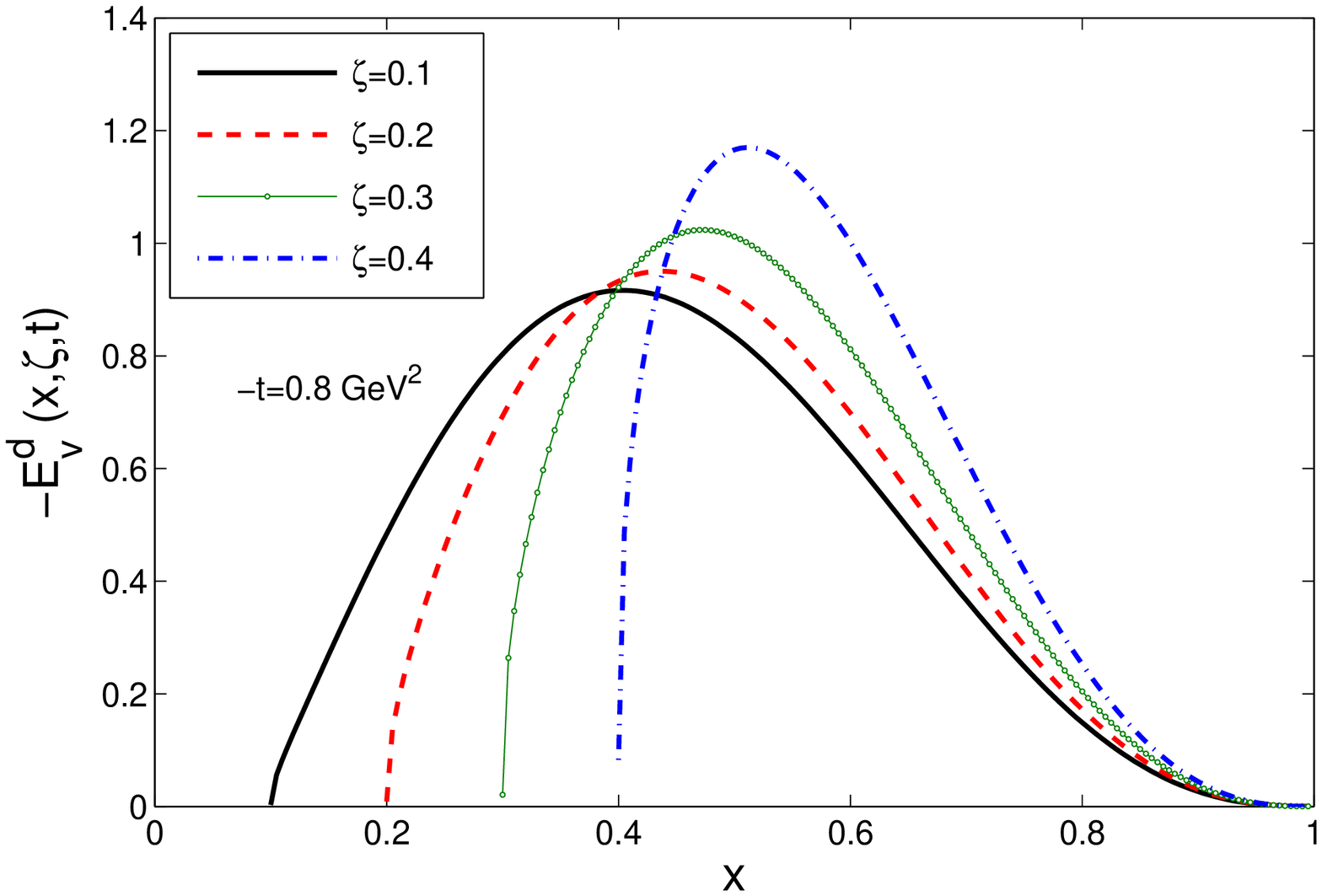}
\end{minipage}
\caption{\label{GPDs_z}(Color online) Plots of (a) $H^u(x,\zeta,t)$ vs $x$ (b) the same as in (a) but for $d$ quark (c) $E^u(x,\zeta,t)$ vs $x$ (d) the same as in (c) but for $d$ quark for fixed $-t=0.8~GeV^2$ and different values of $\zeta$. }
\end{figure*} 

In Fig.\ref{GPDs_t}, we show the skewness dependent GPDs as a function of $x$  and $t$  for a fixed $\zeta=0.15$. The overall behaviors of the GPDs with nonzero $\zeta$   are similar to the zero skewness  GPDs.
The same GPDs for a fixed $x=0.8$ are shown as a function of $\zeta$ for different values of  $-t$  in Fig. \ref{GPDs_vs_zeta}.
 In Fig.\ref{GPDs_z}(a) and \ref{GPDs_z}(b) we have plotted the GPD $H(x,\zeta,t)$ as a function of $x$ for $u$ and $d$ quarks for different values of $\zeta$ with fixed value of $-t=0.8~ GeV^2$. The similar plots of $E(x,\zeta,t)$ for $u$ and $d$ quarks are shown in Fig.\ref{GPDs_z}(c) and \ref{GPDs_z}(d).  In Fig.\ref{GPDs_z},  the peaks of all the distributions move to higher $x$ as $\zeta$ increases and  the amplitudes of the  distributions increase with increasing $\zeta$ for a fixed value of $-t$.   Due to the factor of $\sqrt{1-\zeta}$ in the denominator of the the GPD $H(x,\zeta,t)$ in Eq.(\ref{gpdH}), the increase in the magnitude  with increasing $\zeta$ is more in $H(x,\zeta,t)$ than in $E(x,\zeta,t)$. 
\subsection{\bf GPDs in transverse impact parameter space}\label{impact}
GPDs  in transverse impact parameter space are defined
as \cite{burk,burk2}:
\be
H(x,\zeta,\bfb)&=&{1\over (2 \pi)^2} \int d^2 \bfq e^{-i \bfq \cdot \bfb}
H(x,\zeta,t),\nonumber\\
E(x,\zeta,\bfb)&=&{1\over (2 \pi)^2} \int d^2 \bfq e^{-i \bfq \cdot \bfb}
E(x,\zeta,t).
\ee
Here, $\bfb$ is the transverse impact parameter. For zero skewness, $\bfb$ gives a measure of  the 
transverse distance between
the struck parton and the center of momentum of the hadron.  $\bfb$ satisfies the condition $\sum_i x_i b_{\perp i}=0$, where the sum is over the number of partons. The relative distance between the struck parton and the center of momentum of the
spectator system is given by  ${\mid \bfb \mid\over 1-x}$, which provides us an estimate of the size of the bound state \cite{diehl}. However, the exact estimation of the nuclear size is not possible as the spatial extension of the spectator system is not available from the GPDs. In the DGLAP domain $x>\zeta$, the impact parameter $\bfb$ implies the location where the quark is pulled out and pushed back to the nucleon. In the ERBL region $x<\zeta$, $\bfb$ gives the transverse location of the quark-antiquark pair inside the nucleon. 

\begin{figure*}[htbp]
\begin{minipage}[c]{0.98\textwidth}
\small{(a)}
\includegraphics[width=7.5cm,height=5.5cm,clip]{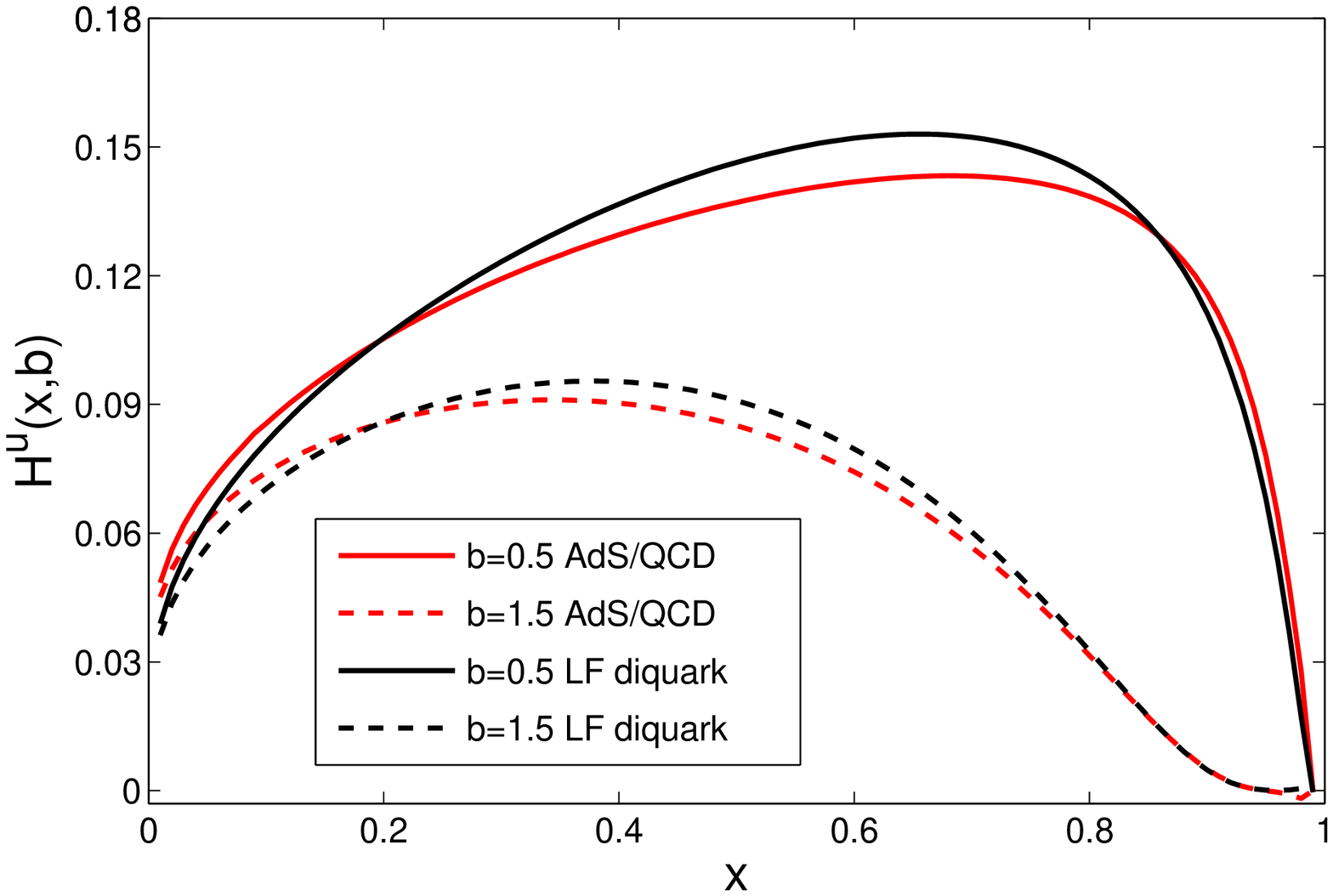}
\hspace{0.1cm}%
\small{(b)}\includegraphics[width=7.5cm,height=5.5cm,clip]{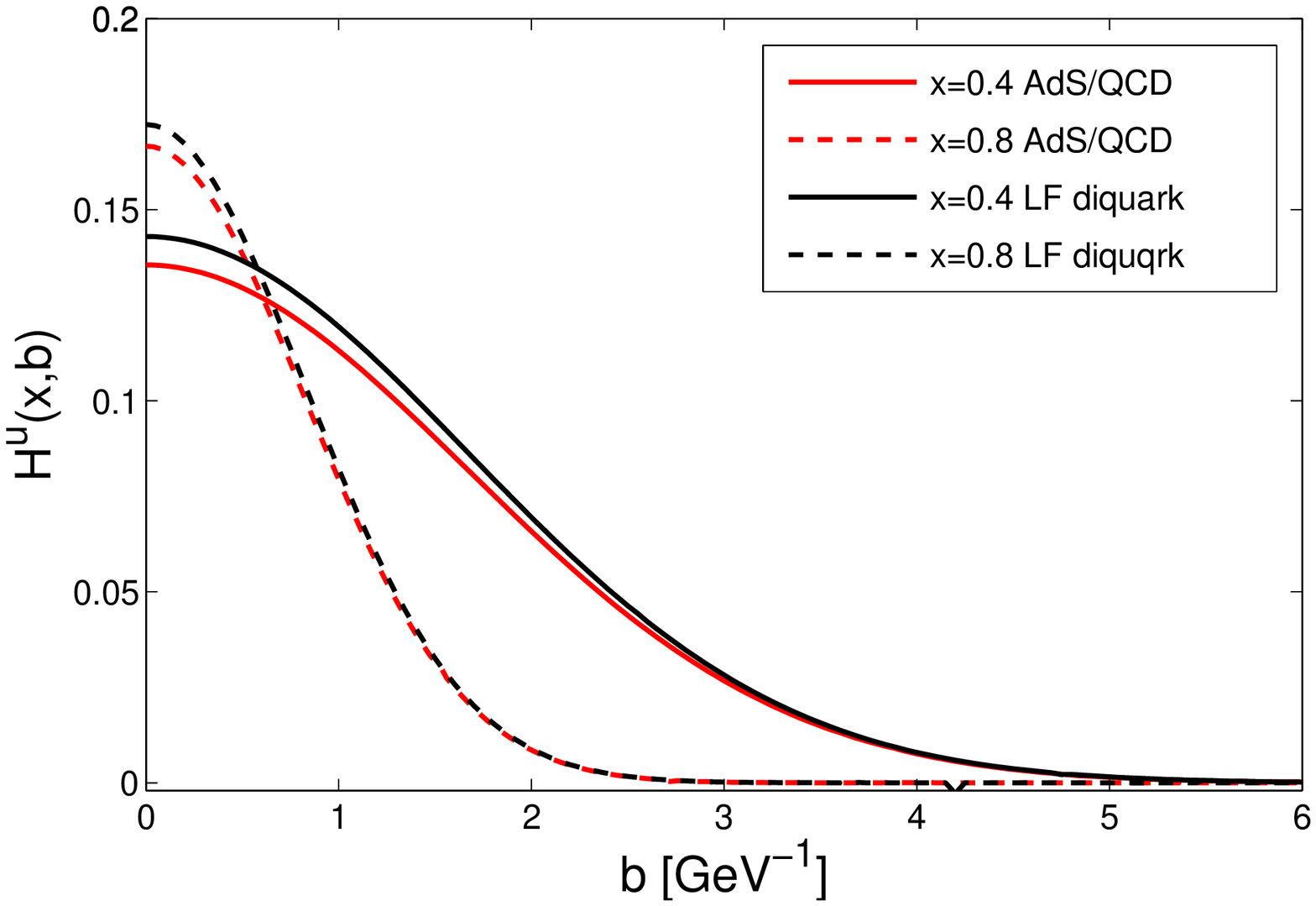}
\end{minipage}
\begin{minipage}[c]{0.98\textwidth}
\small{(c)}
\includegraphics[width=7.5cm,height=5.5cm,clip]{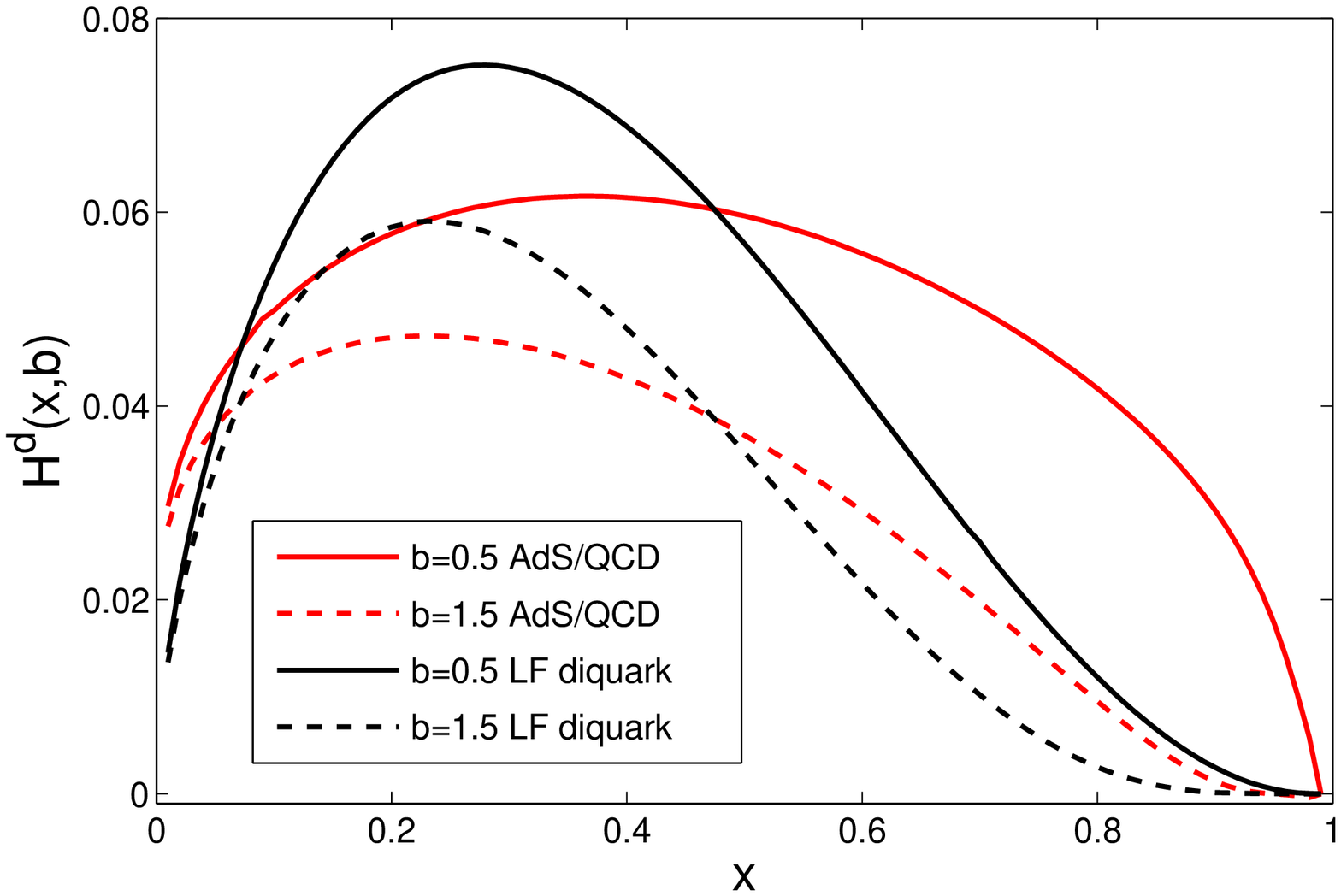}
\hspace{0.1cm}%
\small{(d)}\includegraphics[width=7.5cm,height=5.5cm,clip]{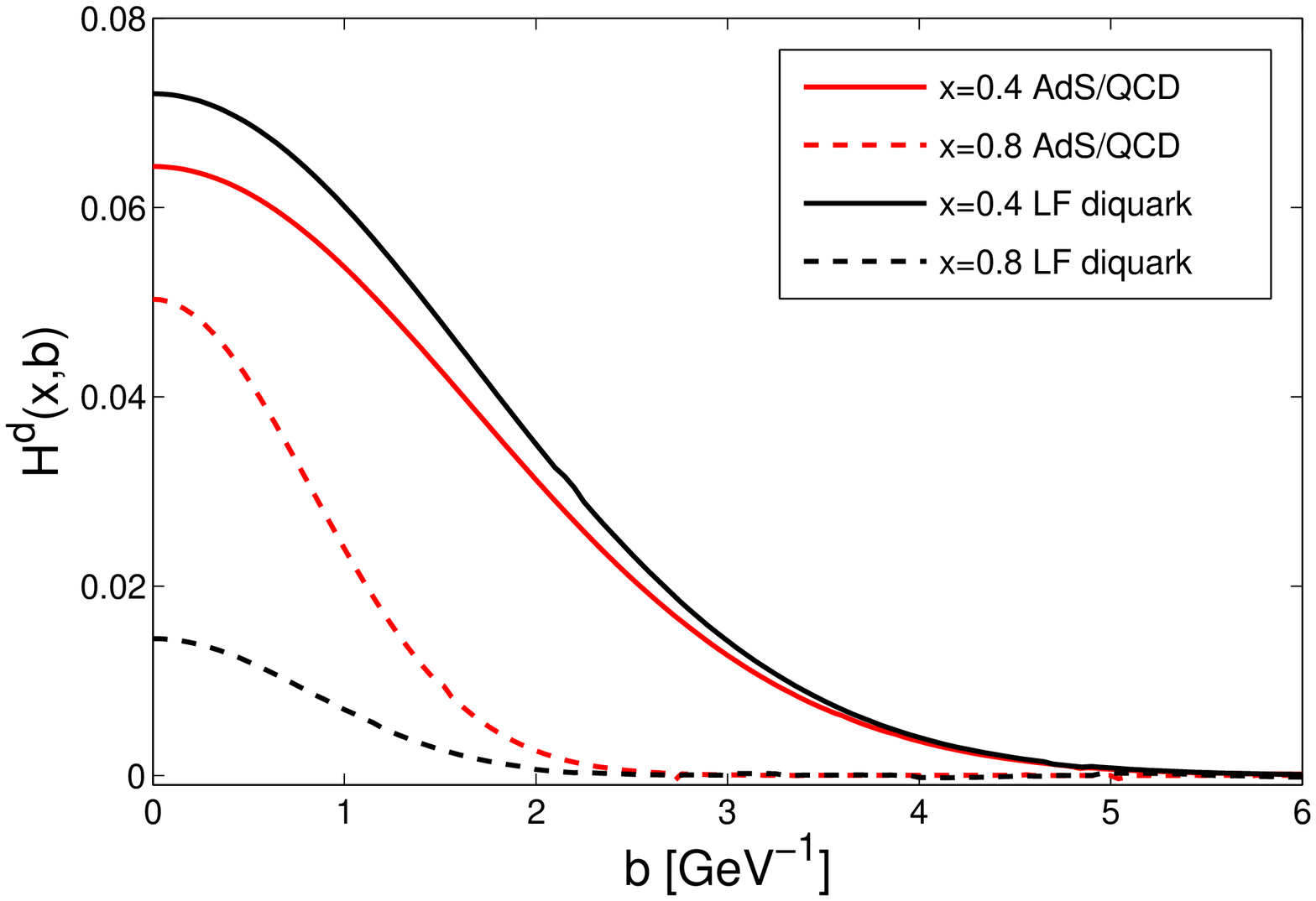}
\end{minipage}
\caption{\label{H_models}(Color online) Plots of (a) $H^u(x,\bfb)$ vs $x$ for fixed values of impact parameter $b = \mid {\bfb} \mid$;    (b) $H^u(x,\bfb)$ vs $b$ for fixed $x$;  (c) and (d) are the same as in (a) and (b)  but for $d$ quarks. $b$ is in $GeV^{-1}$.}
\end{figure*} 

\begin{figure*}[htbp]
\begin{minipage}[c]{0.98\textwidth}
\small{(a)}
\includegraphics[width=7.5cm,height=5.5cm,clip]{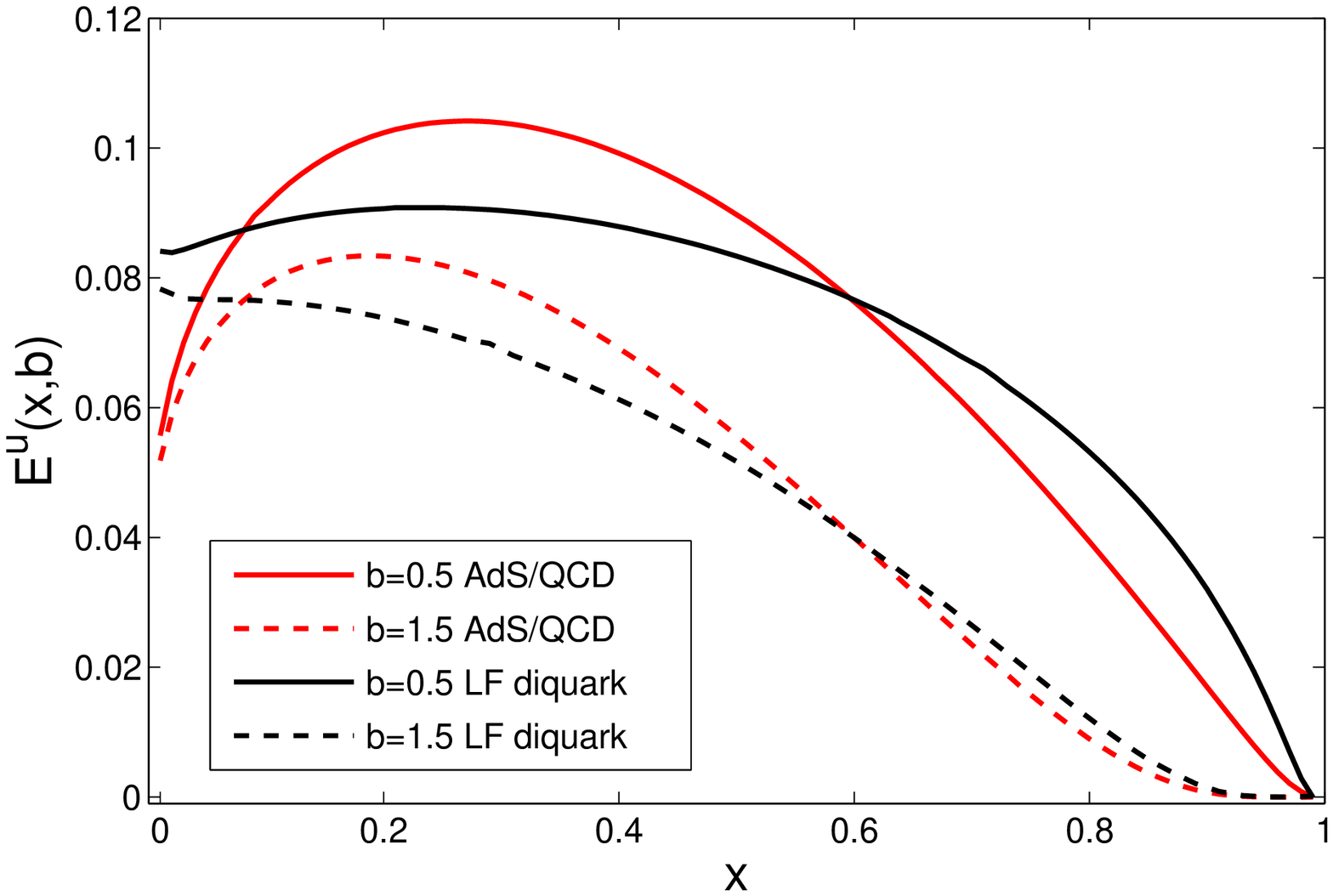}
\hspace{0.1cm}%
\small{(b)}\includegraphics[width=7.5cm,height=5.5cm,clip]{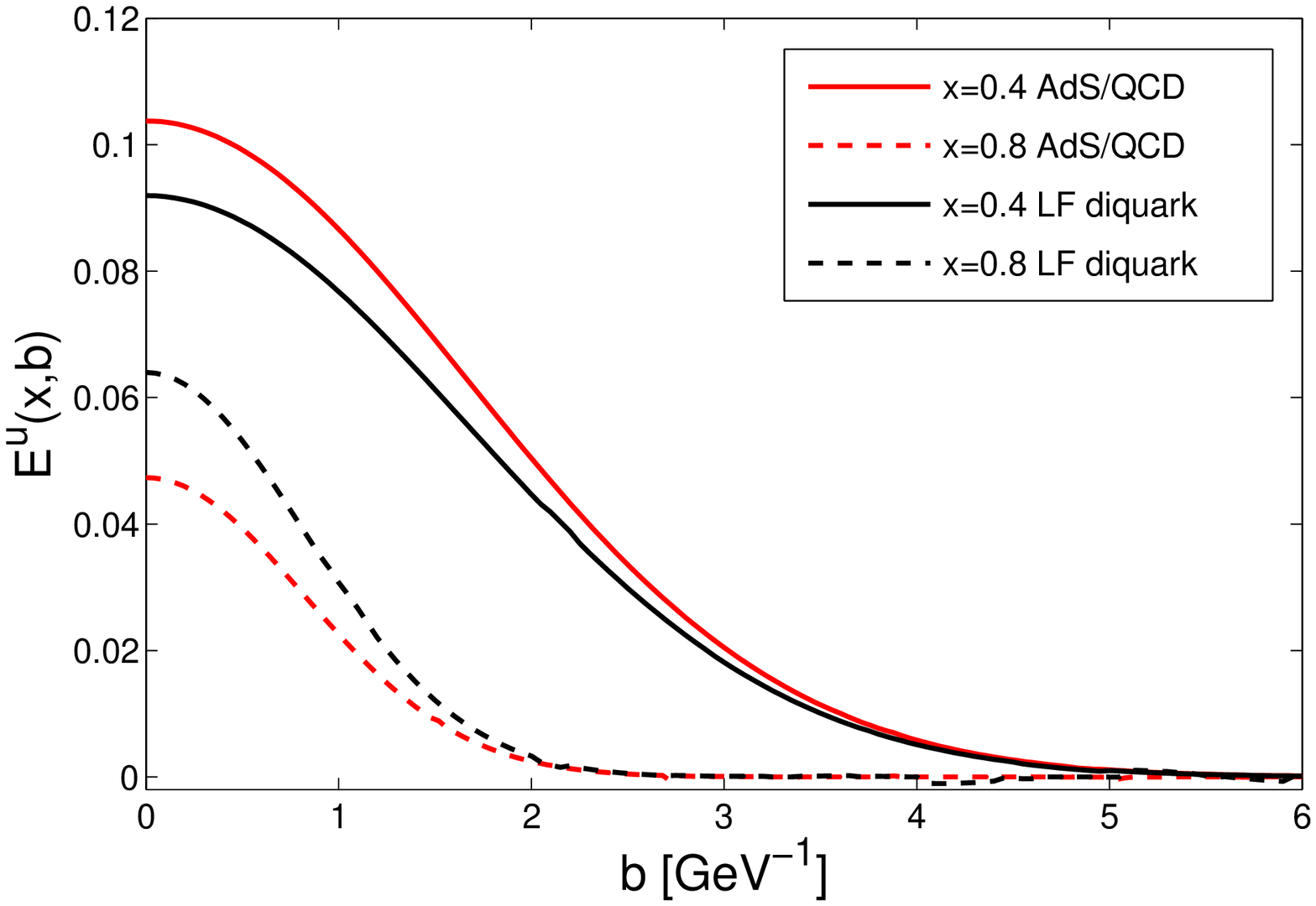}
\end{minipage}
\begin{minipage}[c]{0.98\textwidth}
\small{(c)}
\includegraphics[width=7.5cm,height=5.5cm,clip]{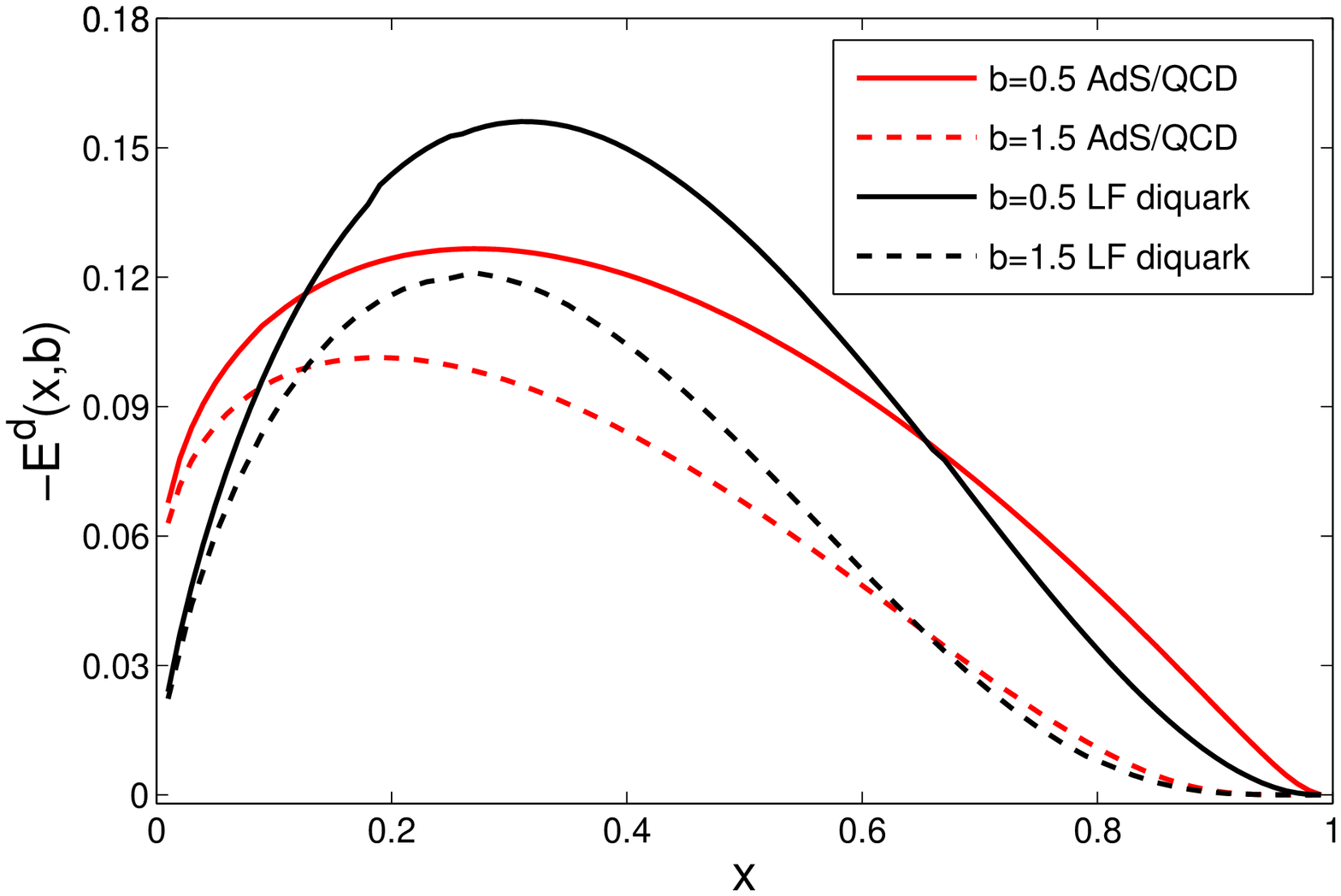}
\hspace{0.1cm}%
\small{(d)}\includegraphics[width=7.5cm,height=5.5cm,clip]{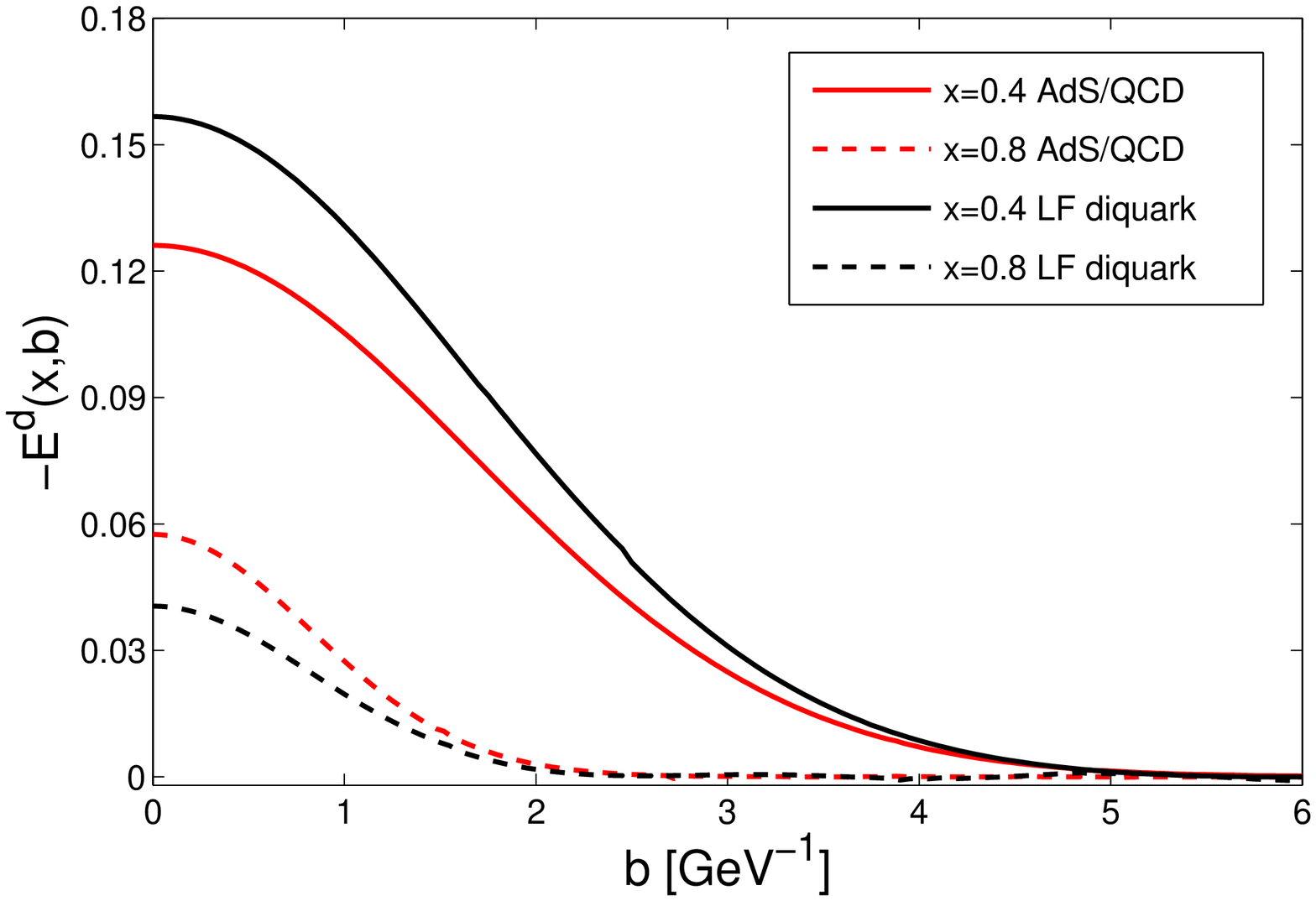}
\end{minipage}
\caption{\label{E_models}(Color online) 
 Plots of (a) $E^u(x,\bfb)$ vs $x$ for fixed values of  $b  = \mid {\bfb}\mid$; (b) $E^u(x,\bfb)$ vs. $b$ for fixed values of $x$;
 (c) the same as in (a) but  for $d$-quarks; and (d)  the same as in (b) but for $d$-quarks. $b$ is in $GeV^{-1}$.}
\end{figure*} 

\begin{figure*}[htbp]
\begin{minipage}[c]{0.98\textwidth}
\small{(a)}
\includegraphics[width=7.5cm,height=5.5cm,clip]{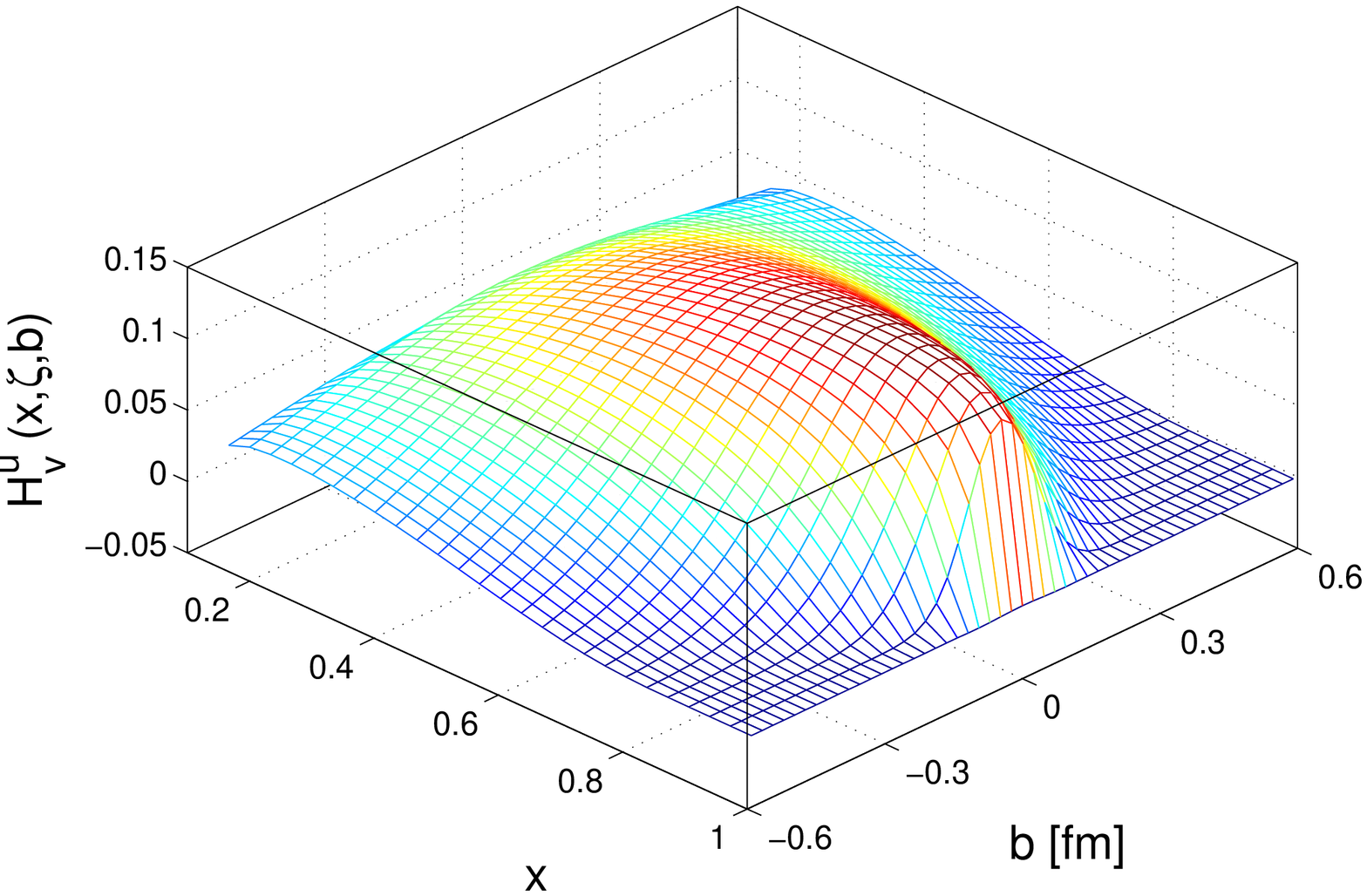}
\hspace{0.1cm}%
\small{(b)}\includegraphics[width=7.5cm,height=5.5cm,clip]{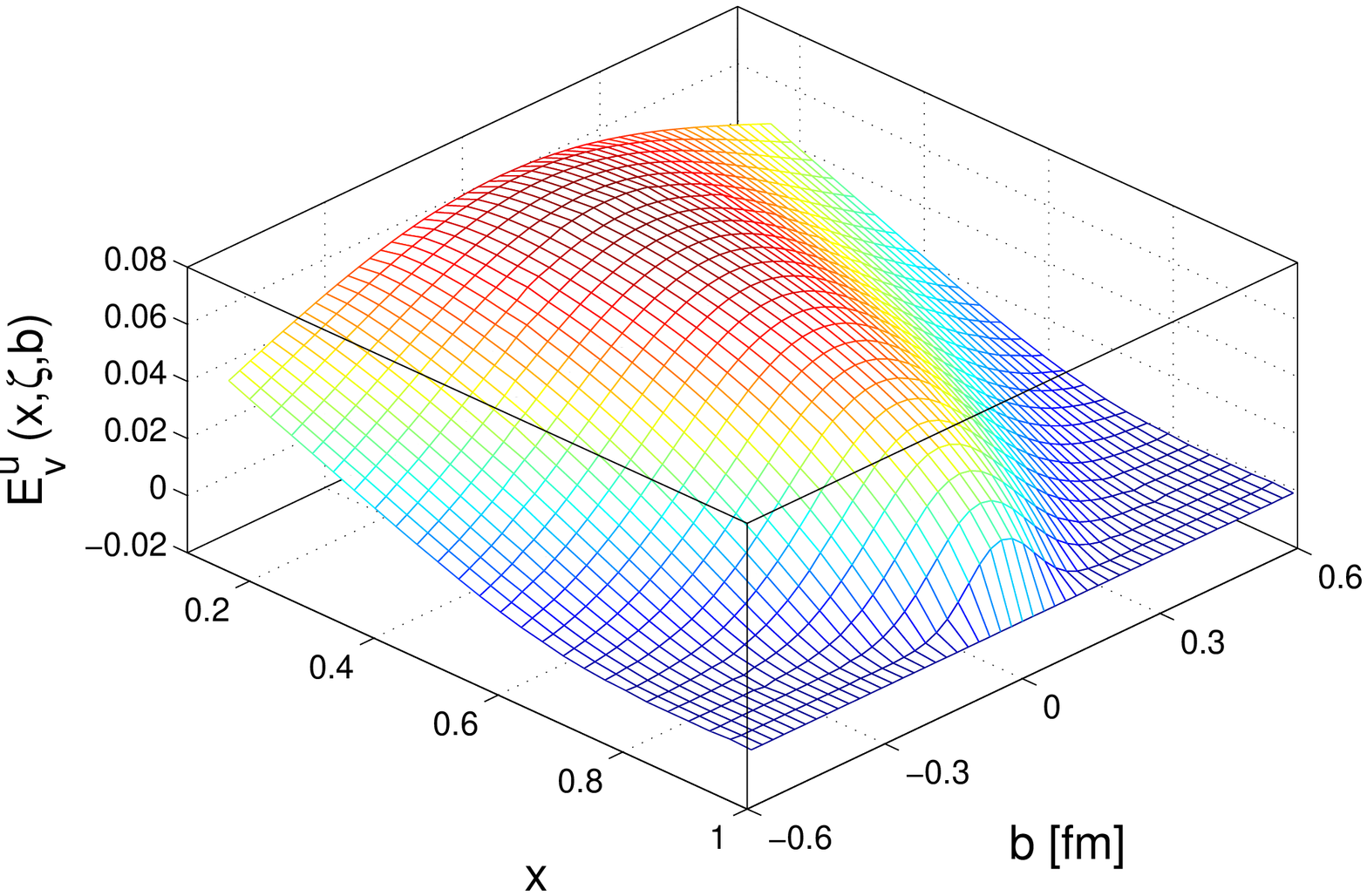}
\end{minipage}
\begin{minipage}[c]{0.98\textwidth}
\small{(c)}
\includegraphics[width=7.5cm,height=5.5cm,clip]{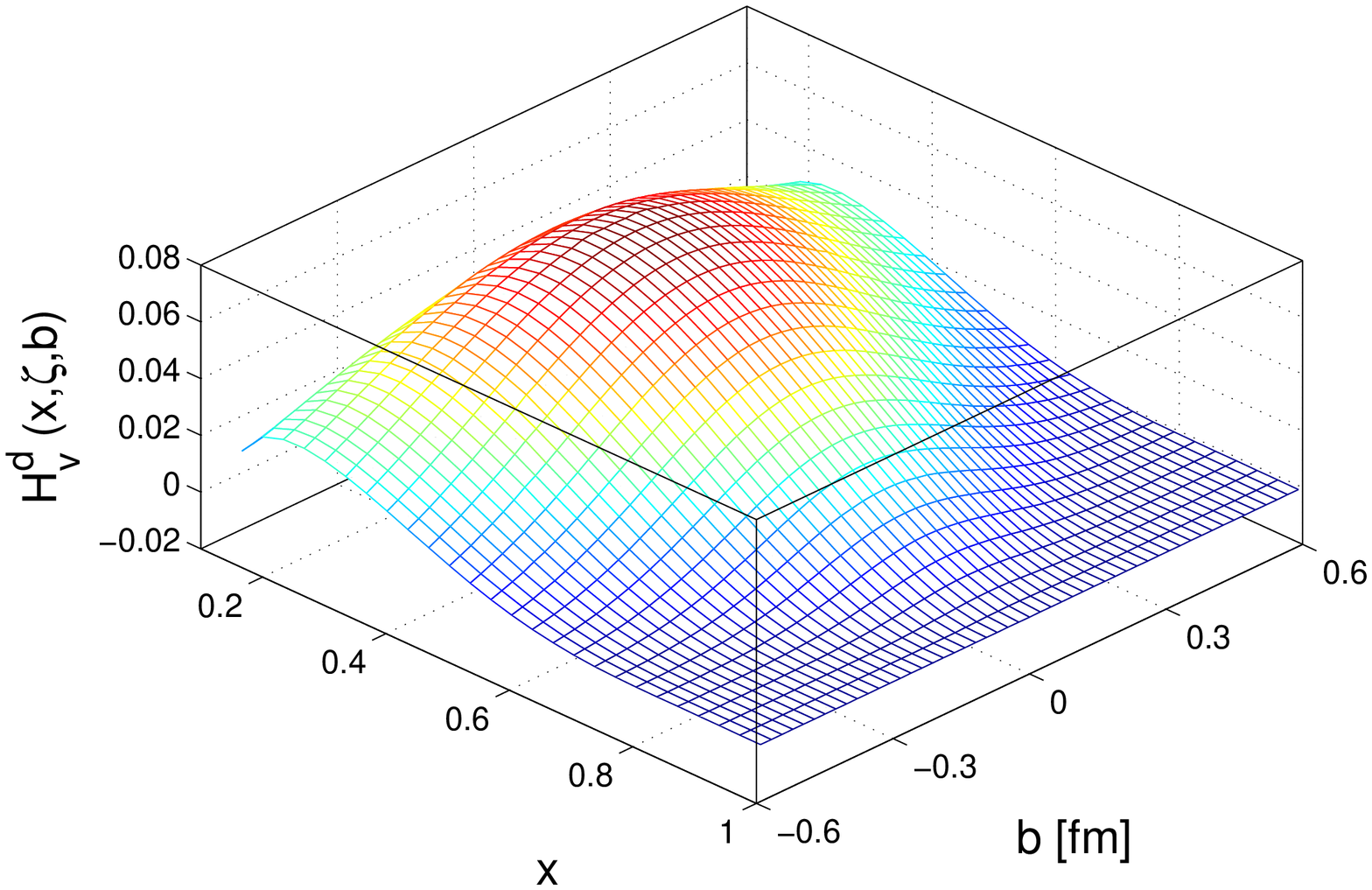}
\hspace{0.1cm}%
\small{(d)}\includegraphics[width=7.5cm,height=5.5cm,clip]{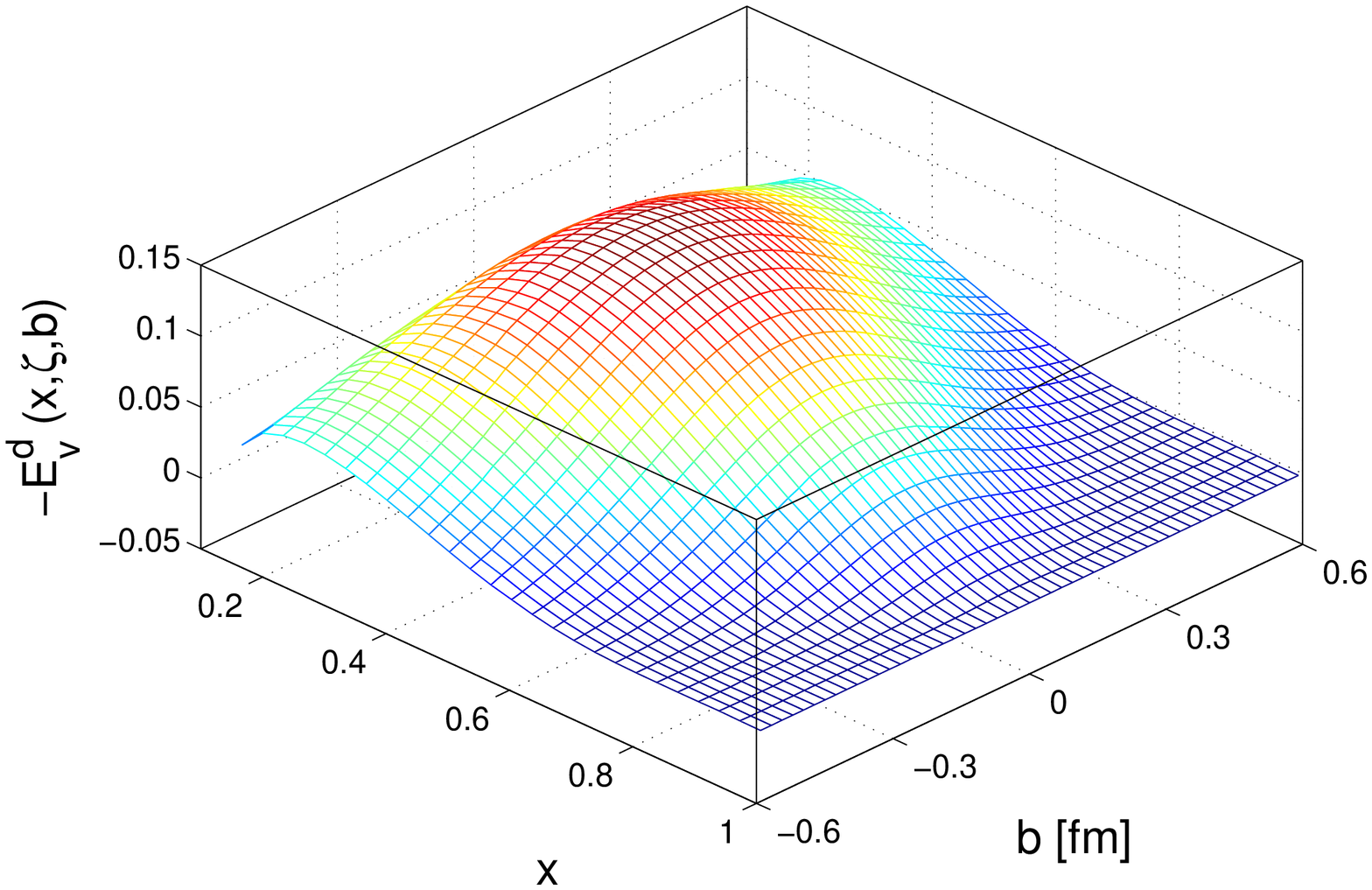}
\end{minipage}
\caption{\label{gpd_xb}(Color online) 
 Plots of (a) $H^u(x,\zeta,\bfb)$ vs $x$ and $b  = \mid {\bfb} \mid$; (b) $E^u(x,\zeta,\bfb)$ vs. $x$ and $b$;
 (c) the same as in (a) but  for $d$-quarks; and (d)  the same as in (b) but for $d$-quarks for fixed value of $\zeta=0.15$. }
\end{figure*} 

\begin{figure*}[htbp]
\begin{minipage}[c]{0.98\textwidth}
\small{(a)}
\includegraphics[width=7.5cm,height=5.5cm,clip]{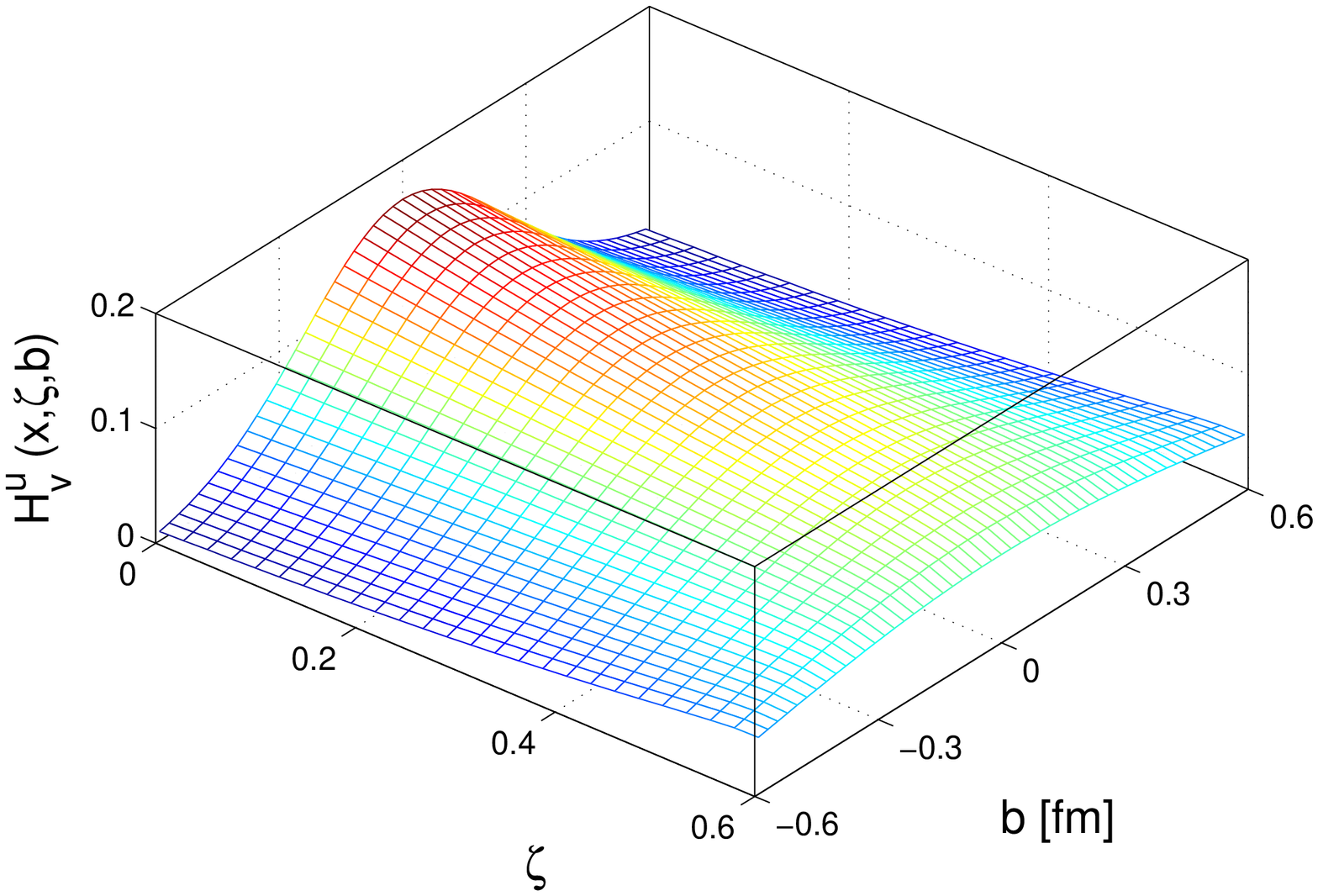}
\hspace{0.1cm}%
\small{(b)}\includegraphics[width=7.5cm,height=5.5cm,clip]{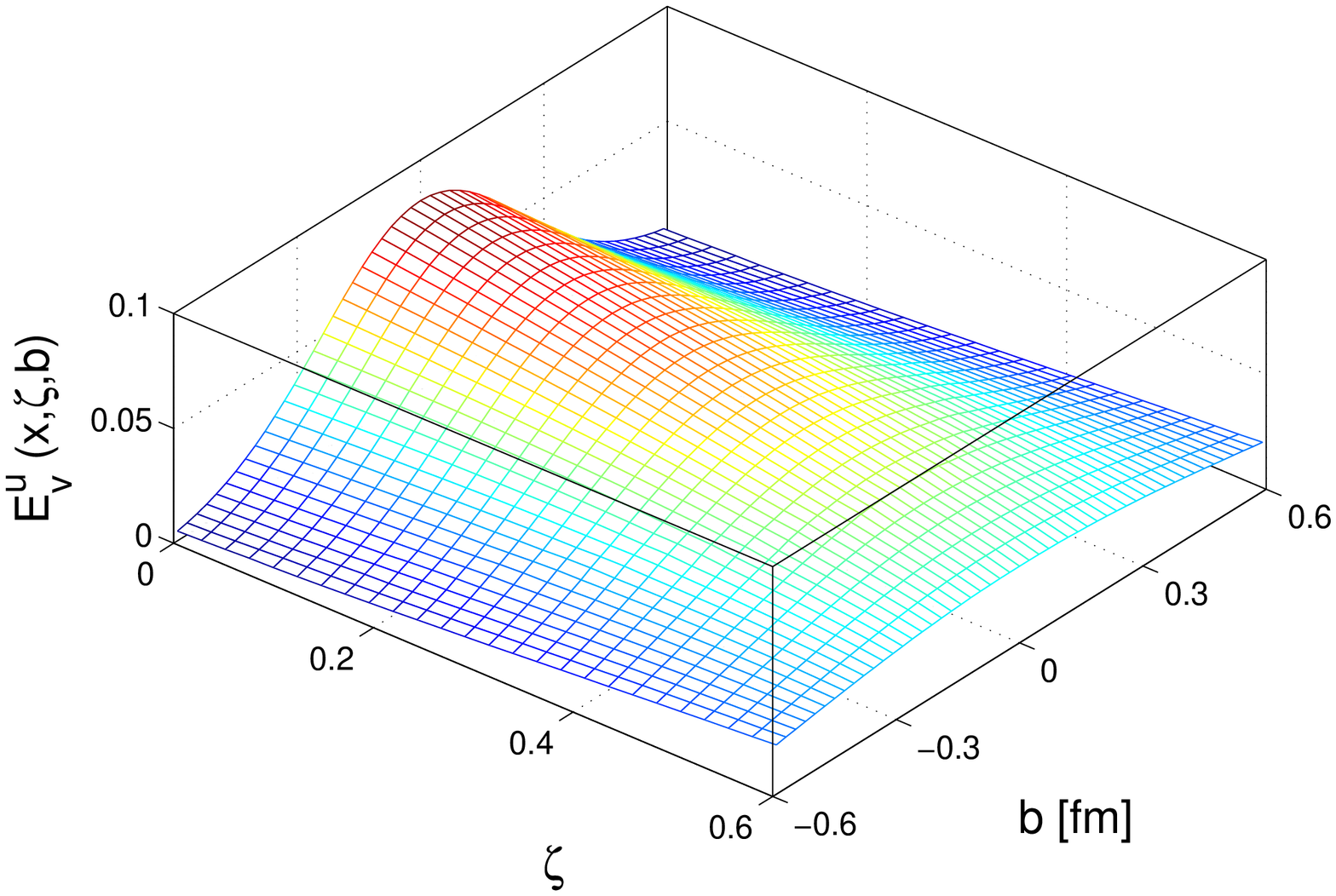}
\end{minipage}
\begin{minipage}[c]{0.98\textwidth}
\small{(c)}
\includegraphics[width=7.5cm,height=5.5cm,clip]{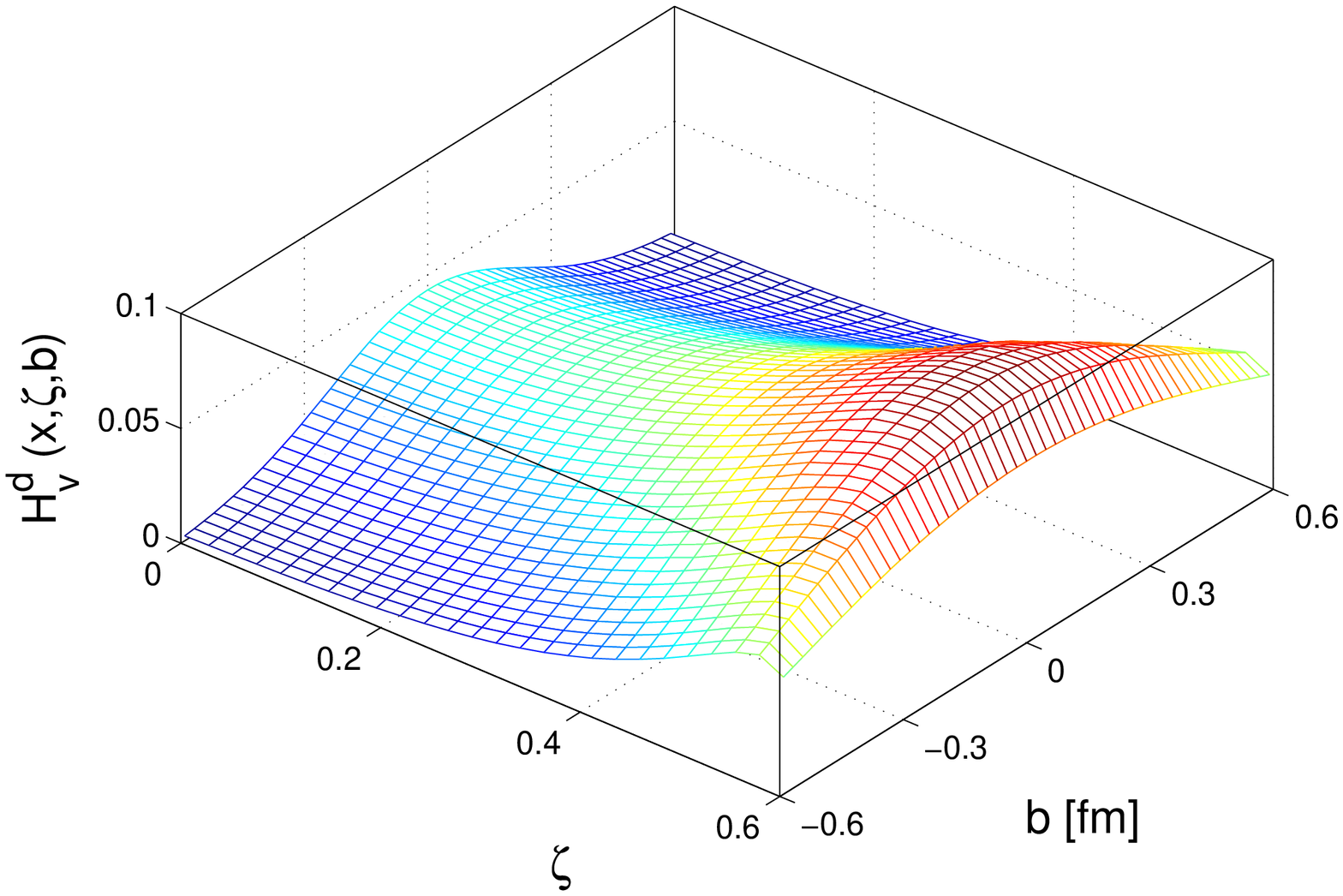}
\hspace{0.1cm}%
\small{(d)}\includegraphics[width=7.5cm,height=5.5cm,clip]{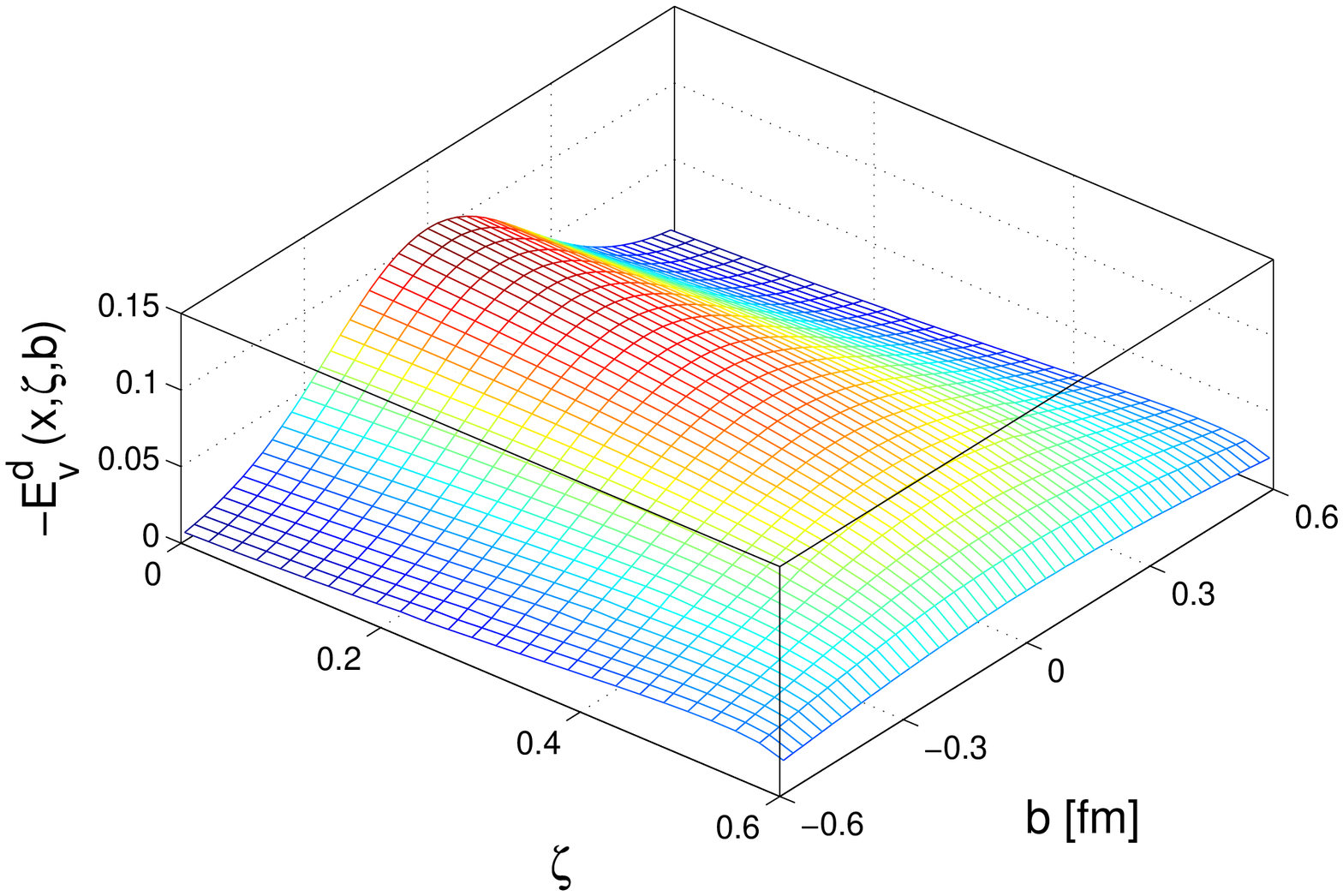}
\end{minipage}
\caption{\label{gpd_zb}(Color online) 
 Plots of (a) $H^u(x,\zeta,\bfb)$ vs $\zeta$ and $b  = \mid {\bfb} \mid$; (b) $E^u(x,\zeta,\bfb)$ vs. $\zeta$ and $b$;
 (c) the same as in (a) but  for $d$-quarks; and (d)  the same as in (b) but for $d$-quarks for fixed value of $x=0.6$. }
\end{figure*}

In Fig .\ref{H_models}, we  compare the transverse impact parameter dependent proton GPD $H(x,\bfb)$ for zero skewness in the light-front quark-diquark model and in the soft-wall AdS/QCD. Unlike $H^d(x,\bfb)$, the diquark model results for $H^u(x,\bfb)$ are  in good agreement with AdS/ QCD. The GPD $H^d(x,\bfb)$ fall off slowly for large $x$ in AdS/QCD compared to the diquark model while the fall-off of $H^u(x,\bfb)$ in both models is same. The reason of the disagreement in $H^d(x,\bfb)$ is that the AdS/QCD model is unable to reproduce $F_1^d$ to match with experimental data whereas the form factor in the diquark model  agrees well with the data (see Fig.\ref{FF_flavors}(b)).
The overall shapes of the curves in Fig.\ref{H_models}(a) and (c) are due to the fact that the  two particle LFWFs are effectively functions of  ``$x$-weighted transverse variable" \cite{BT}
 $z=\sqrt{x(1-x)}  | b_\perp |$ which is true for both the AdS/QCD  and the diquark models. Both the models lack the symmetry about $x\to (1-x)$,  the asymmetry is more prominent for $d$ quark, in the the diquark model  due to the  parameters   $a_q^{(i)} \ne b_q^{(i)}$ in the diquark wave functions.
 In Fig. \ref{E_models} we have compared the two models for the proton GPD $E(x,\bfb)$. The GPD $E(x,\bfb)$ in impact parameter space in the models are similar in behavior for both $u$ and $d$ quarks, though the agreements in the magnitudes are not exact. In AdS/QCD, the nature of $E(x,\bfb)$ for $u$ and $d$ quarks is almost similar when 
plotted against $x$ for fixed values of impact parameter 
$b=\mid\bfb\mid$, whereas in the diquark model, it shows quite different behavior for both $u$ and $d$ quarks. But the behaviors of $H(x,\bfb)$  with respect to $x$ are distinctly different for  $u$ and $d$ quarks in both the models as can be seen in Fig. \ref{H_models}. 
It is interesting to note that in both cases, the GPD $H(x,\bfb)$ is larger for $u$-quarks than $d$-quarks  whereas the magnitude of  the GPD $E(x,\bfb)$ is marginally  larger for $d$-quarks than the that for $u$-quarks at small values of impact parameter $b$. The similar behavior of the GPDs of a phenomenological model was observed in \cite{CMM}. Another interesting behavior of all the GPDs is that the width of all the distributions in transverse impact parameter space  decreases as $x$ increases, which implies that the distributions are more localized near the center of momentum for higher values of $x$. 

The skewness dependent GPDs in transverse impact parameter space for $u$ and $d$ quarks as functions of $x$ and $b$  are shown in Fig.\ref{gpd_xb} for a fixed value of $\zeta =0.15$. Similarly, all GPDs as functions of $\zeta$ and $b$ for a fixed value of $x=0.6$ are shown in Fig.\ref{gpd_zb}. 
 Though there is no divergence at $x=\zeta$, in the numerical computations, the exact value of $x=\zeta$ has been omitted for technical reason. At small value of $b$, $H(x,\zeta,\bfb)$ decreases for $u$ quark but increases for $d$ quark with increasing $\zeta$,
 while $E(x,\zeta,\bfb)$ decreases with increasing $\zeta$  for both $u$ and $d$  for a fixed value of $x$. For the fixed values $x$ and low $\zeta$, the peak of $u$ quark is sufficiently large compare to $d$ for $H(x,\zeta,\bfb)$ but for $E(x,\zeta,\bfb)$, $d$ quark is marginally large compare to $u$ quark. Substantial differences are observed in both GPDs between $u$ and $d$ quarks when the GPDs are plotted against $x$ for fixed values of $\zeta$ and $b$.  $H^d(x,\zeta,b)$ seems to be nonzero at $x=\zeta$  in Fig.\ref{gpd_zb}(c), but  it is due to the fact that $x=\zeta$ is not included in the plot.  It goes to zero as  $x\to \zeta$.
 It is interesting to note that the peaks of all the distributions also become broader as $\zeta$ increases for a fixed value of $x$.  This  means that the probability of hitting the active quark  at a larger transverse impact parameter $b$  increases as the  momentum transfer in the longitudinal direction increases.  

\begin{figure*}[htbp]
\begin{minipage}[c]{0.98\textwidth}
\small{(a)}
\includegraphics[width=7.5cm,height=5.5cm,clip]{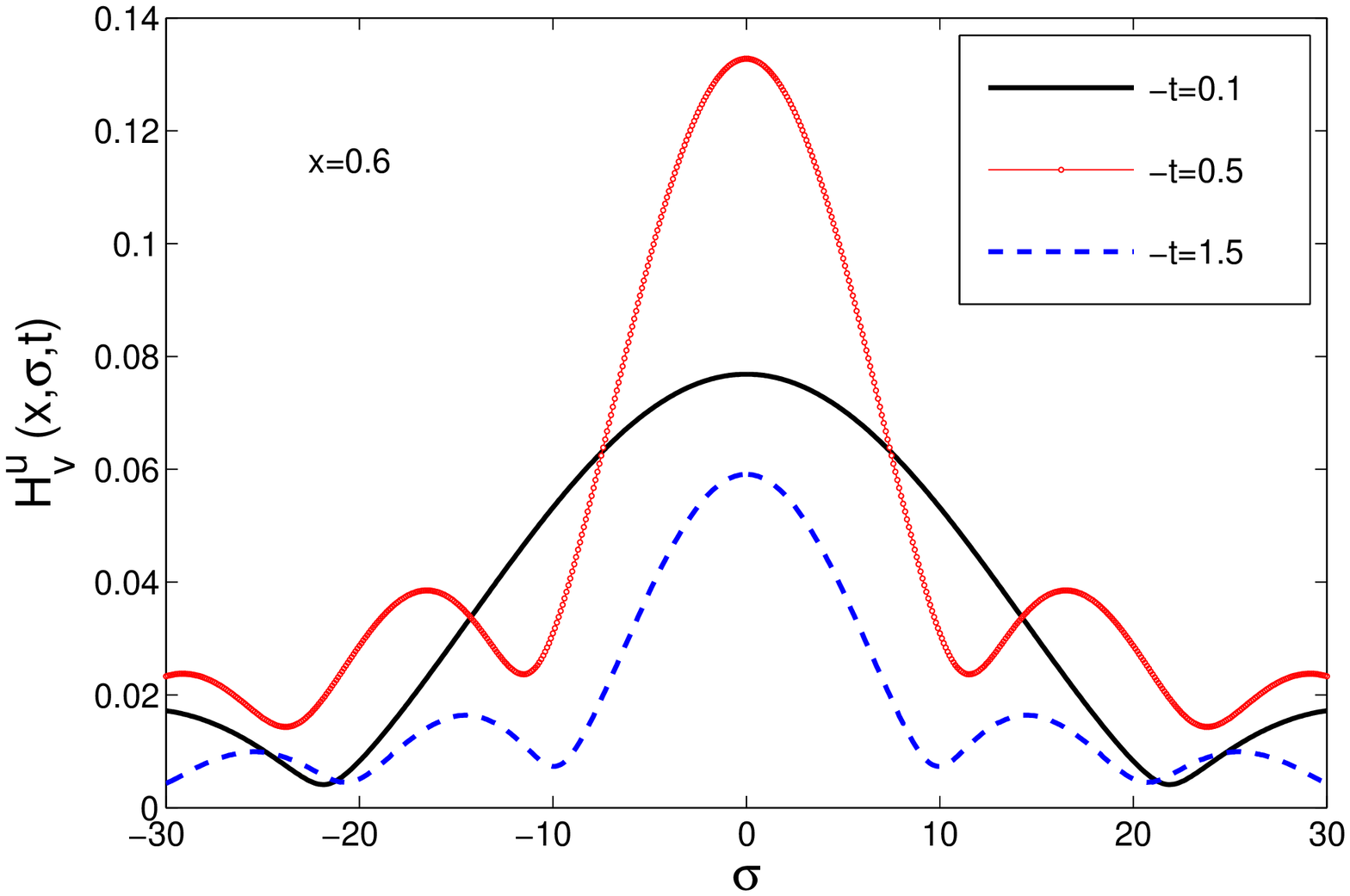}
\hspace{0.1cm}%
\small{(b)}\includegraphics[width=7.5cm,height=5.5cm,clip]{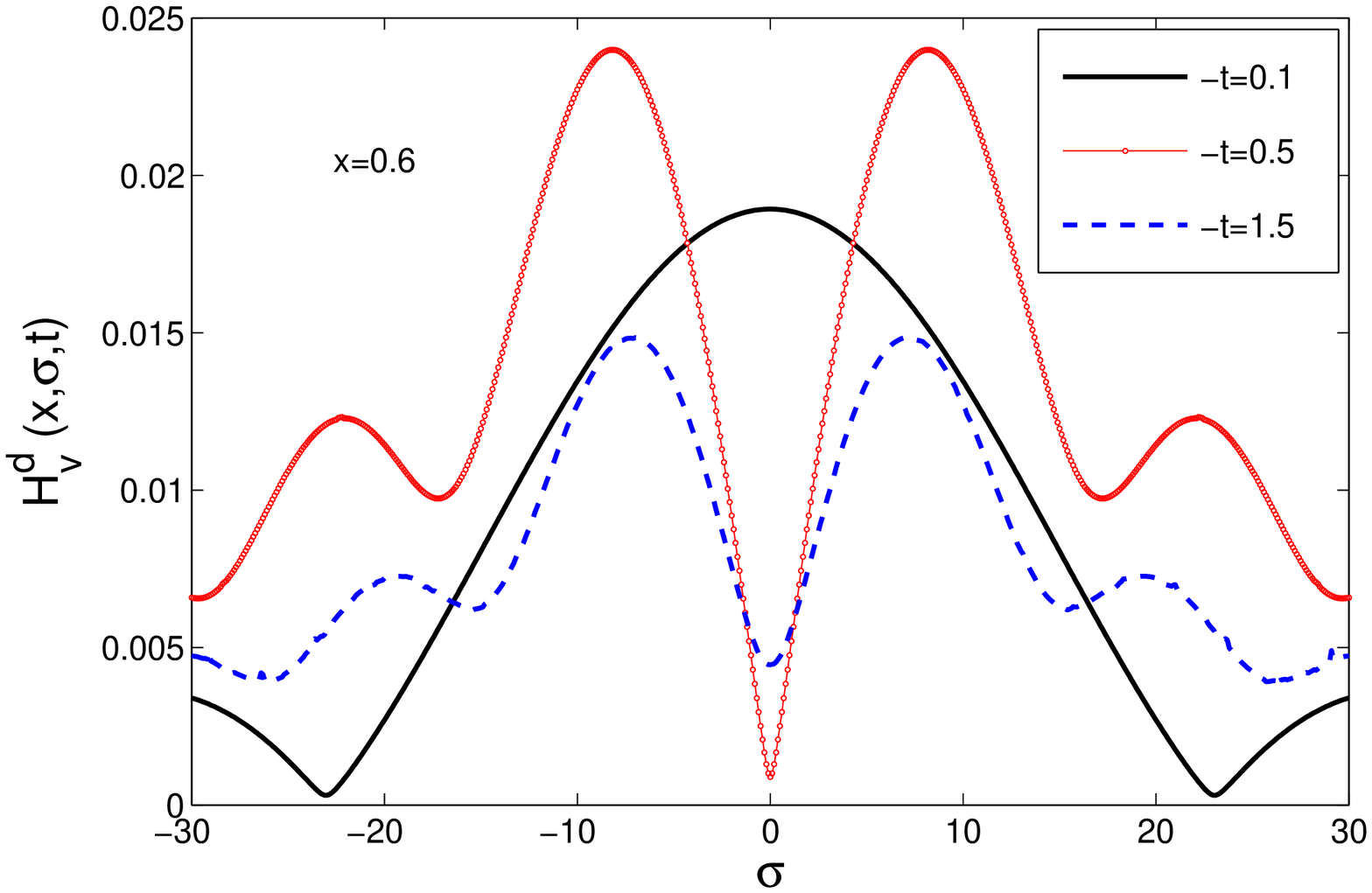}
\end{minipage}
\begin{minipage}[c]{0.98\textwidth}
\small{(c)}
\includegraphics[width=7.5cm,height=5.5cm,clip]{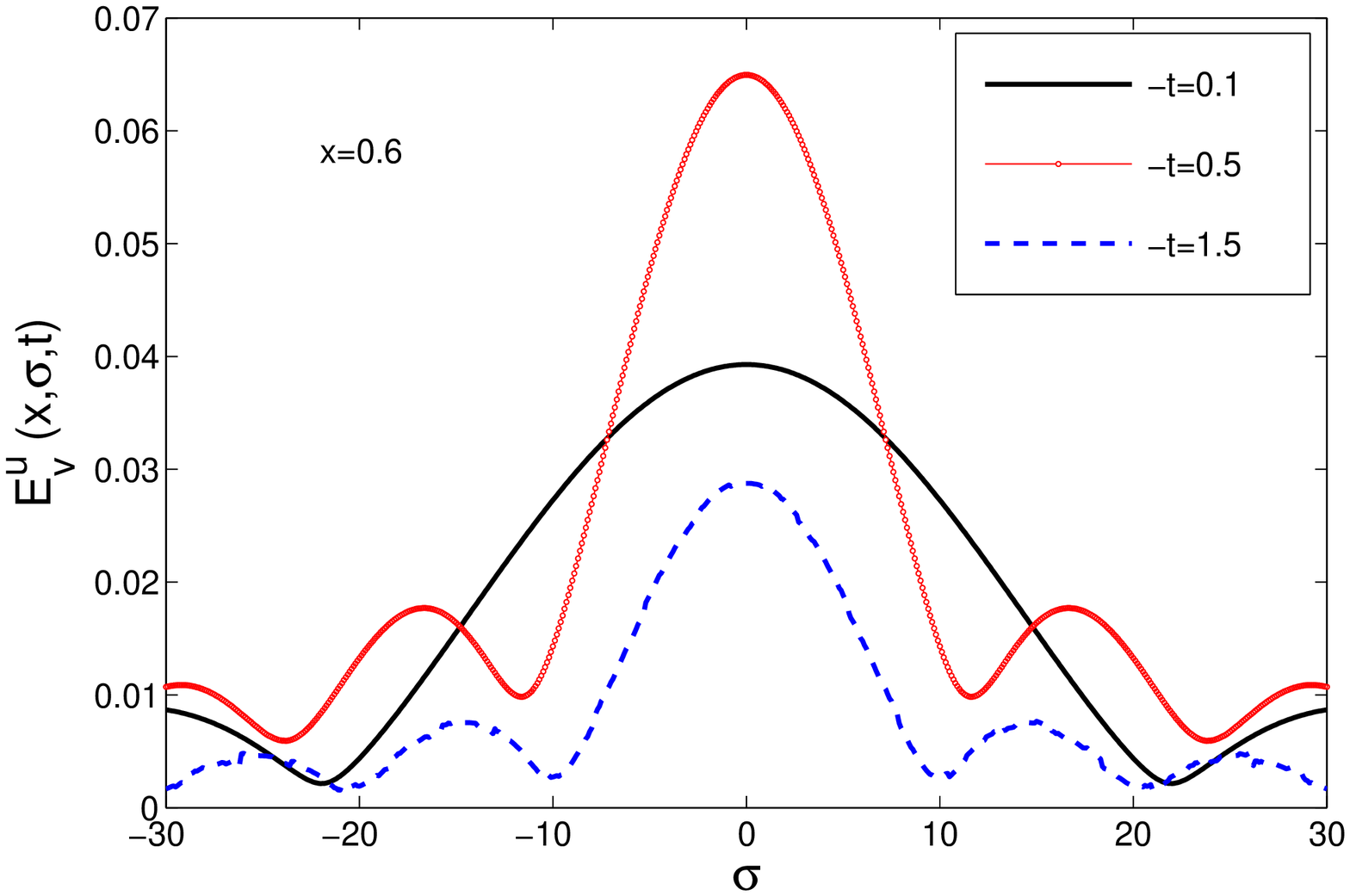}
\hspace{0.1cm}%
\small{(d)}\includegraphics[width=7.5cm,height=5.5cm,clip]{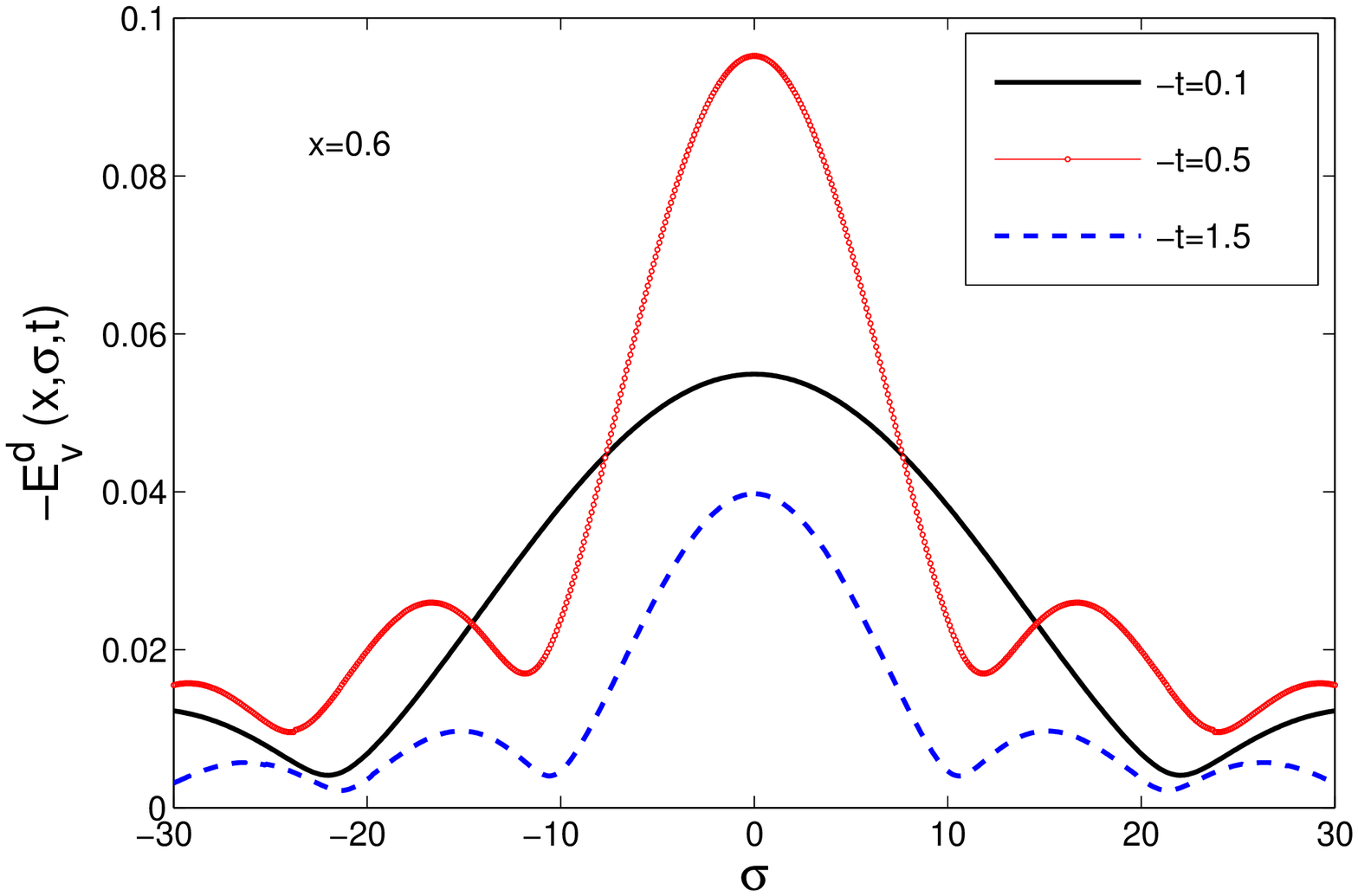}
\end{minipage}
\caption{\label{GPDs_longi}(Color online) Plots of (a) $H^u(x,\sigma,t)$ vs $\sigma$ (b) the same as in (a) but for $d$ quark (c) $E^u(x,\sigma,t)$ vs $\sigma$ (d) the same as in (c) but for $d$ quark for fixed $x$ and different values of $-t$ in $GeV^2$. }
\end{figure*} 
\subsection{\bf GPDs in longitudinal impact parameter space}\label{long_impact}

The boost invariant longitudinal impact parameter is defined as $\sigma=\frac{1}{2}b^-P^+$ which is conjugate to the skewness $\zeta$,  the measure of longitudinal momentum transfer. The parameter $\sigma$ was first Introduced in \cite{BDHAV} and it was shown that the DVCS amplitude in a QED model of a dressed electron  shows an interesting diffraction pattern in the longitudinal impact parameter space. Since Lorentz boosts are kinematical in the light front, the correlation defined in the three dimensional position space $b_\perp$ and $\sigma$ is frame independent.  It was shown in the same simple relativistic spin half system of an electron dressed  with a photon that  the GPDs also exhibit the similar diffraction pattern  in the longitudinal impact parameter space \cite{CMM1}. Similar diffraction pattern was also observed in a phenomenological model for proton GPDs\cite{CMM2}.  So, it is very interesting to investigate if the similar pattern is also observed in this light front quark model. The GPDs in 
longitudinal position space are defined as:
\be
H(x,\sigma,t)&=&{1\over 2 \pi} \int_0^{\zeta_f} d \zeta e^{i\zeta P^+b^-/2}H(x,\zeta,t),\nonumber\\
&=&{1\over 2 \pi} \int_0^{\zeta_f} d \zeta e^{i\zeta \sigma}H(x,\zeta,t),\nonumber\\
E(x,\sigma,t)&=&{1\over 2 \pi} \int_0^{\zeta_f} d \zeta e^{i\zeta P^+b^-/2}E(x,\zeta,t),\nonumber\\
&=&{1\over 2 \pi} \int_0^{\zeta_f} d \zeta e^{i\zeta \sigma}E(x,\zeta,t).
\ee
Since we are considering the region $\zeta<x<1$, the upper limit of $\zeta$ integration $\zeta_f$ is given by $\zeta_{max}$ if $x$ is larger than $\zeta_{max}$, otherwise by $x$ if $x$ is smaller than $\zeta_{max}$ where the maximum value of $\zeta$ for a fixed $-t$ is given by
\be
\zeta_{max}=\frac{-t}{2M_n^2}\bigg(\sqrt{1+\frac{4M_n^2}{(-t)}}-1\bigg).
\ee
In Fig.\ref{GPDs_longi}, we show the GPDs in longitudinal position space $\sigma$ considering the DGLAP region. We observe that the GPDs show diffraction pattern in longitudinal impact parameter space, similar to the nature of a dressed electron in QED or in a holographic model for the meson \cite{BDHAV}. This effect has also been observed for the GPDs of a phenomenological model \cite{CMM2} as well as for the chiral odd GPDs of light-front QED model \cite{CMM1}. Except for $H^d(x,\sigma,t)$, all the distributions have a primary maximum at $\sigma=0$ followed by a series of secondary maxima. $H^d(x,\sigma,t)$ has a peculiar behavior having a  maximum at $\sigma=0$ for very  small $-t$ and shows diffraction pattern while for relatively larger values of   $-t$  it shows  a minimum at $\sigma=0$. The  minima in $E(x,\sigma,t)$ occur at the same positions for both $u$ and $d$ quarks. In all cases, the position of the first minimum moves in to smaller values of $\sigma$ as $-t$ increases. The characteristics of $E(x,\sigma,t)$ for both $u$ and $d$ quarks is almost same whereas for $H(x,\sigma,t)$, the nature of $u$ and $d$ quark changes as 
$-t$ increases.  In \cite{BDHAV},  similar diffraction patterns were observed for DVCS amplitudes  with both $2\to 2$ and $3\to1$ contributions, so we expect that the pattern will survive if higher Fock sectors are included in the model.

\section{Transverse charge and magnetization densities}\label{density}
The two dimensional Fourier transform of the Dirac form factor gives the transverse charge density in the transverse plane for the unpolarized nucleons,
\be
\rho_{ch}(b)
&=&\int \frac{d^2q_{\perp}}{(2\pi)^2}F_1(q^2)e^{iq_{\perp}.b_{\perp}}\nonumber\\
&=&\int_0^\infty \frac{dQ}{2\pi}QJ_0(Qb)F_1(Q^2),
\ee
where $b$ represents the impact parameter and $J_0$ is the cylindrical Bessel function of order zero. We can write 
a similar formula for charge density for flavor $\rho_{fch}^q (b)$ with $F_1$ is replaced
by $F_1^q$. In a similar fashion, one defines the magnetization density in the transverse plane by the Fourier transform of the Pauli form factor,
\be
\widetilde{\rho}_{M}(b) &= & \int \frac{d^2q_{\perp}}{(2\pi)^2}F_2(q^2)e^{iq_{\perp}.b_{\perp}}\nonumber\\
&=&\int_0^\infty \frac{dQ}{2\pi}QJ_0(Qb)F_2(Q^2),
\ee
whereas,
\be
\rho_m(b)= -b\frac{\partial \widetilde{\rho}_M(b)}{\partial b}
=b\int_0^\infty \frac{dQ}{2\pi}Q^2J_1(Qb)F_2(Q^2),
\ee
\begin{figure*}[htbp]
\begin{minipage}[c]{0.98\textwidth}
\small{(a)}
\includegraphics[width=7.5cm,height=6cm,clip]{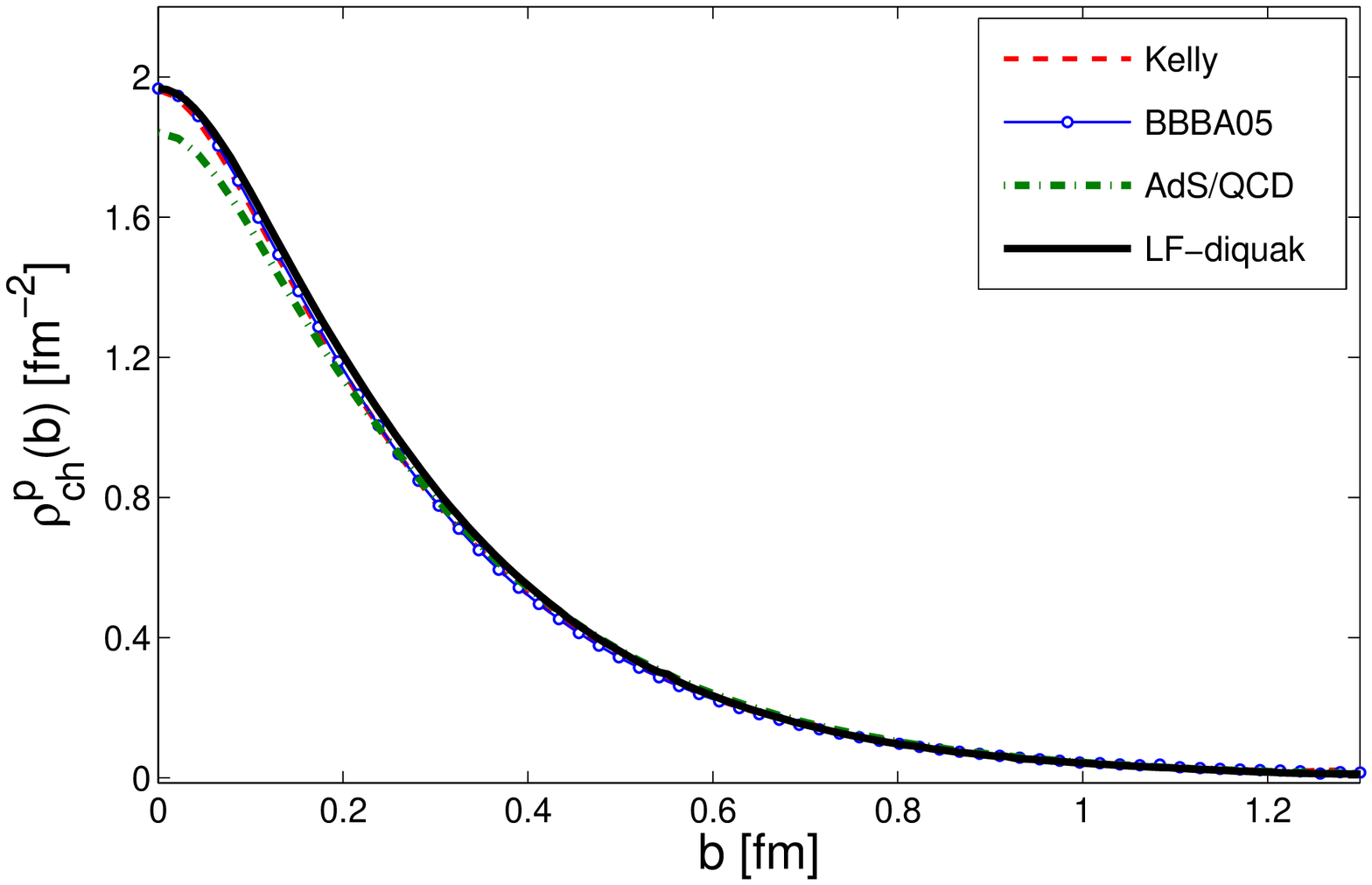}
\hspace{0.1cm}%
\small{(b)}\includegraphics[width=7.5cm,height=6cm,clip]{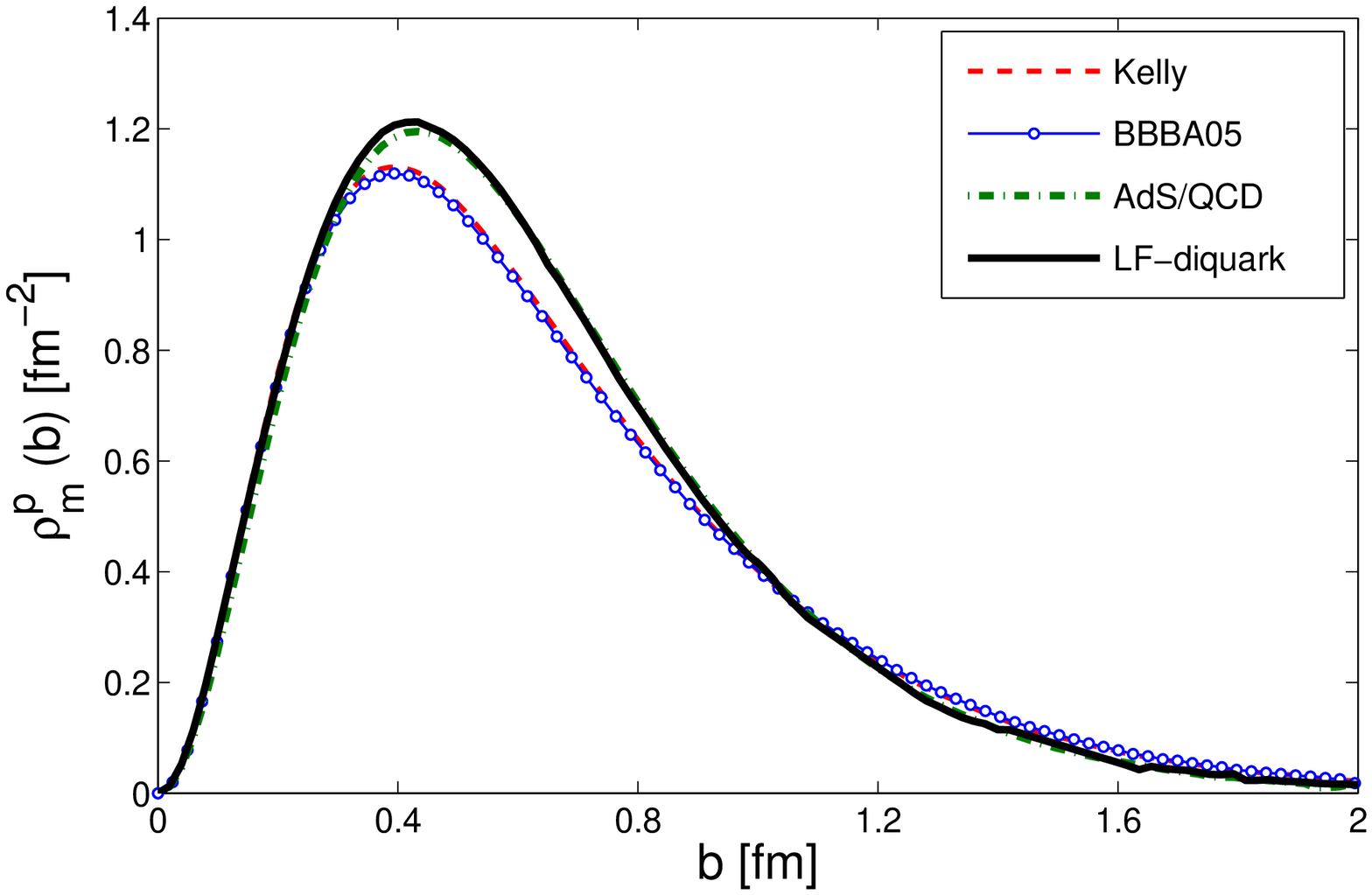}
\end{minipage}
\begin{minipage}[c]{0.98\textwidth}
\small{(c)}\includegraphics[width=7.5cm,height=6cm,clip]{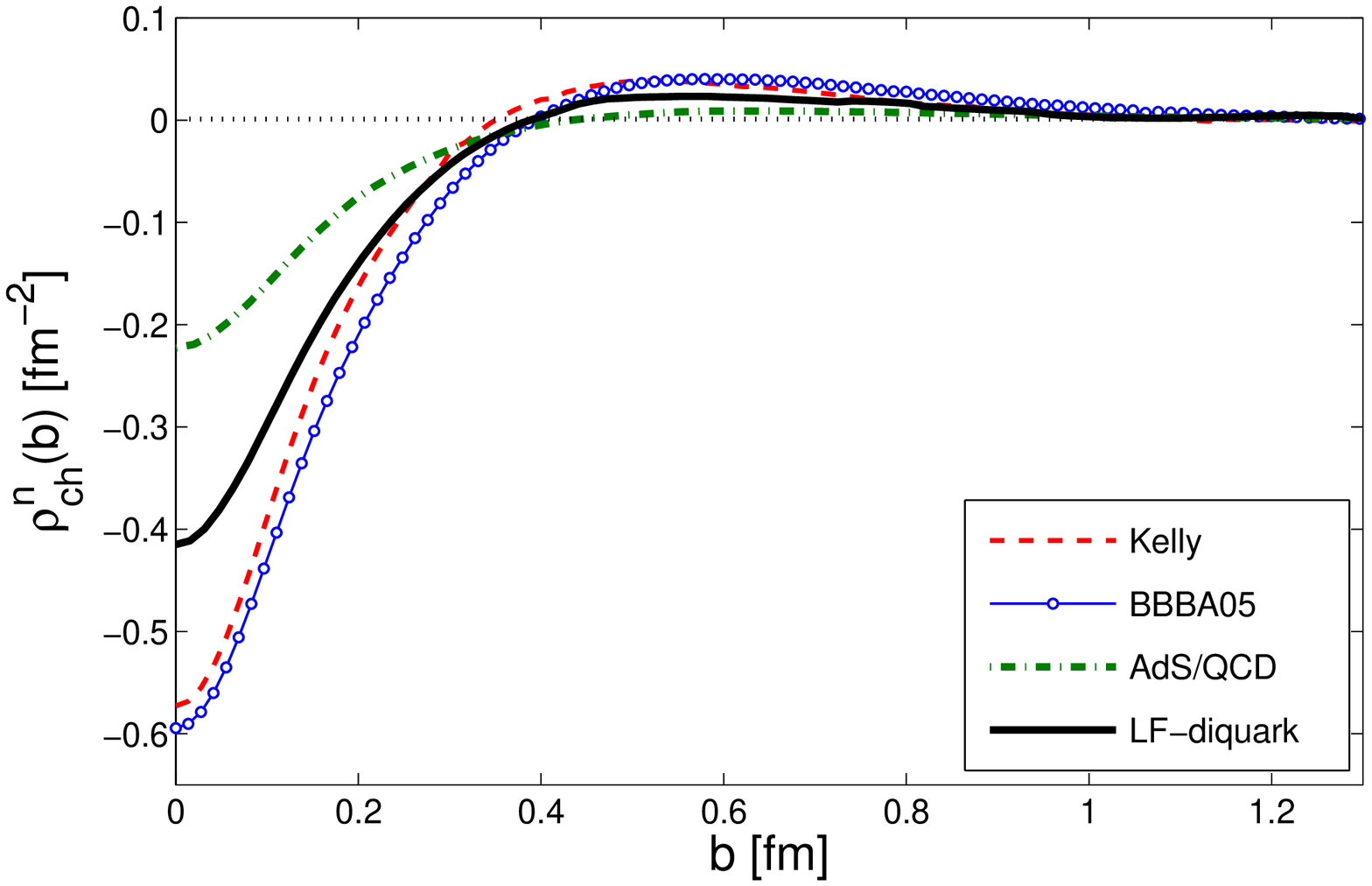}
\hspace{0.1cm}%
\small{(d)}\includegraphics[width=7.5cm,height=6cm,clip]{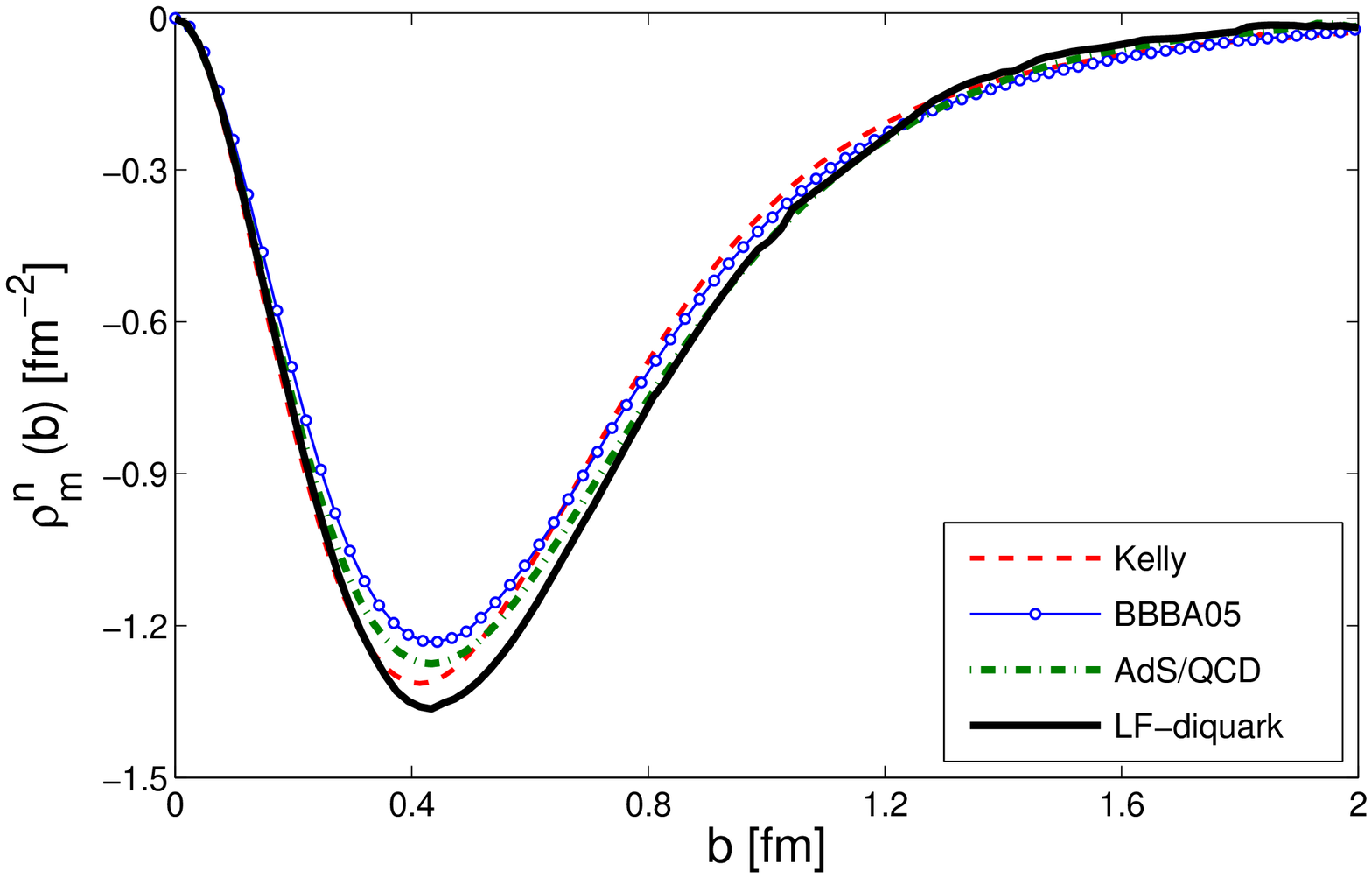}
\end{minipage}
\caption{\label{nucleon_den}(Color online) Plots of transverse charge and anomalous magnetization densities for nucleon. (a) and (b) represent $\rho_{ch}$ and ${\rho}_m$ for the proton. (c) and (d) same as proton but for neutron.  Dashed lines represent the parameterization of  Kelly \cite{kelly04},  and  the lines with circles represent the parameterization of  Bradford $at~el$ \cite{brad}, dot-dashed lines are for soft-wall model \cite{CM3}.
The solid lines represent the  light-front scalar diquark model.
}
\end{figure*}
\begin{figure*}[htbp]
\begin{minipage}[c]{0.98\textwidth}
\small{(a)}
\includegraphics[width=7.5cm,height=6cm,clip]{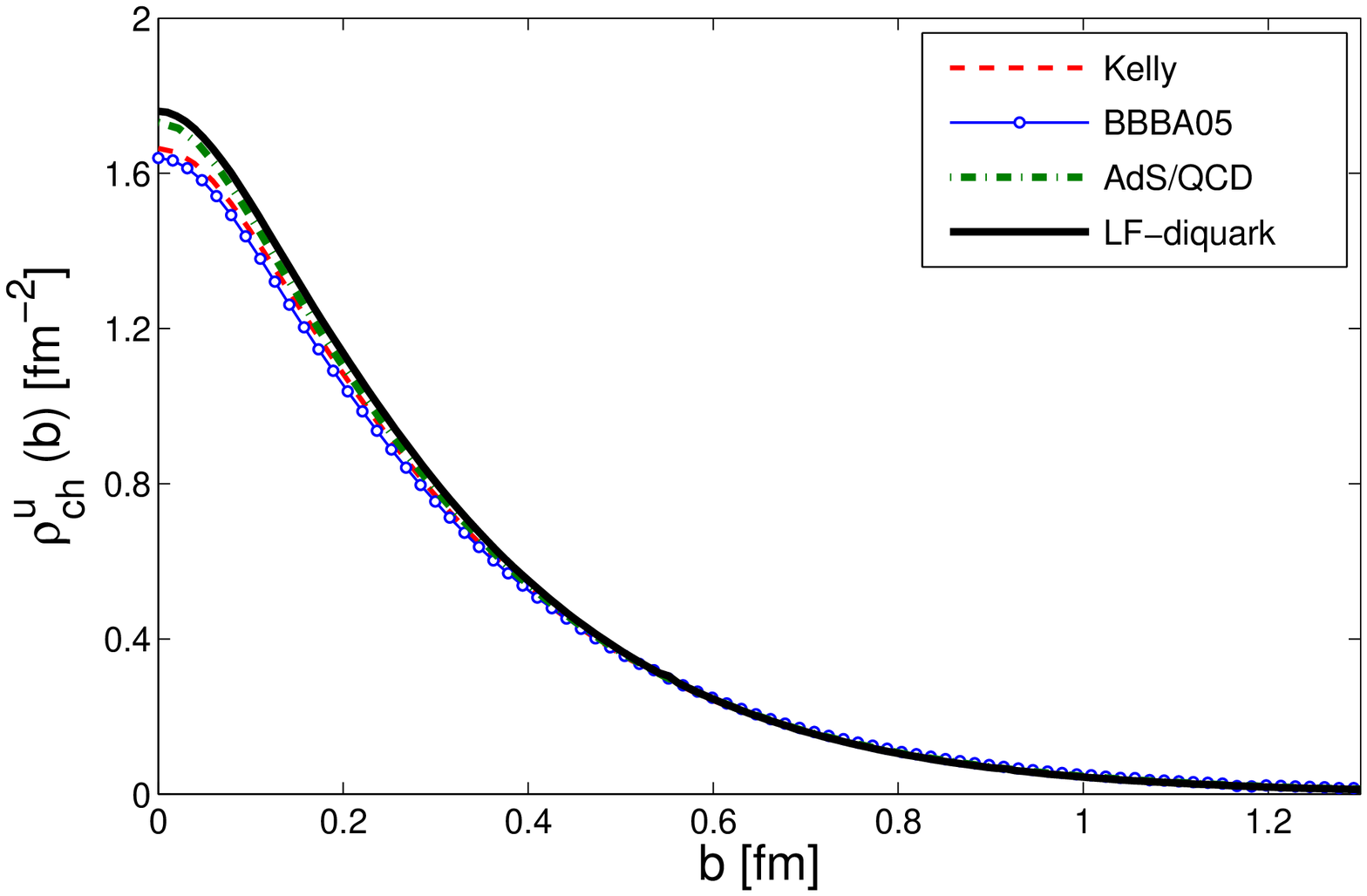}
\hspace{0.1cm}%
\small{(b)}\includegraphics[width=7.5cm,height=6cm,clip]{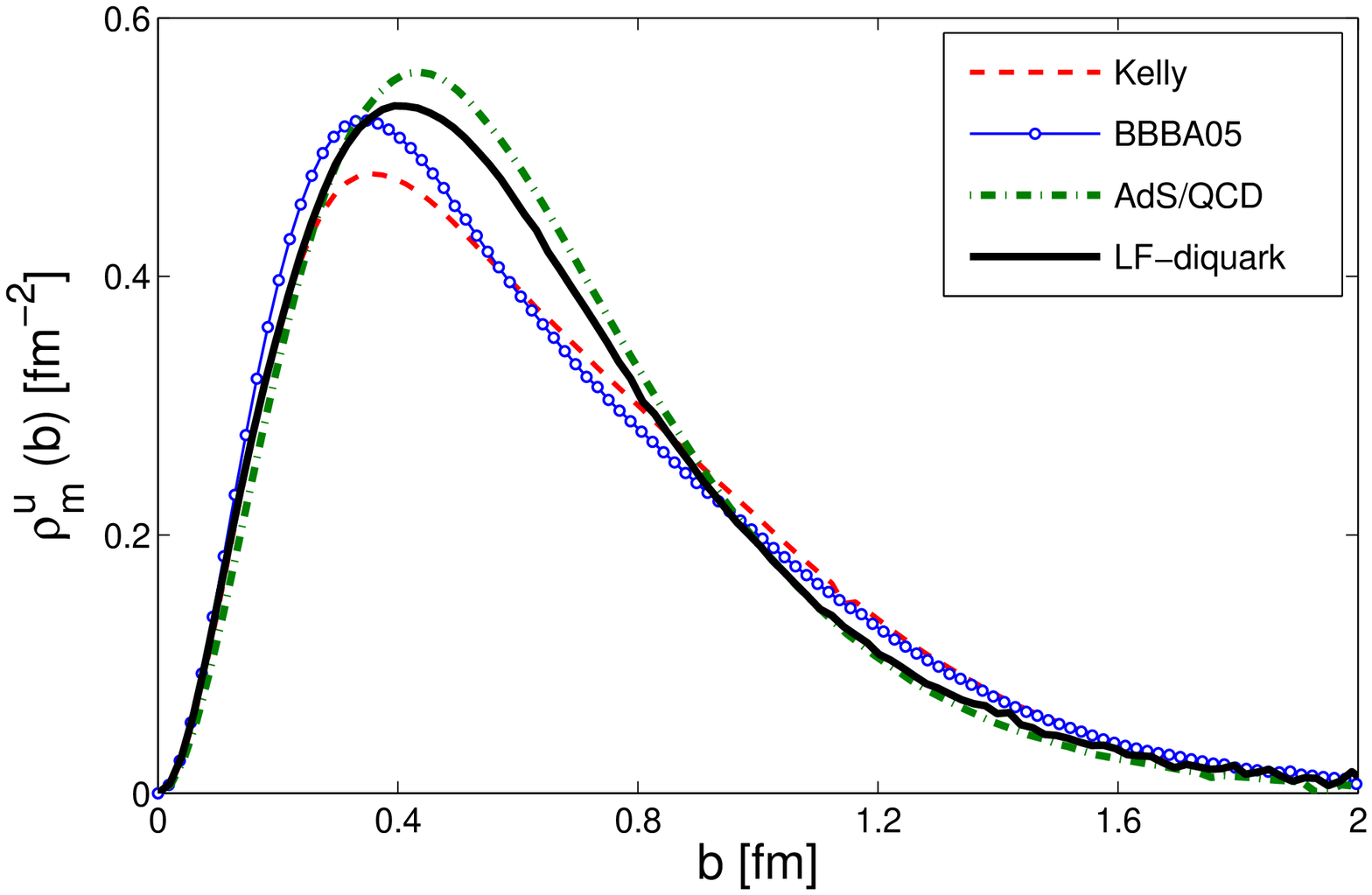}
\end{minipage}
\begin{minipage}[c]{0.98\textwidth}
\small{(c)}\includegraphics[width=7.5cm,height=6cm,clip]{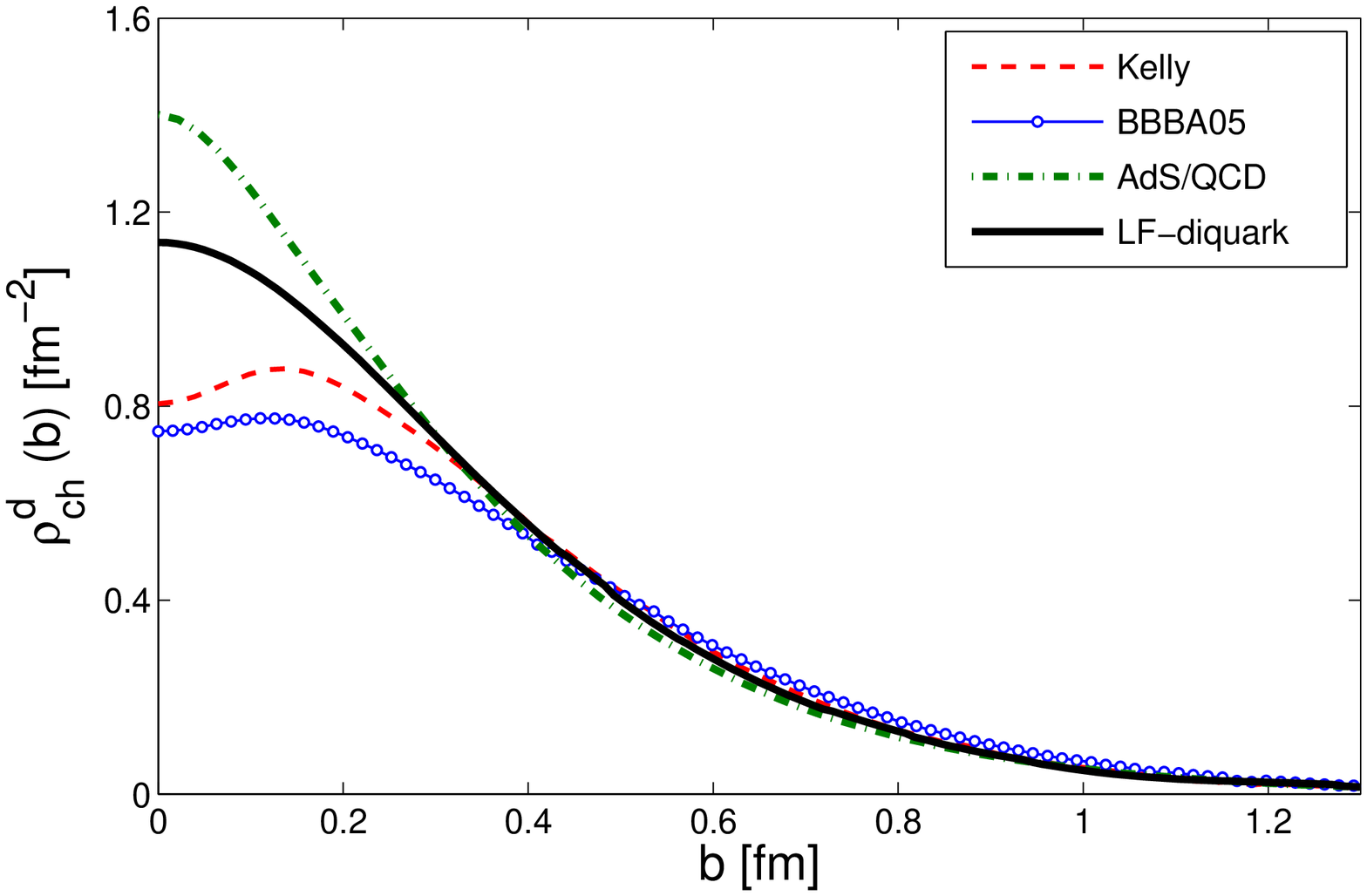}
\hspace{0.1cm}%
\small{(d)}\includegraphics[width=7.5cm,height=6cm,clip]{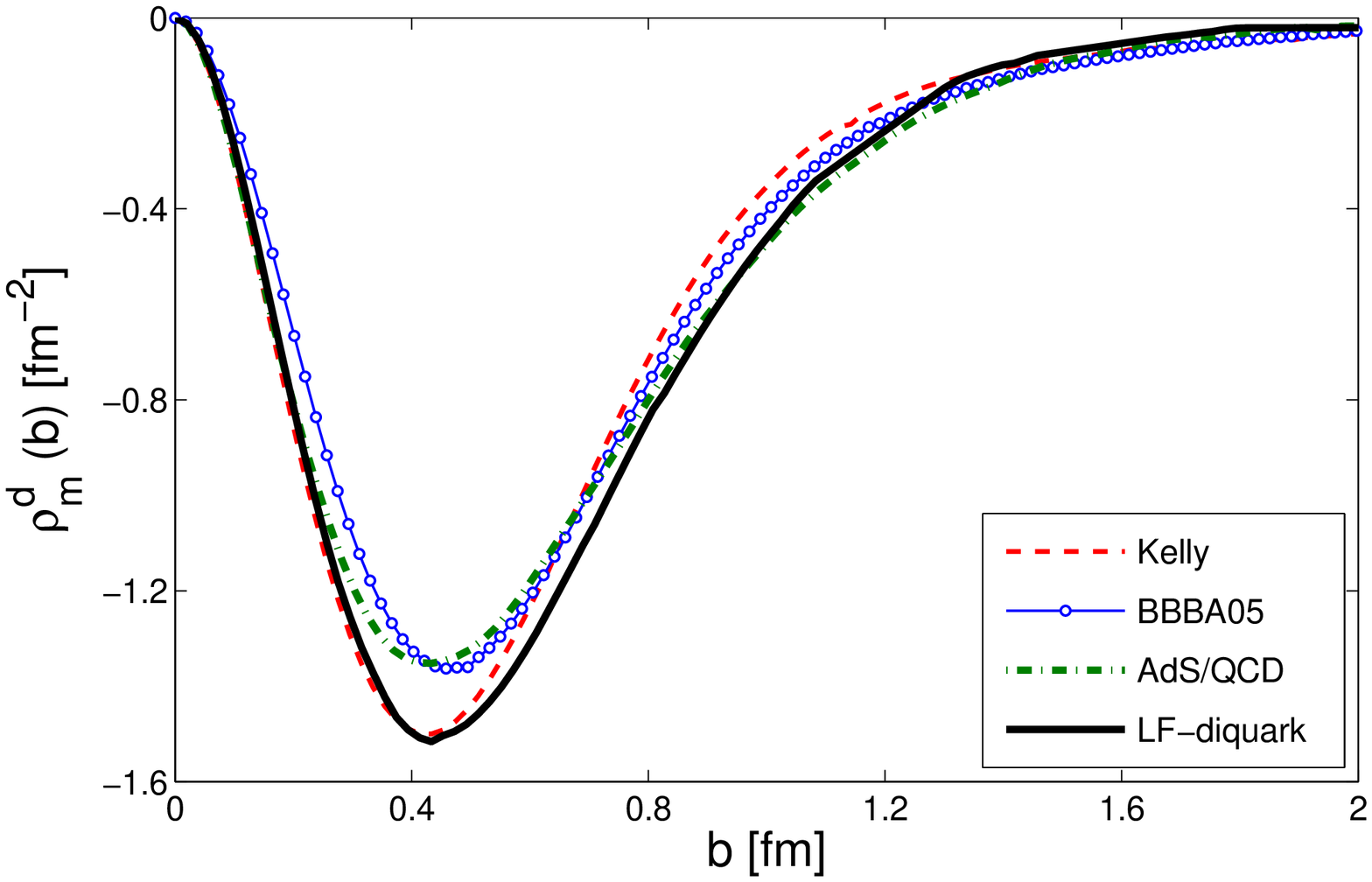}
\end{minipage}
\caption{\label{quark_den}(Color online) Plots of transverse charge and anomalous magnetization densities for quark. (a) and (b) represent $\rho_{ch}$ and ${\rho}_m$ for the $u$ quark. (c) and (d) same as $u$ but for $d$ quark.  Dashed lines represent the parameterization of  Kelly \cite{kelly04},  and  the lines with circles represent the parameterization of  Bradford $at~el$ \cite{brad}, dot-dashed lines are soft-wall model \cite{CM3}.
The solid lines represent the light-front scalar diquark model.
}
\end{figure*}
has the interpretation of anomalous magnetization density \cite{miller10}. Since these quantities are not
directly measured in experiments, actual experimental data are not available. In Ref.\cite{venkat}, an
 estimation of the proton charge and magnetization densities has been done from
experimental data of electromagnetic form factors. To get an insight into the contributions of the different
 flavors, we evaluate the charge and anomalous magnetization densities for the $u$ and
$d$ quarks. 

We can define the decompositions of the transverse charge and magnetization densities for nucleons in the similar way as electromagnetic form factors. The charge densities decompositions in terms of two flavors can be written as
\be
 \rho_{ch}^p&=& e_u \rho_{fch}^u+e_d \rho_{fch}^d,\nonumber\\\label{ch_mag1}
 \rho_{ch}^n&=& e_u \rho_{fch}^d+e_d \rho_{fch}^u,
 \ee
where $e_u$ and $e_d$ are charge of $u$ and $d$ quarks respectively.  Due to the charge and isospin symmetry the $u$ and  $d$ quark densities in the proton are the same as the $d$ and $u$ densities in the neutron \cite{miller07,CM3}. Under the charge and isospin symmetry, we can write
\be
 \rho_{ch}^u(b)&=&  \rho_{ch}^p+ \frac{\rho_{ch}^n}{2}=\frac{ \rho_{fch}^u}{2},\nonumber\\\label{ch_mag2}
 \rho_{ch}^d(b)&=&  \rho_{ch}^p+2 \rho_{ch}^n= \rho_{fch}^d,
 \ee
where $\rho_{ch}^q(b)$ is the charge density of each quark and $\rho^q_{fch}$ is the charge density for each flavor. We can similarly decompose  $\rho_m$ into magnetization densities for each flavor. 
The flavor contributions  to proton charge and magnetization densities are $e_{u/d}\rho_{fch}^{u/d}$ and $e_{u/d}{\rho}_{fm}^{u/d}$. Similarly for neutron, the flavor contributions are $(e_{d/u} $ $\rho_{fch}^{u/d}) $ and $(e_{d/u} {\rho}_{fm}^{u/d})$.  
\begin{figure*}[htbp]
\begin{minipage}[c]{0.98\textwidth}
\small{(a)}
\includegraphics[width=7.5cm,height=6cm,clip]{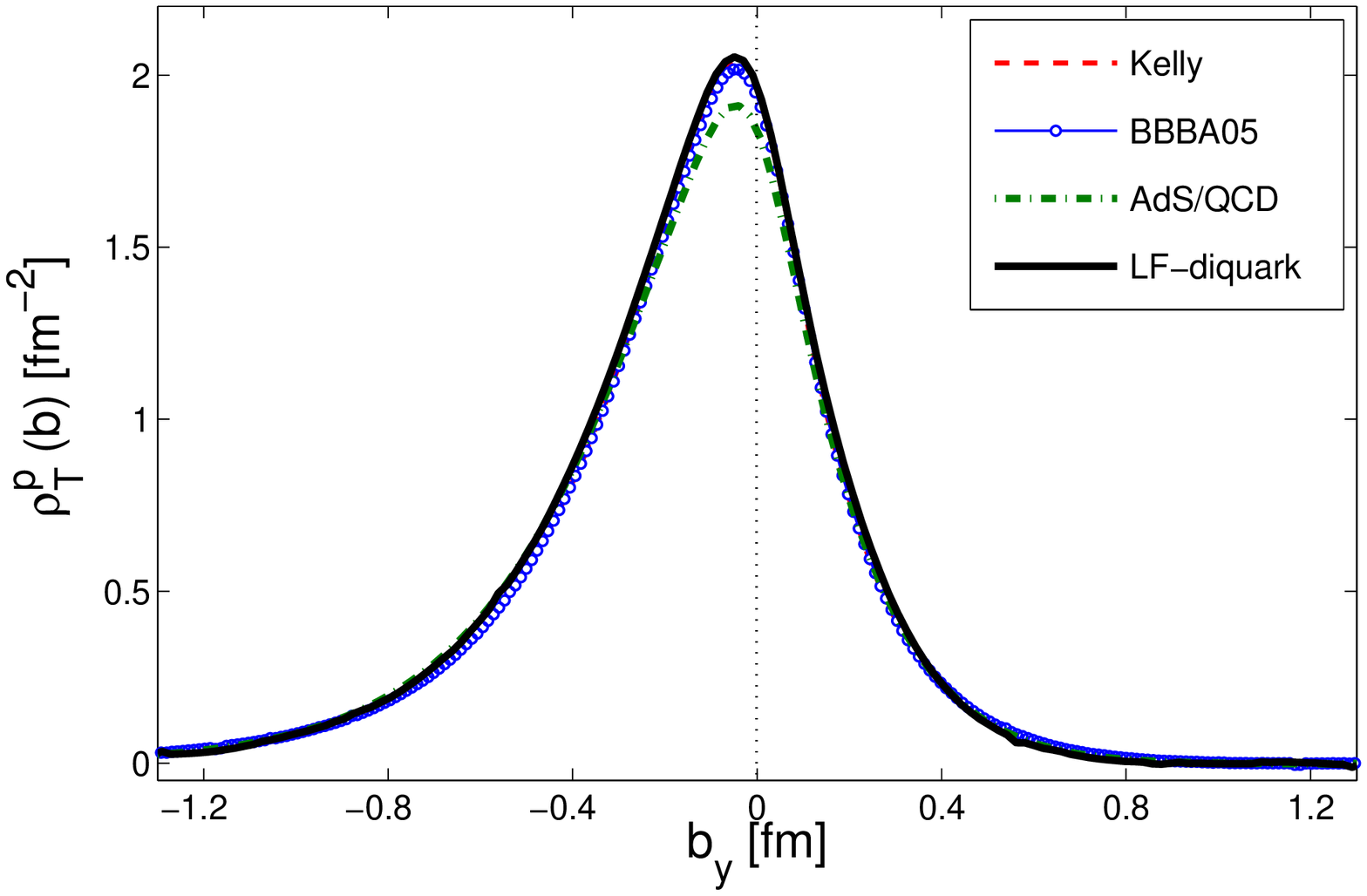}
\hspace{0.1cm}%
\small{(b)}\includegraphics[width=7.5cm,height=6cm,clip]{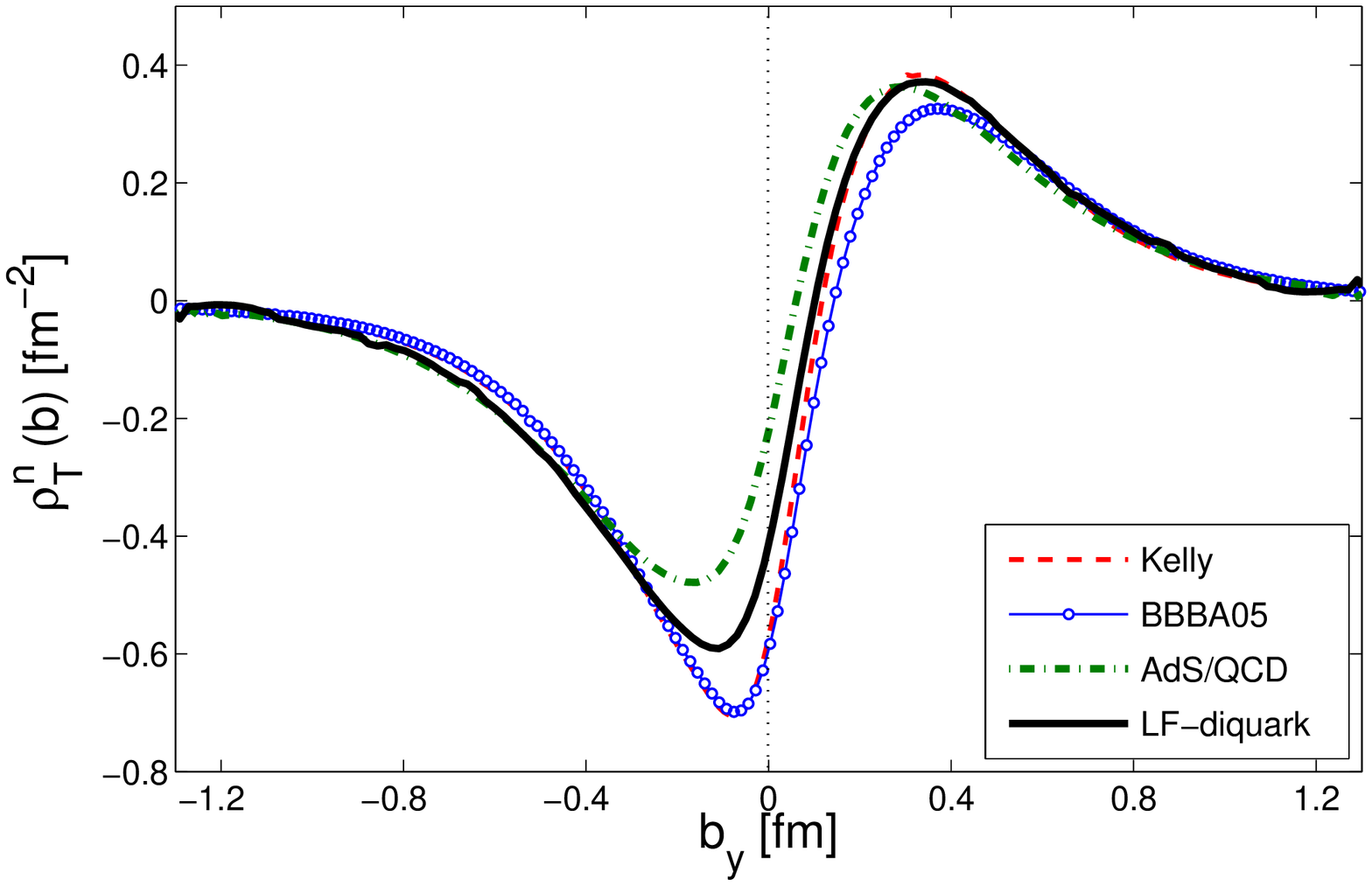}
\end{minipage}
\begin{minipage}[c]{0.98\textwidth}
\small{(c)}
\includegraphics[width=7.5cm,height=6cm,clip]{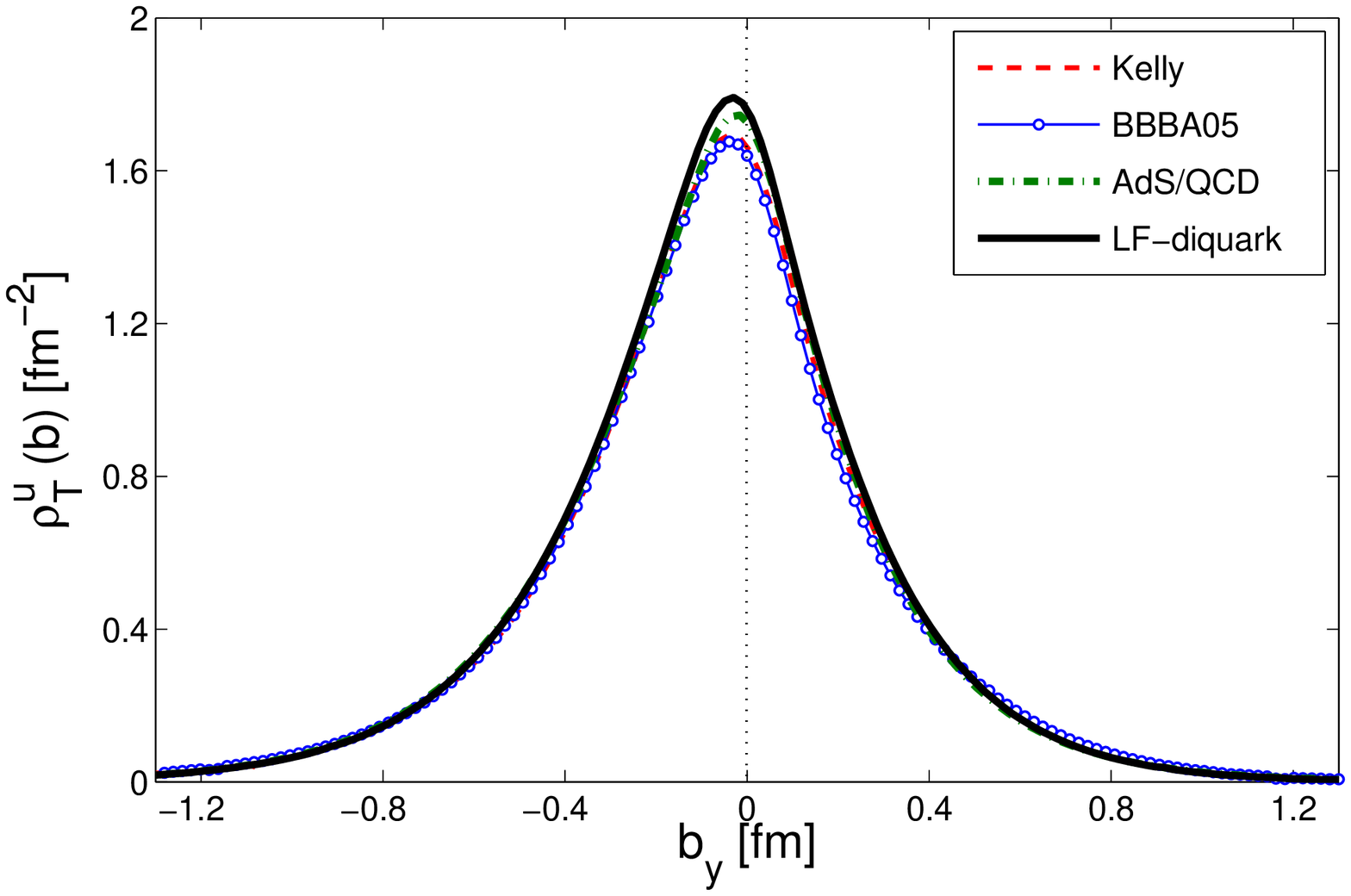}
\hspace{0.1cm}%
\small{(d)}\includegraphics[width=7.5cm,height=6cm,clip]{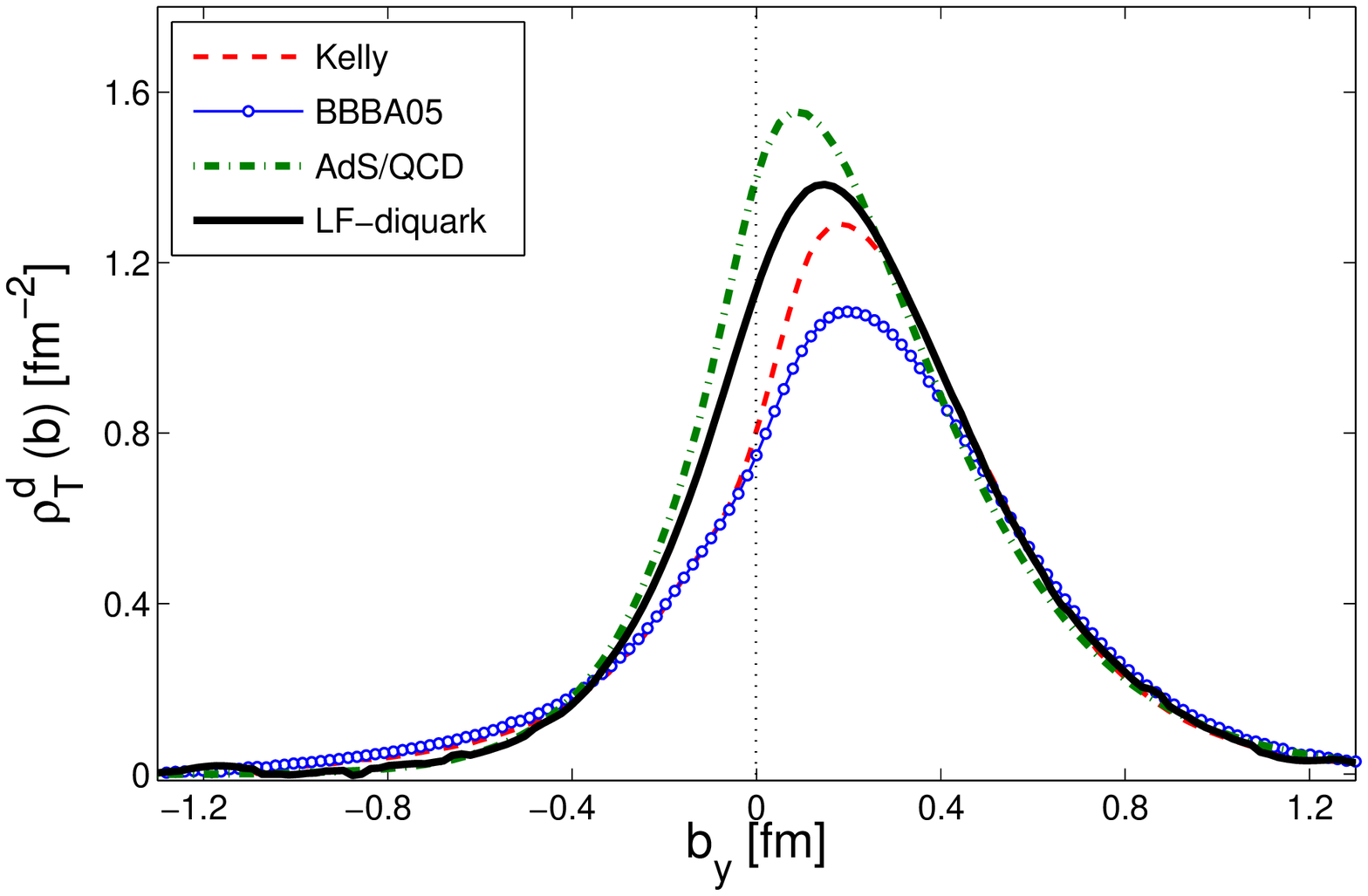}
\end{minipage}
\caption{\label{flavor_T}(Color online) The charge  densities for the transversely polarized (a) proton (b) neutron and (c) up (d) down quark charge  densities for the transversely polarized nucleon. 
Dashed line represents the parameterization of  Kelly \cite{kelly04},  and  line with circles represents the parameterization of  Bradford $at~el$ \cite{brad}, dot-dashed line is soft-wall Model-I in \cite{CM3}.
The solid line represents this work for light-front scalar diquark model.}
\end{figure*}

\begin{figure*}[htbp]
\begin{minipage}[c]{0.98\textwidth}
\small{(a)}
\includegraphics[width=7.5cm,height=6cm,clip]{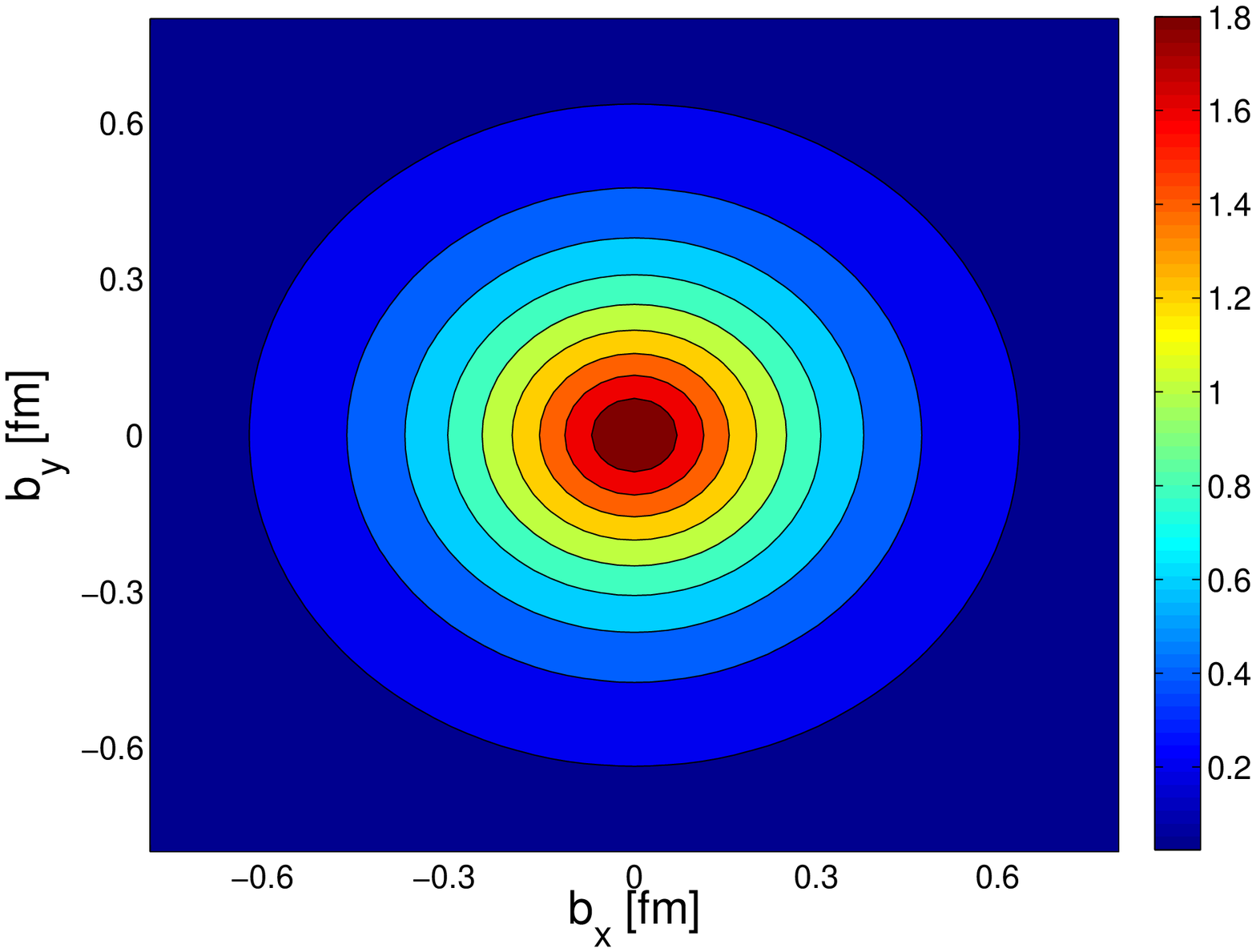}
\hspace{0.1cm}%
\small{(b)}\includegraphics[width=7.5cm,height=6cm,clip]{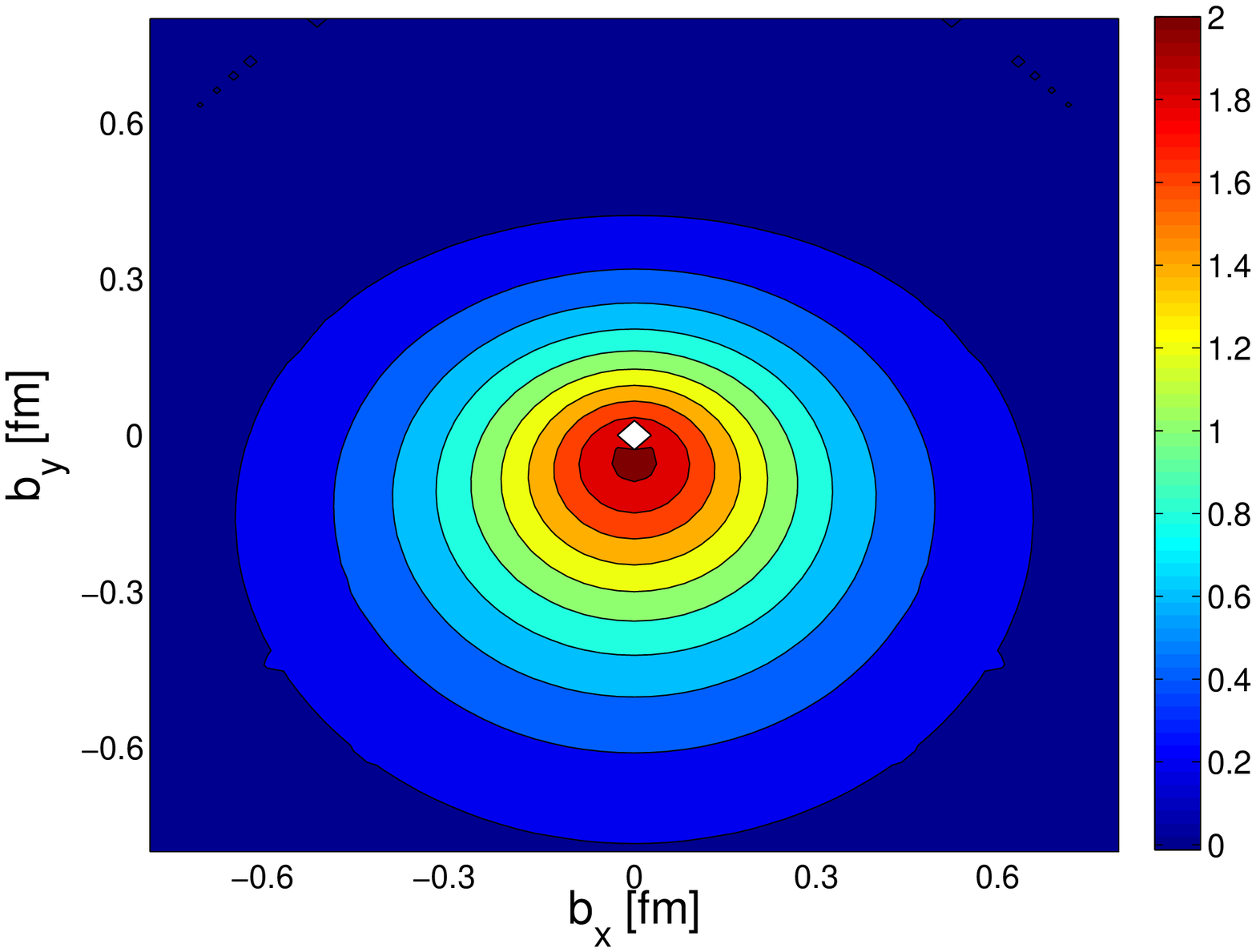}
\end{minipage}
\begin{minipage}[c]{0.98\textwidth}
\small{(c)}
\includegraphics[width=7.5cm,height=6cm,clip]{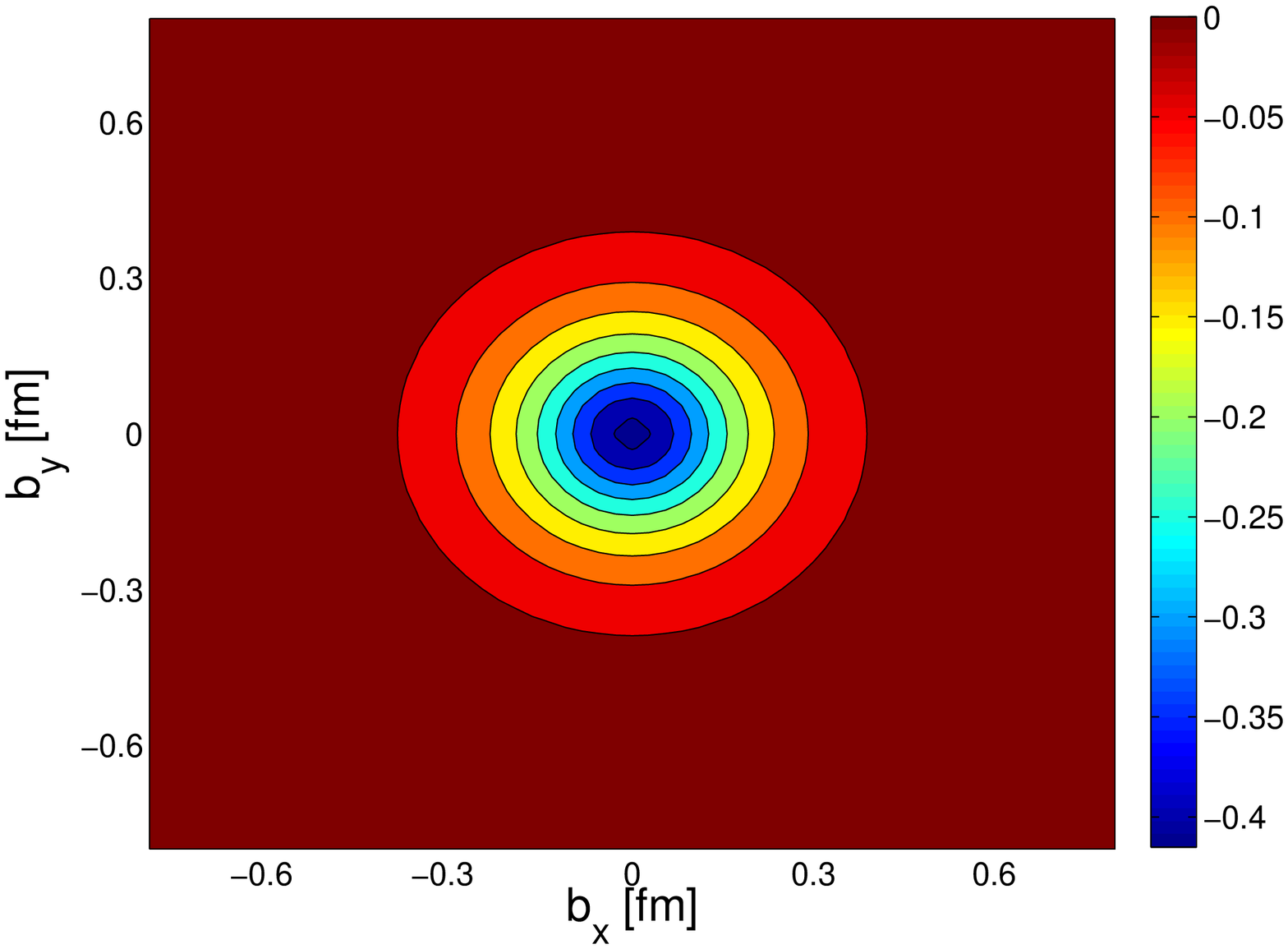}
\hspace{0.1cm}%
\small{(d)}\includegraphics[width=7.5cm,height=6cm,clip]{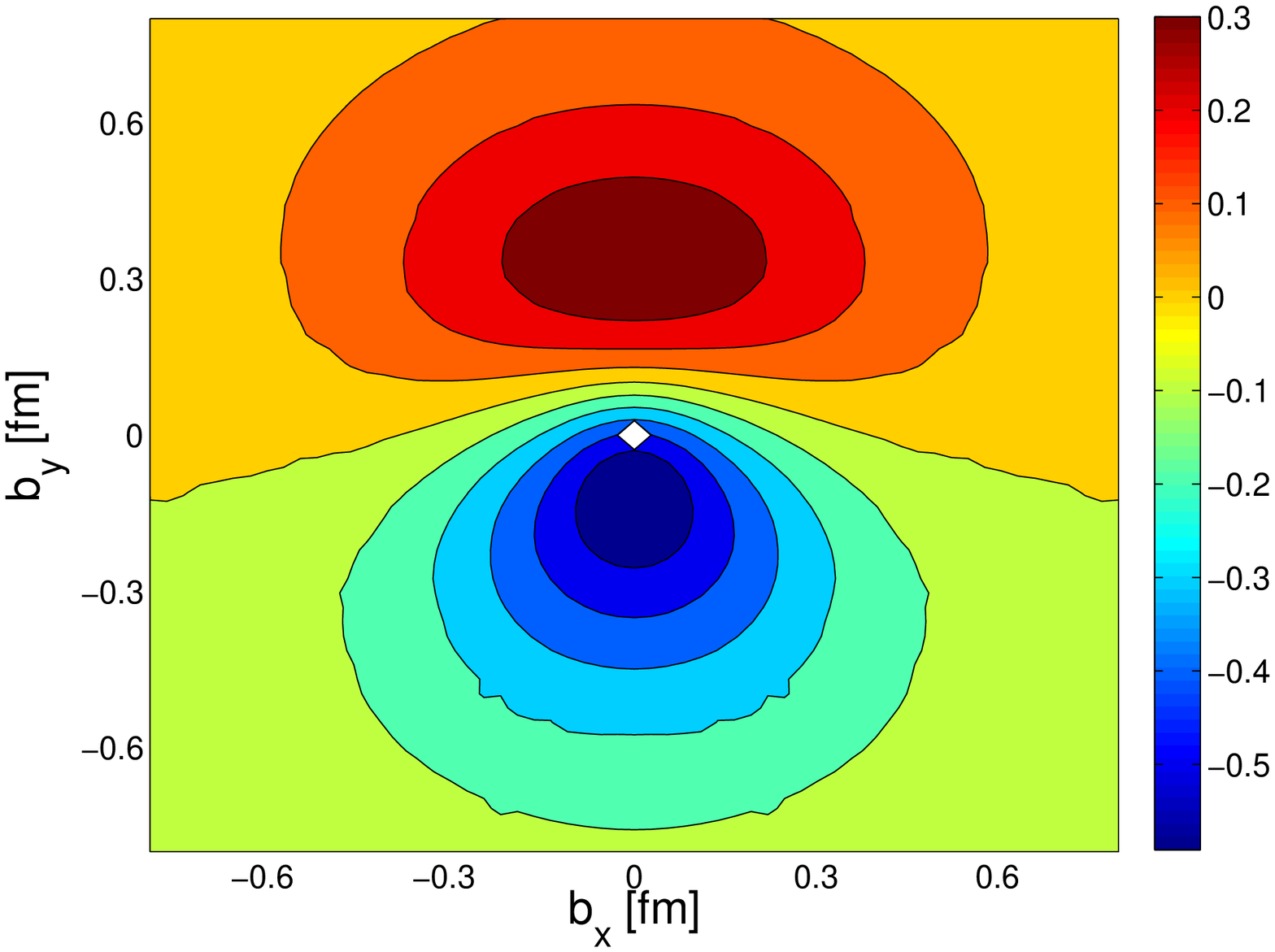}
\end{minipage}
\caption{\label{nucleons_top}(Color online) The charge  densities in the transverse plane for  the (a) unpolarized  proton (b) transversely polarized  proton and  (c) unpolarized neutron and  (d) transversely polarized neutron. Transverse polarization is along $x$ direction. }
\end{figure*} 
\begin{figure*}[htbp]
\begin{minipage}[c]{0.98\textwidth}
\small{(a)}
\includegraphics[width=7.5cm,height=6cm,clip]{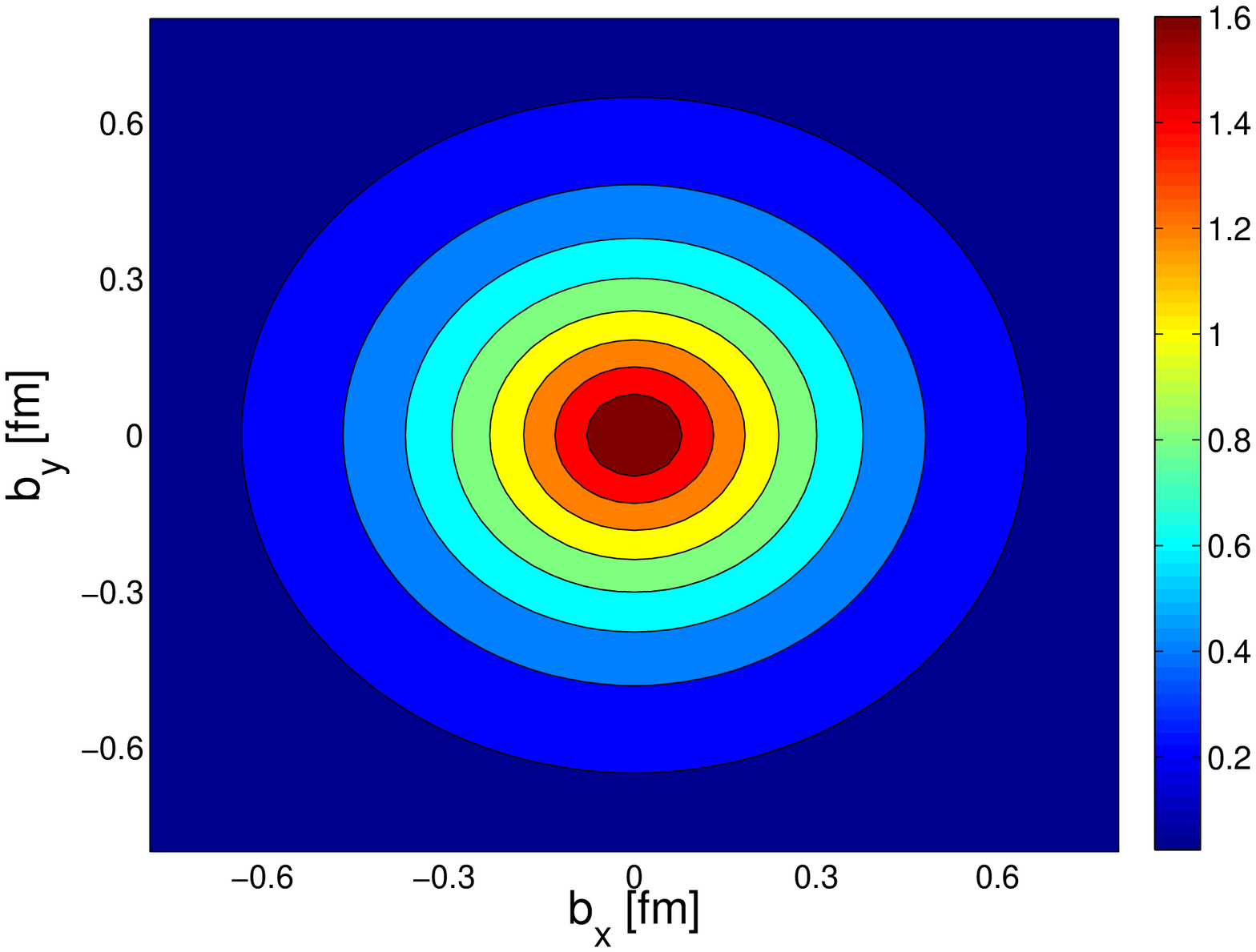}
\hspace{0.1cm}%
\small{(b)}\includegraphics[width=7.5cm,height=6cm,clip]{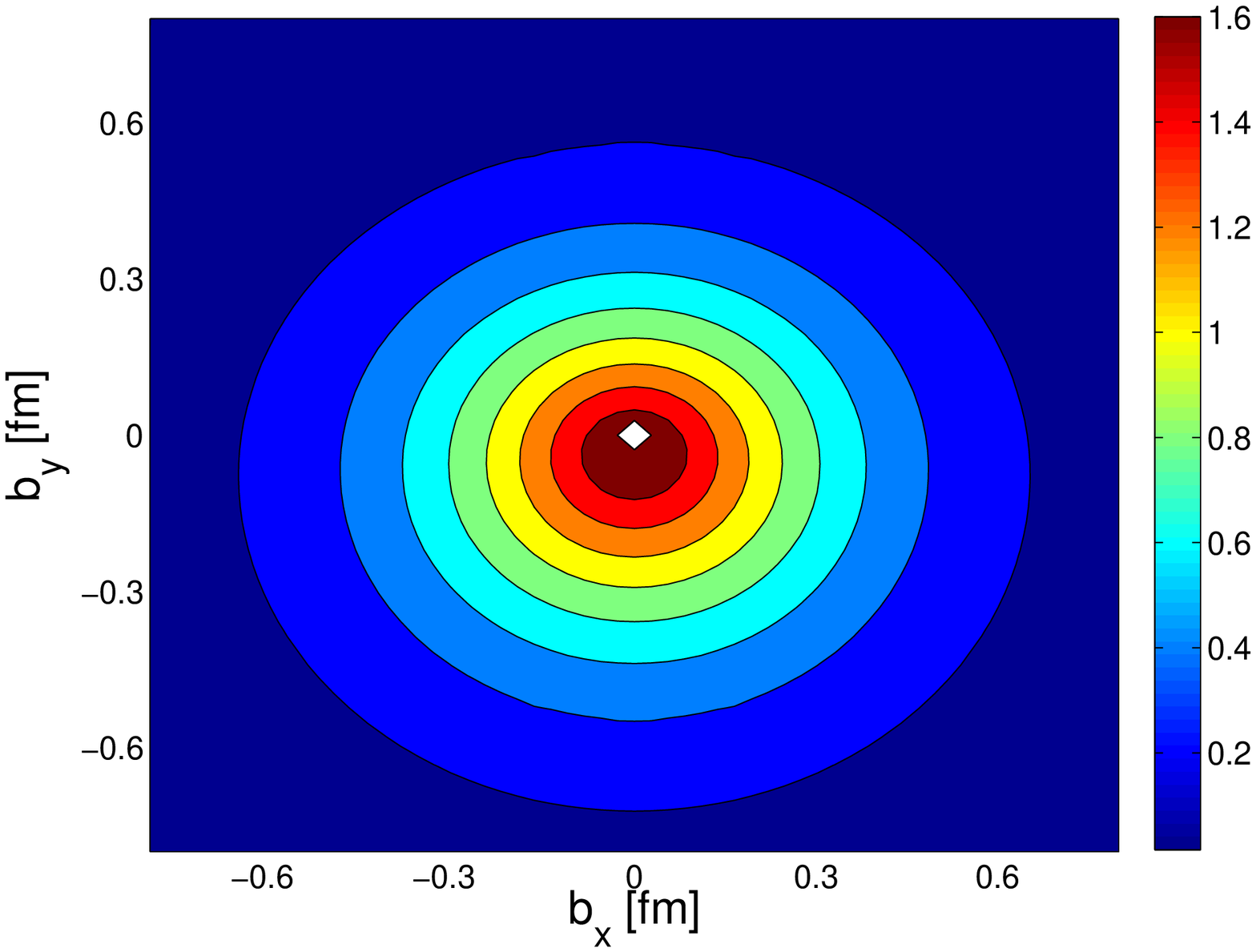}
\end{minipage}
\begin{minipage}[c]{0.98\textwidth}
\small{(c)}
\includegraphics[width=7.5cm,height=6cm,clip]{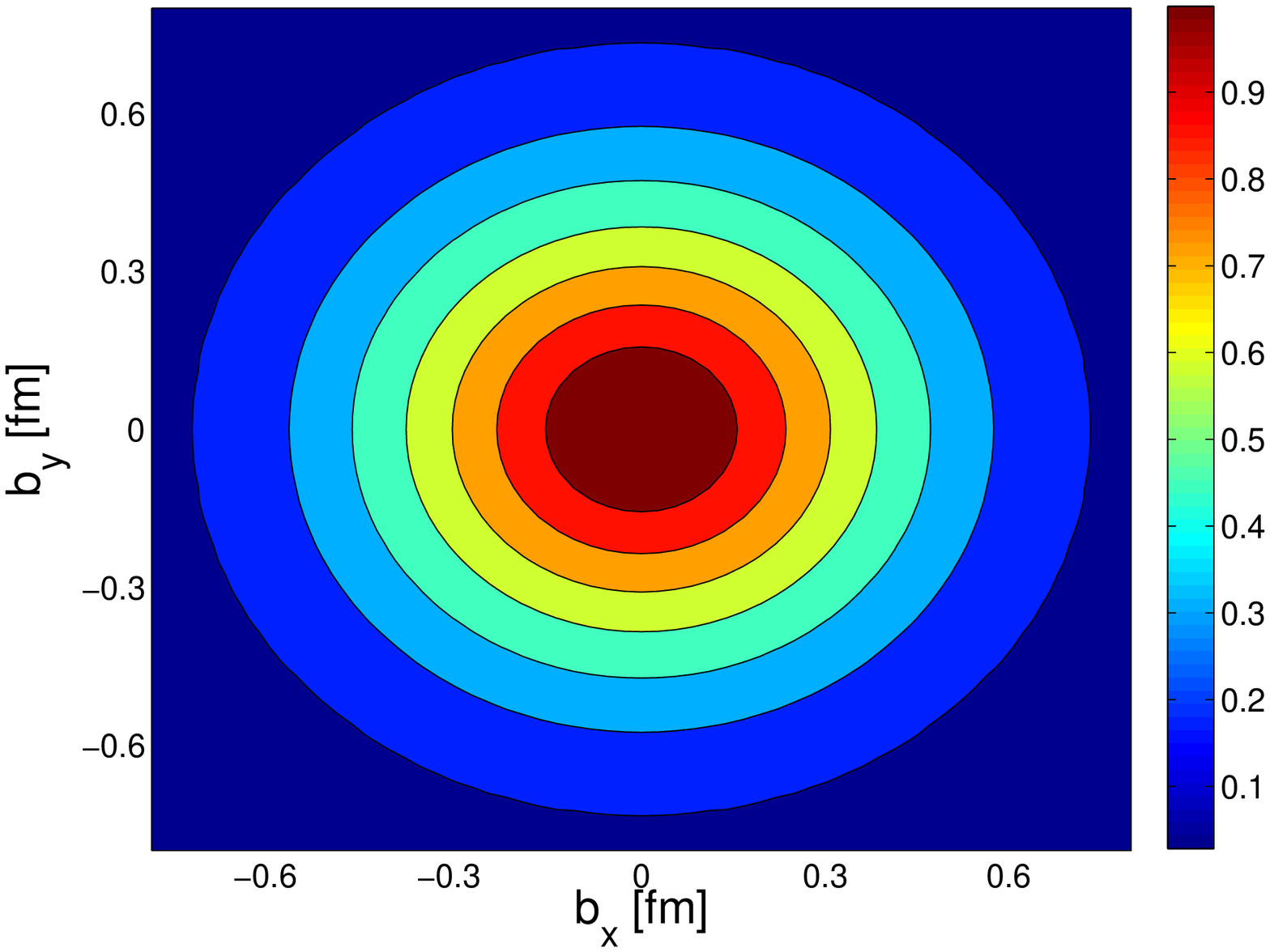}
\hspace{0.1cm}%
\small{(d)}\includegraphics[width=7.5cm,height=6cm,clip]{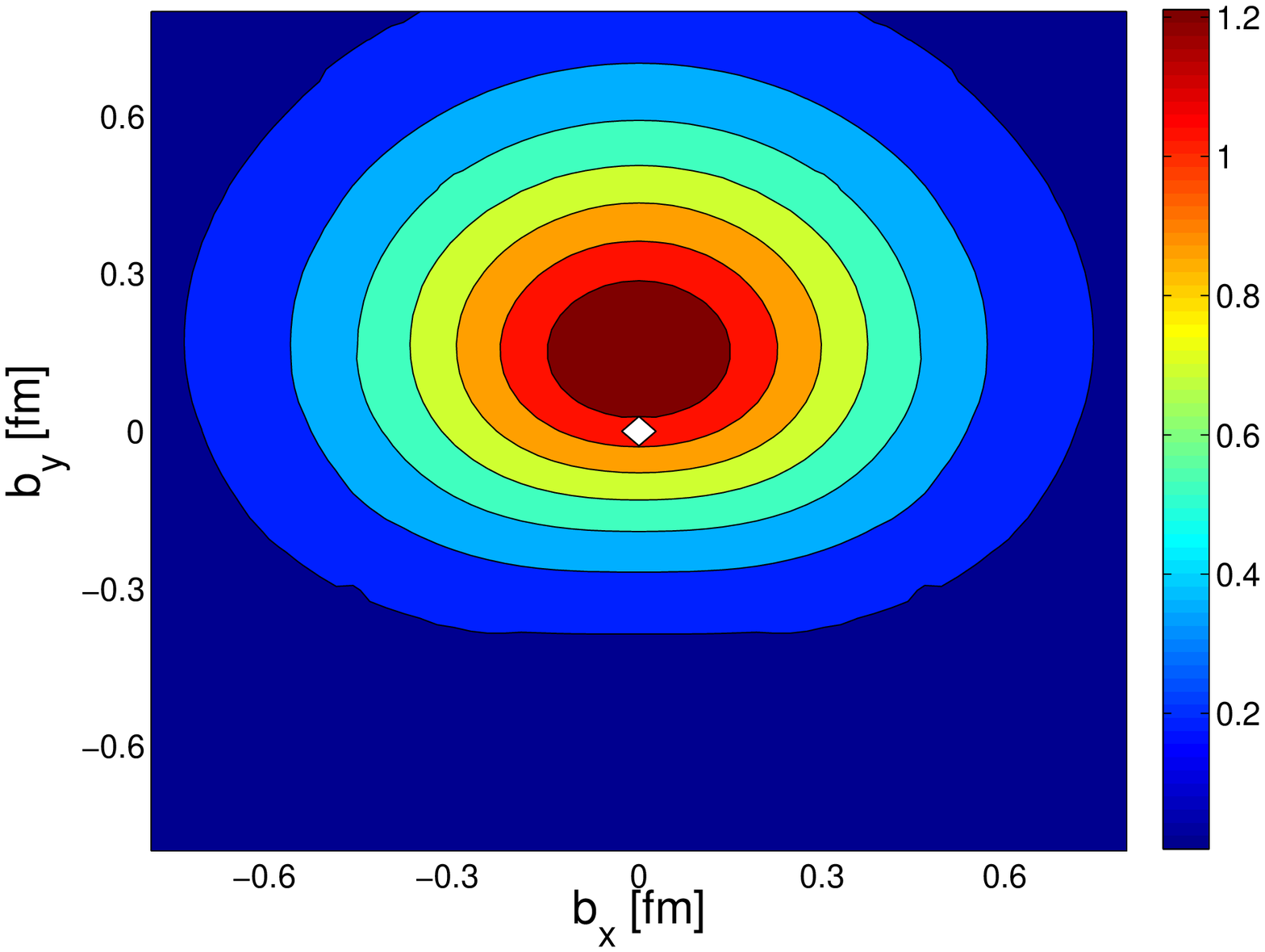}
\end{minipage}
\caption{\label{flavor_top}(Color online) The charge  densities in the transverse plane of $u$ quark for (a) unpolarized (b) transversely polarized nucleon and $d$ quark for (c) unpolarized (d) transversely polarized nucleon. Polarization is along $x$-direction.}
\end{figure*} 
In Fig. \ref{nucleon_den}, we show the charge and anomalous magnetization densities for proton and neutron. The plots suggest that the light-front diquark model's results for the charge and magnetization density of proton and the magnetization density of neutron are in excellent agreement with the two different global parameterizations of Kelly \cite{kelly04} and Bradford $at$ $el$ \cite{brad}. Though the diquark model is unable to reproduce the data for the neutron charge density at small $b$, still it is better than the AdS/QCD Model-I predictions presented in Ref.\cite{CM3}.  In Fig.\ref{nucleon_den}(c), one can notice a negatively charged core surrounded by a  ring of positive charge density   for neutron. 
In Figs. \ref{quark_den}(a) and \ref{quark_den}(b), we show the charge and anomalous magnetization densities for $u$ quark. Similarly for the $d$ quark, the transverse densities are shown in Figs. \ref{quark_den}(c) and \ref{quark_den}(d). The charge density for  $d$ quark in diquark model deviates at small $b$ form the two global parameterizations of Kelly and Bradford but is in excellent agreement for $u$ quark. Again, diquark model provides better result than the AdS/QCD Model-I results presented in Ref.\cite{CM3}. The anomalous magnetization densities in both $u$ and $d$ quarks in the LF diquark model match very well with the parameterizations

For transversely polarized nucleon, the charge density in the transverse plane is given by \cite{vande}
\be
\rho_T(b)=\rho_{ch}-\sin(\phi_b-\phi_s)\frac{1}{2M b}\rho_m\label{trans_pol},
\ee
where $M$ is the mass of nucleon and the transverse polarization of the nucleon is given by
 $S_\perp=(\cos\phi_s \hat{x}+\sin\phi_s\hat{y})$ and the transverse impact parameter $b_\perp=b(\cos\phi_b \hat{x} +\sin\phi_b\hat{y})$. Without loss of generality, the polarization of the nucleon is taken along $x$-axis ie., $\phi_s=0$. The second term in Eq.(\ref{trans_pol}), provides the deviation from circular symmetry of the unpolarized charge density \cite{vande}. The charge densities for the transversely polarized proton and neutron have been shown in Fig. \ref{flavor_T}(a) and \ref{flavor_T}(b). We show the $u$ and $d$ quark charge  densities for the transversely polarized nucleon in Fig.\ref{flavor_T}(c) and \ref{flavor_T}(d). Again, the densities in the LF diquak model are in  good agreement with the global parameterizations. 
 The comparison of charge densities for the transversely polarized and unpolarized 
proton is shown in Fig. \ref{nucleons_top}( a) and (b) and the similar plots for neutron are shown in Fig.\ref{nucleons_top} (c) and (d).  For the nucleons polarized along the $x$ direction, the 
charge densities get shifted towards negative $b_y$ direction for proton. The deviation is much larger for the neutron compared to the proton. Due to large anomalous magnetic moment which produces an induced electric dipole moment in $y$-direction, the distortion shows a dipolar pattern in the case of neutron \cite{vande}. The behaviors are in agreement with  the results reported in Refs. \cite{vande,miller10,silva}.

We compare the up quark charge densities for the transversely polarized and unpolarized nucleon 
in Fig.\ref{flavor_top} (a) and (b) and the similar plots for $d$ quark are shown in Fig. \ref{flavor_top} (c) and (d). The deviation or distortion from the symmetric unpolarized density is more for down quark than the up quark.  For the nucleons polarized in $x$ direction,  the charge density shifts  towards positive $b_y$ direction for $d$ quark but in opposite direction  for the $u$ quark.  

\section{Summary and  conclusions}\label{con}
The parameters in a light front quark-diquark model of nucleons \cite{Gut} are found to be inconsistent with the experimental data. We have re-evaluated the parameters in this model for the AdS/QCD scale parameter $\kappa=0.4066$ GeV which was previously obtained by fitting the nucleon form factors in soft wall AdS/QCD\cite{CM1,CM2}. 
The new parameters reproduce the experimental data for the nucleon form factor quite well for a wide range of $Q^2$ values. We have compared our results with AdS/QCD soft wall model.
Then,
 we have evaluated the GPDs for $u$ and $d$ quark in proton for both zero and nonzero skewness in the light front quark-diquark model.  
We observed that  all the GPDs  in  the momentum space as well as  in the transverse impact parameter space are more or less in agreement with the results of AdS/QCD. We have calculated the GPDs for nonzero skewness in the DGLAP region(i.e., for $x>\zeta$).  
The peaks of the distributions move to higher values of $x$ for fixed $\zeta$  with increasing  $-t$. In the model, the behaviors of the GPD $H$ in impact parameter space for $u$ and $d$ quarks are quite different when plotted in both $x$ and $b$. The difference in the behaviors of  $E(x,\zeta,t)$ for  $u$ and $d$ quarks are clearly observed  when plotted against $x$.
For nonzero skewness, we have also shown the GPDs in longitudinal impact parameter space $\sigma$. We found that both the GPDs $H$ and $E$ for $u$ and $d$ quarks in $\sigma$ space show diffraction patterns. Similar diffraction patterns also have been  observed  in some other models. In case of $E(x,\sigma,t)$, the qualitative nature of the diffraction pattern is same for both $u$ and $d$ quarks. For $H^d(x,\sigma,t)$, the diffraction pattern is  observed only for small $-t$ values; as $-t$ increases, a dip appears at the centre( at $\sigma=0$).

Finally, we have presented the transverse charge and magnetization densities for nucleon and also for individual  quarks.  The results are consistent with two phenomenological parameterizations\cite{kelly04,brad}. The unpolarized densities are axially symmetric whereas the charge densities get distorted for transversely polarized nucleon. The charge density is shifted along $y$ direction if the nucleon is polarized along $x$ direction. The charge density for transversely polarized neutron shows a dipole pattern. The shift of charge density of $u$ quark for transversely polarized nucleon from the symmetric unpolarized density is smaller than $d$ quark and in opposite direction. 

\begin{figure*}[bhtp]
\begin{minipage}[c]{0.98\textwidth}
\small{(a)}
\includegraphics[width=7.5cm,height=5.5cm,clip]{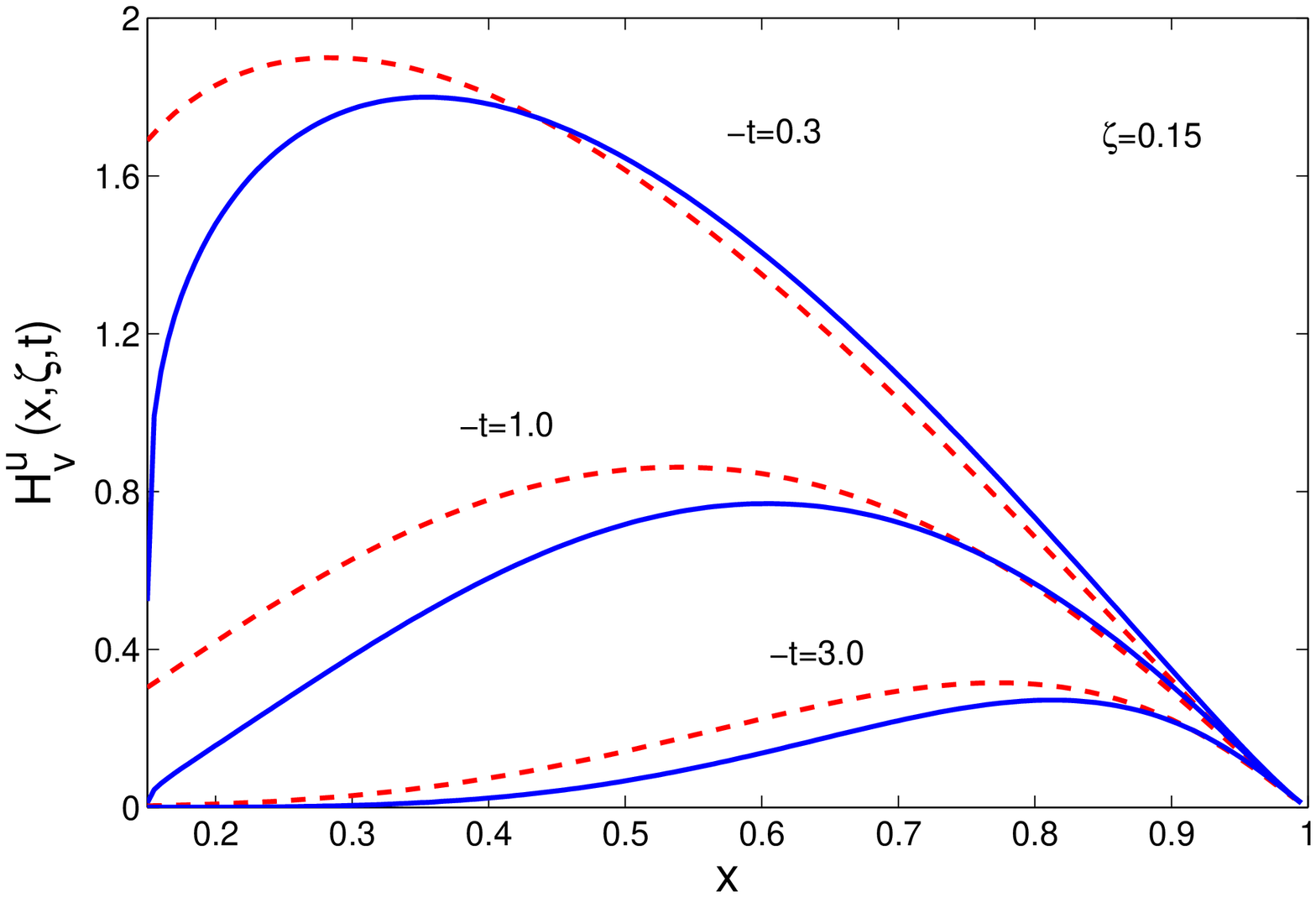}
\hspace{0.1cm}%
\small{(b)}\includegraphics[width=7.5cm,height=5.5cm,clip]{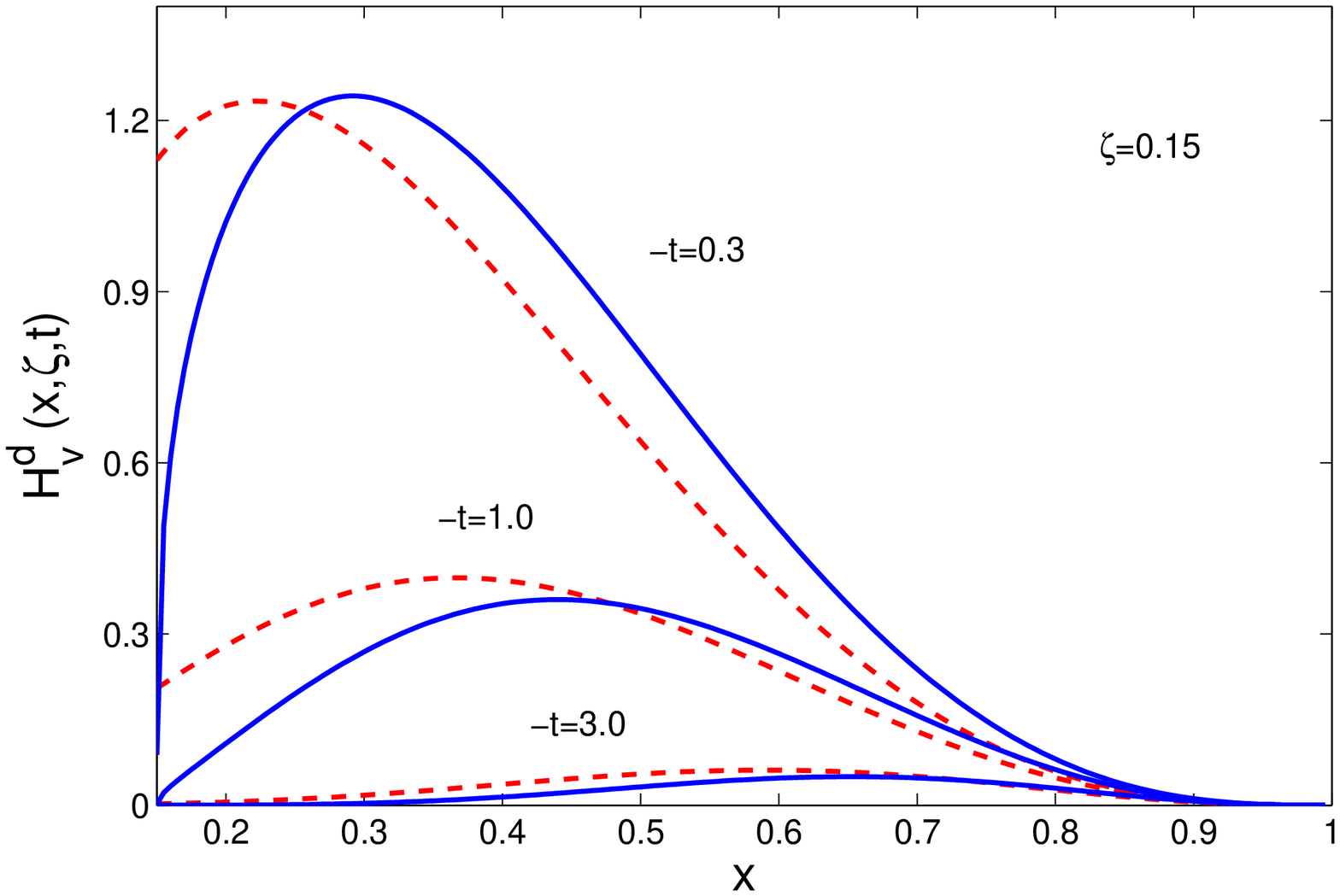}
\end{minipage}
\begin{minipage}[c]{0.98\textwidth}
\small{(c)}
\includegraphics[width=7.5cm,height=5.5cm,clip]{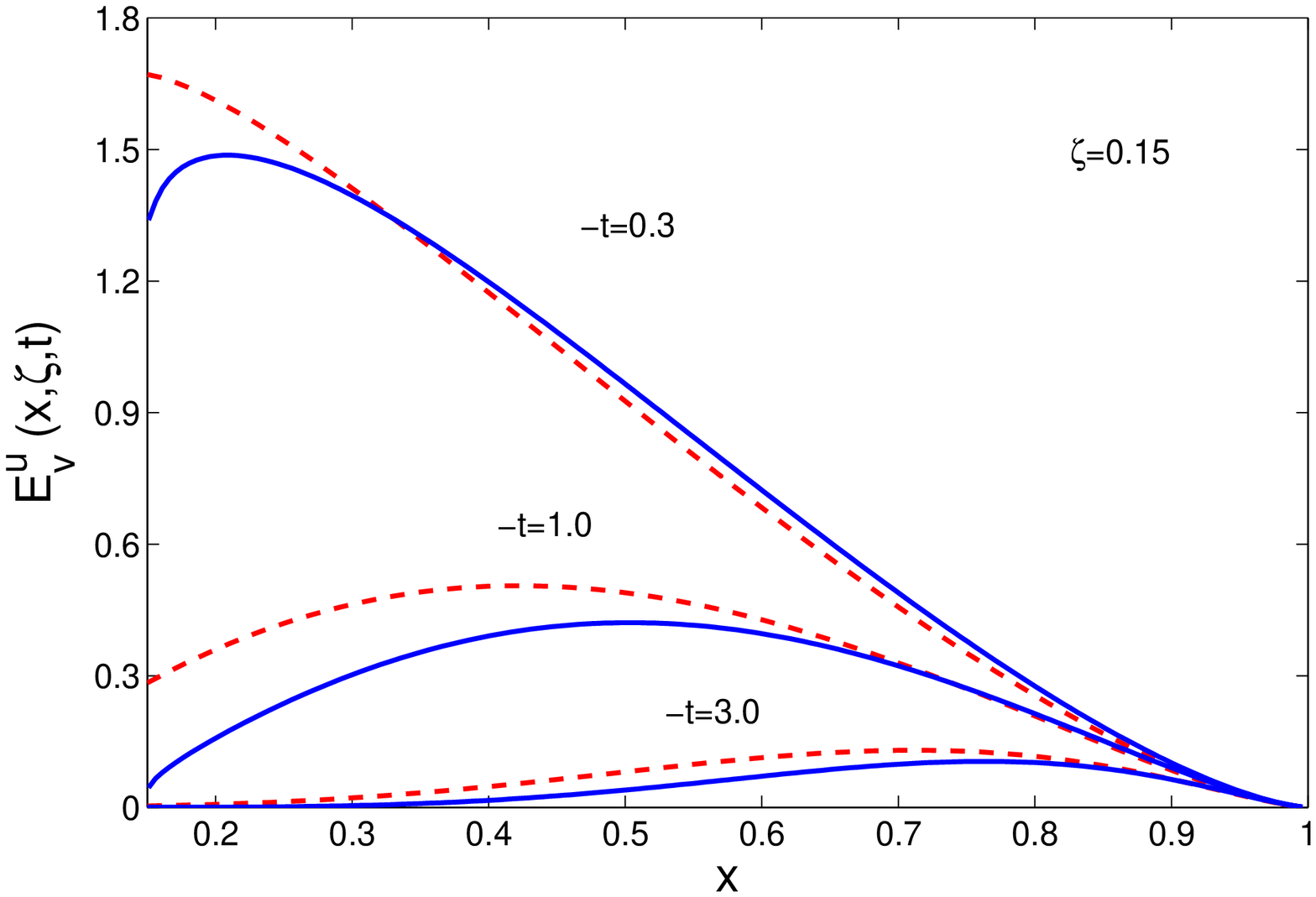}
\hspace{0.1cm}%
\small{(d)}\includegraphics[width=7.5cm,height=5.5cm,clip]{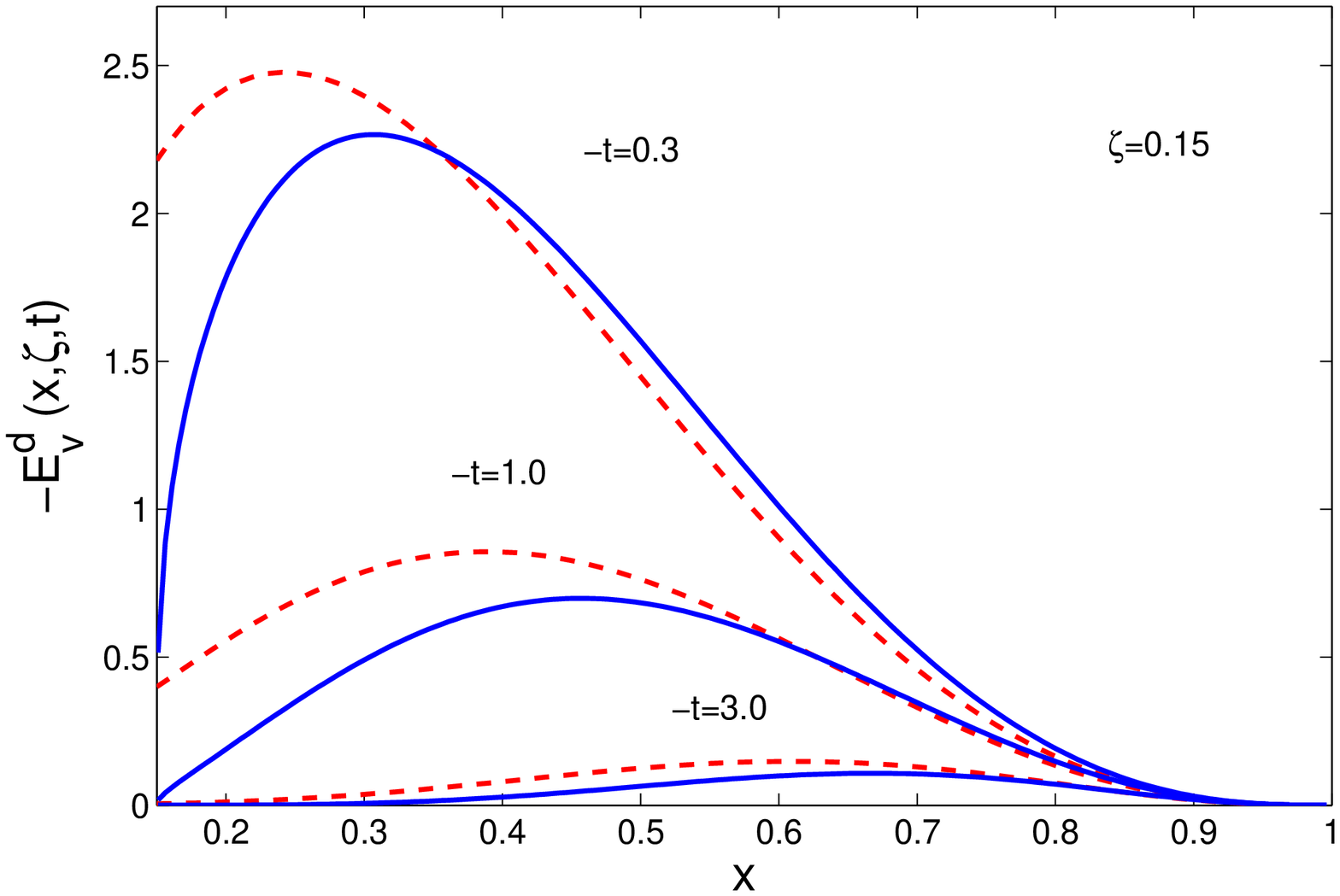}
\end{minipage}
\caption{\label{GPDs_z_com1}(Color online) Plots of (a) $H^u(x,\zeta,t)$ vs $x$ (b) the same as in (a) but for $d$ quark (c) $E^u(x,\zeta,t)$ vs $x$ (d) the same as in (c) but for $d$ quark for fixed $\zeta=0.15$ and three different values of $ -t=0.3 {\rm GeV^{2}}, ~1.0  {\rm GeV^{2}} ~{\rm and}~ 3.0  {\rm GeV^{2}}$. Red dashed lines represent the results calculated using double distribution for the same values of $\zeta$ and $-t$. }
\end{figure*} 
\begin{figure*}[htbp]
\begin{minipage}[c]{0.98\textwidth}
\small{(a)}
\includegraphics[width=7.5cm,height=5.5cm,clip]{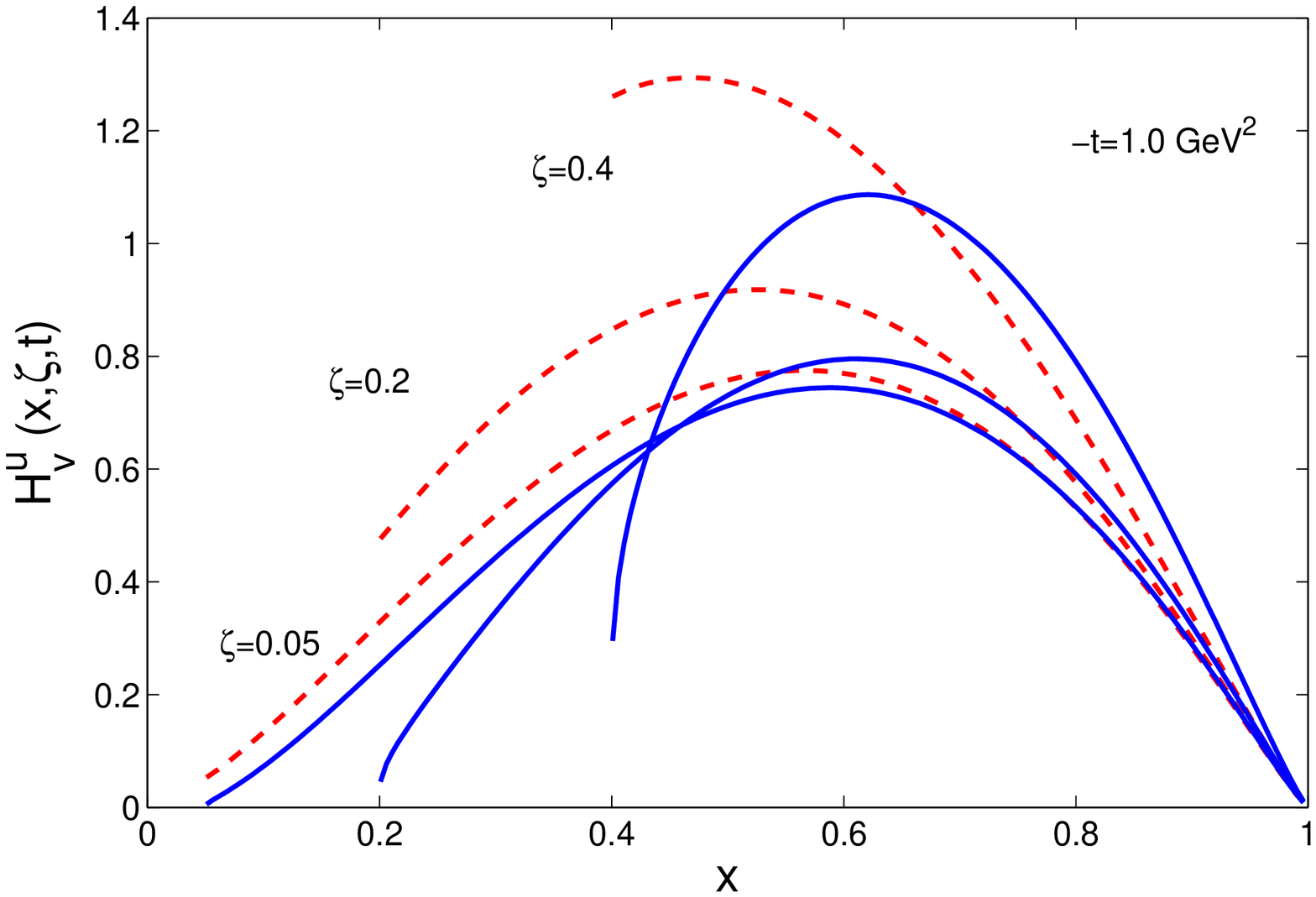}
\hspace{0.1cm}%
\small{(b)}\includegraphics[width=7.5cm,height=5.5cm,clip]{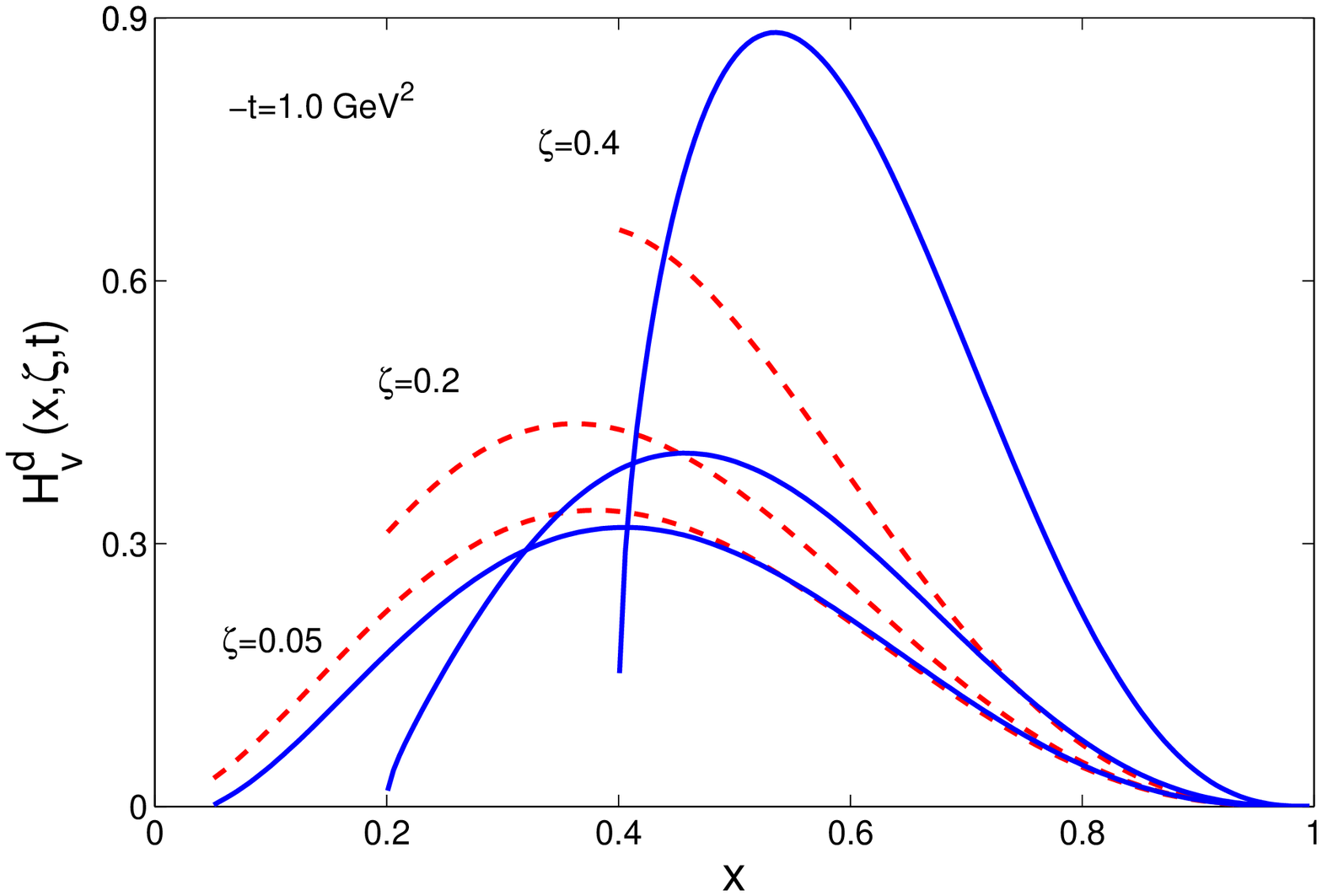}
\end{minipage}
\begin{minipage}[c]{0.98\textwidth}
\small{(c)}
\includegraphics[width=7.5cm,height=5.5cm,clip]{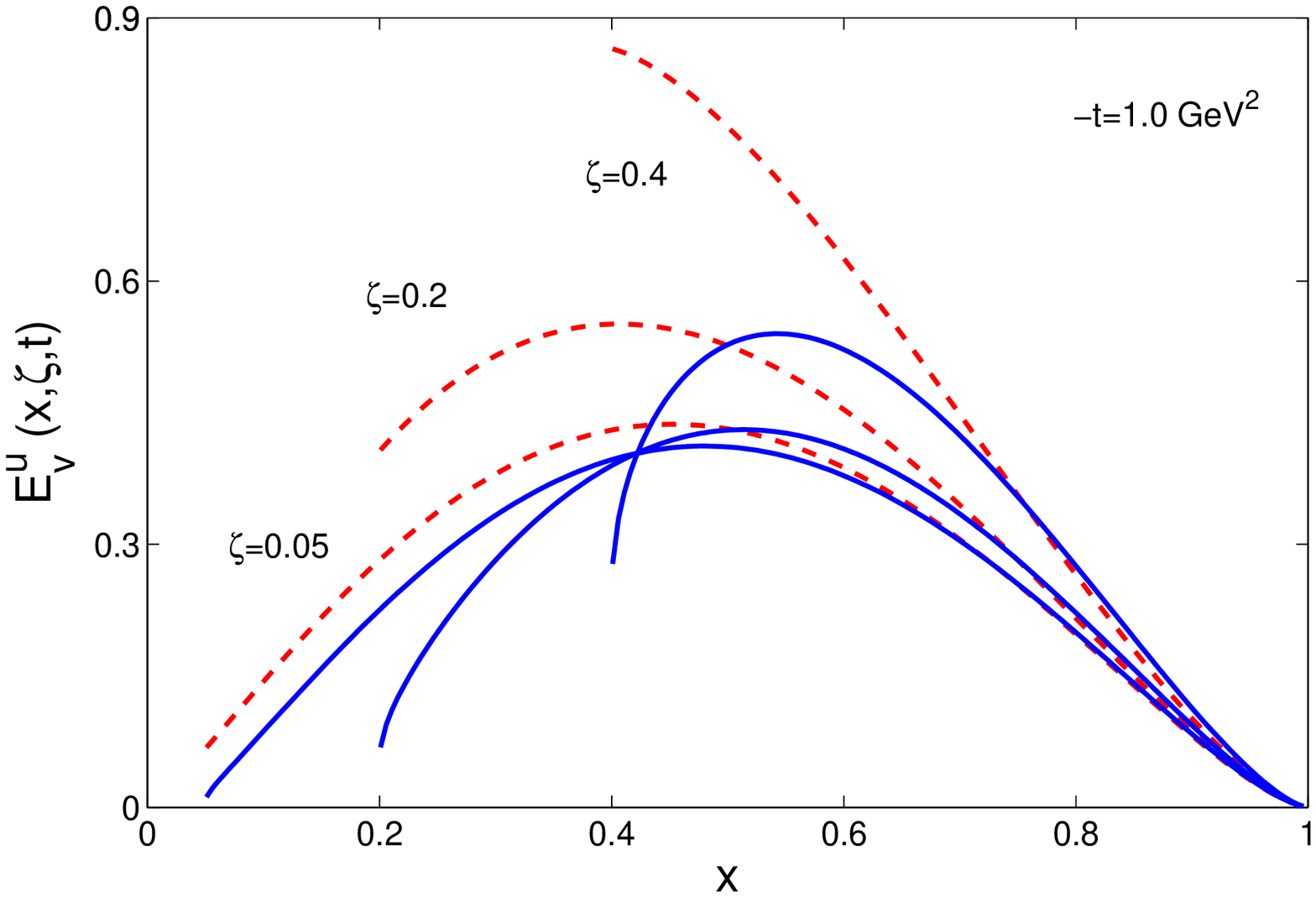}
\hspace{0.1cm}%
\small{(d)}\includegraphics[width=7.5cm,height=5.5cm,clip]{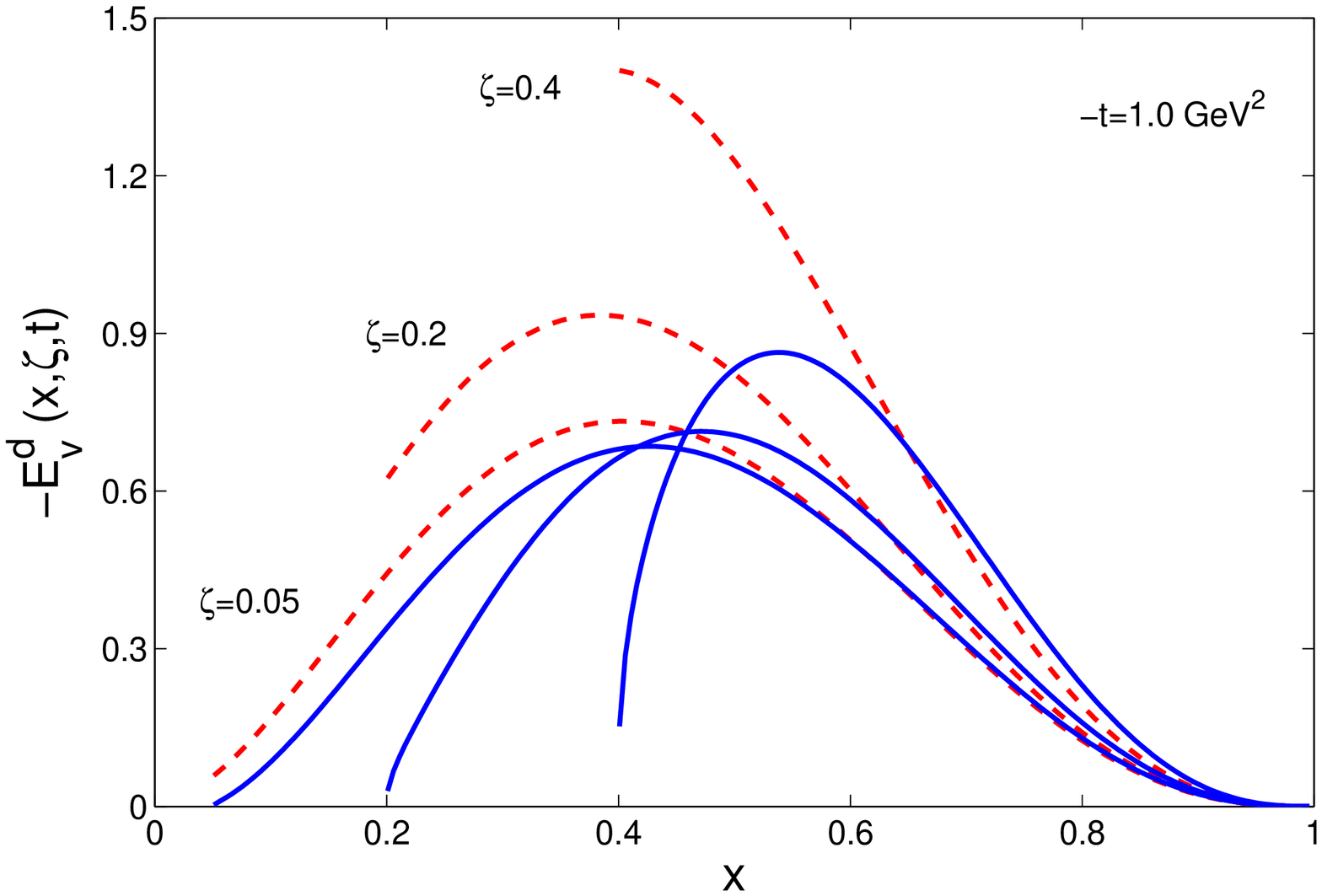}
\end{minipage}
\caption{\label{GPDs_z_com2}(Color online) Plots of (a) $H^u(x,\zeta,t)$ vs $x$ (b) the same as in (a) but for $d$ quark (c) $E^u(x,\zeta,t)$ vs $x$ (d) the same as in (c) but for $d$ quark for fixed $-t=1.0~ {\rm GeV^2}$  and three different values of $\zeta=0.05,~0.2~,~0.4$. Red dashed lines represent the  results calculated using double distribution for same values of $-t$ and  $\zeta$. }
\end{figure*} 
\appendix
\section{ comparison of GPDs in quark-diquark model  with a double distribution(DD) model }
The GPDs for $\zeta=0$ admit a density interpretation when one takes the Fourier transform to the impact parameter space but in experiments, $\zeta$ is always nonzero. In recent past, there have been a lot of works to model GPDs with nonzero skewness by modeling relevant DDs \cite{rad1,rad2}.
 In this section, we compare our results for nonzero skewness with the GPDs  modeled from the Double Distributions (DD)\cite{mueller,ji,rad}. 
 The GPDs have an integral representation in terms of the double distributions $f(\beta,\alpha,t)$. For the valance quarks, the GPDs can be written as
\be
F_v^q(x,\zeta,t)=\int_0^1 d\beta\int_{\beta-1}^{1-\beta}d\alpha~\delta(x-\beta-\zeta \alpha)~f_v^q(\beta,\alpha,t),\nonumber\\\label{skew}
\ee
where $F_v^q=H_v^q$, $E_v^q$. Here, we use  the factorized DD ansatz for the GPDs as suggested by 
Musatov and Radyushkin \cite{musa}  
\be
f_v^q(\beta,\alpha,t)=F_v^q(\beta,0,t)h(\beta,\alpha),\label{pro}
\ee
where the weight function $h(\beta,\alpha)$ generates the skewness dependence of the GPDs and satisfies the normalization condition
\be
\int_{-1+|\beta|}^{1-|\beta|}h(\beta,\alpha)d\alpha=1.
\ee
The general form of the profile function is given by \cite{musa}
\be
h^{(N)}(\beta,\alpha)=\frac{\Gamma(2N+2)}{2^{2N+1}\Gamma^2(N+1)}\frac{[(1-|\beta|)^2-\alpha^2]^N}{(1-|\beta|)^{2N+1}},
\ee
where the parameter $N$ governs the width of the function. We use the $N=2$ profile function. The similar profile function for $N=2$ has been used in many phenomenological model of DVCS and exclusive meson production \cite{rev,kroll1,kroll2,diehldd}.
Inserting Eq. \ref{pro} in the Eq. \ref{skew}, with the help of delta-function one can perform the integral over $\alpha$ and obtains
\be
F_v^q(x,\zeta,t)&=&\frac{3}{4\zeta^3}\int_{\beta_{min}}^{\beta_{max}}\frac{d\beta}{1-\beta}~F_v^q(\beta,0,t)\nonumber\\
&&\Big(1+\zeta-\frac{1-x}{1-\beta}\Big)\Big(\frac{1-x}{1-\beta}-1+\zeta\Big),
\ee
for $x>\zeta$, the integration boundaries are
\be
\beta_{min}&=&x-\frac{\zeta}{1-\zeta}(1-x)\nonumber\\
\beta_{max}&=&x+\frac{\zeta}{1-\zeta}(1-x).
\ee

In Fig. \ref{GPDs_z_com1} and Fig. \ref{GPDs_z_com2}, we show the skewness dependent GPDs calculated using double distribution parameterization and compare with the results directly calculated in the quark-diquark model.  Fig.\ref{GPDs_z_com1} suggests that  for small $\zeta$ and large $-t$, the results of double distribution are more or less in agreement with the diquark model results,  while Fig.\ref{GPDs_z_com2} shows that at moderate or high values of skewness $\zeta$, only at higher $x$ the two models agree but otherwise the agreement is  lost.



\end{document}